\documentclass{aastex61}
\usepackage{amsmath}
\usepackage{rotating}
\usepackage{multirow}
\usepackage[normalem]{ulem}

\newcommand\aastex{AAS\TeX}

\received{\today}

\submitjournal{ApJ}
\shorttitle{\aastex\ Photopolarimetric characteristics of uniformly cloudy BDs}
\shortauthors{Sanghavi et al.}

\begin{document}

\title{Photopolarimetric characteristics of brown dwarfs, Part I: uniform cloud decks}

\correspondingauthor{Suniti Sanghavi}
\email{sanghavi@jpl.nasa.gov}

\author[0000-0003-0754-9154]{Suniti Sanghavi}
\affil{Jet Propulsion Laboratory/Caltech, 4800 Oak Grove Dr, Pasadena, CA 91109, USA}
\affil{California Institute of Technology, 1200 E California Blvd, Pasadena, CA 91125, USA}


\author{Avi Shporer}
\affil{MIT Kavli Institute, Massachusetts Institute of Technology,
77 Massachusetts Ave, Cambridge, MA 02139, USA}
\begin{abstract}
This work is a theoretical exploration for facilitating the interpretation of polarimetric observations in terms of cloudiness, rotational velocities and effective temperatures of brown dwarfs (BDs). An envelope of scatterers like free-electrons, atoms/molecules, or haze/clouds affects the Stokes-vector of radiation emitted by oblate bodies. Due to high rotation rates, BDs can be considerably oblate. We present a conics-based radiative transfer (RT) scheme for computing the disc-resolved and disc-integrated polarized emission of an oblate BD or extrasolar giant planet (EGP) bearing homogenous or patchy clouds. Assuming a uniform grey atmosphere, we theoretically examine the photopolarimetric sensitivity to its scattering properties like cloud optical thickness and grain-size, concurrently with BD properties, like oblateness, inclination and effective temperature, which are all treated as free parameters. Additionally, we examine the potential effects of gravitational darkening (GD), revealing that it could significantly amplify disc-integrated polarization. GD imparts a non-linear inverse temperature-dependence to the resulting polarization.

Photopolarimetric observations are sensitive to oblateness and inclination. The degree-of-polarization (DoP) increases in response to both, making it potentially useful for assessing the spatial orientation of the BD.
Under our model assumptions, increasing droplet size in optically thick clouds causes a blue-ward shift in near-infrared (NIR) colors of BDs --- interesting in view of the observed J-K brightening in L/T transition. For large cloud grains, polarization decreases sharply, while transmitted intensity shows a steady increase. BD polarization is thus a potential indicator not only of the presence of clouds but also provides information on cloud grain size.


\end{abstract}




\section{Introduction} \label{sec_intro}

{
The identification of clouds in brown dwarf (BD) atmospheres and the characterization of cloud properties is vital for gaining a better understanding of phenomena like the L/T transition \citep{geballe2002toward, geballe2003transition, burgasser2006hubble, burrows2006and}, which is hypothesized to be caused by the gradual fragmentation and sedimentation of clouds \citep{burgasser2002evidence, tsuji2004dust, marley2010patchy} as BDs transition from L- to T-dwarfs. This change in the nature of clouds, combined with concurrent changes in atmospheric composition are expected to lead to a relative brightening of bluer colors in the near-infrared (NIR) spectra of BDs in L/T-transition \citep{knapp2004near, stephens20090}.

The hypothesis of cloud fragmentation has been supported by observations of photometric variability \citep{artigau2009photometric, apai2013hst, radigan2012large, radigan2014strong} { and Doppler imaging and spectroscopy \citep{crossfield2014global}} on L/T-transition dwarfs. Even more information on clouds can be obtained from spectro/photopolarimetric measurements such as those made by \cite{menard2002optical, osorio2005optical, goldman2009polarisation, Zap2011ApJ, miles2013linear, 2017MNRAS.466.3184M, manjavacas2017testing}. However, for BDs, the interpretation of such measurements using radiative transfer models (RTMs) is complicated by the fact that the polarimetric signature is affected not only by patchy clouds but also the rotationally flattened (oblate) shape of BDs, as shown by \cite{de2011characterizing}. 
{Until recently,} RTMs \citep{sengupta2009multiple, marley2011probing, de2011characterizing} were able to model either cloud patches on spherical BDs or uniformly cloudy oblate BDs. { \cite{stolker2017polarized} presented a Monte-Carlo based approach to deal with both oblateness and patchy clouds.} {In this work, we present the first analytic framework to simultaneously simulate oblateness and patchiness for any given viewing geometry for BDs and EGPs. This is essential for a full understanding of polarimetric observations of BDs, for which both oblateness and patchiness due to clouds are important features.}

We employ a conics-based method that allows the simultaneous simulation of patchy clouds on oblate BDs at any given inclination angle. For the computation of radiances on the BD, we use the 1D analytic RTM vSmartMOM \citep{sanghavi2014vsmartmom}, allowing a fast and accurate, full microphysical representation of cloud scattering using Mie theory \citep{sanghavi2014revisiting, sanghavi2015adaptation}. We use this to examine the photopolarimetric response of BDs to cloud properties like optical thickness and effective grain size (for a given log-normal distribution) in addition to geometric properties of the BD like oblateness and inclination angle.}

{This paper is the first in a series in which we aim to develop an RTM framework} for interpreting spectropolarimetric observations of BDs, which vary as a function of geometric characteristics like oblateness and inclination, but also of physical properties of the BD atmosphere like opacity and cloudiness. {We currently assume a grey, conservatively scattering atmosphere, and treat all BD and cloud parameters as free parameters without assuming any implicit correlation between them. While this provides a clear understanding of the underlying RT processes connecting the parameter state space to the observations, we forfeit the ability to make conclusive comparisons 
with BD observations. 

In reality,} opacity due to gaseous absorption in a BD atmosphere depends on its effective temperature,  surface gravity and metallicity \citep{molliere2015model, mordasini2015global, allard1997model, kirkpatrick2005new}. Since these opacities vary with wavelength, they affect the photospheric depth in different parts of the spectrum \citep{charnay2017self, freytag2010role, freedman2008line}. In the absence of clouds, the effect is thus to truncate the atmospheric column differently at different wavelengths, giving rise to the extreme cases where 
\begin{enumerate}
    \item The whole { gaseous} column extending from the top-of-atmosphere (TOA) to the bottom-of-atmosphere (BOA) of the BD is non-absorbing, and contributes to the photopolarimetric signal exclusively by scattering the radiation emitted at the BOA,  
    \item No light emitted at the {BOA} escapes the TOA. {The atmosphere is fully opaque due to gaseous absorption, causing the entire gaseous column to heat up to $\sim T_\mathrm{eff}$ and emit radiation all the way up to nearly the TOA.}
\end{enumerate}
The true photosphere at a given wavelength lies between these extremes, and has {been determined} for purposes of RT modeling to lie between 0.1-5 bar \citep{sengupta2009multiple, sengupta2010observed,de2011characterizing, marley2011probing}. While changes in photospheric depth may not greatly affect Rayleigh scattering in the IR, the photopolarimetric signal at optical and shorter NIR wavelengths is sensitive to it due to the $\lambda^{-4}$ dependence {\citep{bohren2008absorption}} of the Rayleigh scattering cross-section, and hence must be accounted for.

Since the cutoff depth varies from case to case, and with wavelength, we consider it useful to start with a simplified representation by means of a free parameter $p_\mathrm{cutoff}$ representing the pressure-level for which the BD atmosphere can be assumed to be completely opaque when $p>p_\mathrm{cutoff}$. {Above this atmospheric level, i.e., when $p<p_\mathrm{cutoff}$, we assume no gaseous or particulate absorption.} Instead of covering the full range of $p_\mathrm{cutoff}$, we currently examine the two extreme cases of opacity enumerated above. This assumption informs the Rayleigh scattering optical thickness attributable to gaseous species that make up the atmosphere. We make the further simplification of considering only clouds that occur above the photosphere at all wavelengths. { This ensures that the geometric thickness of the cloud is spectrally constant, so that only grain size affects the spectral dependence of the optical thickness of the cloud.} To characterize BD clouds, we consider several monochromatic wavelengths ranging from the visible to the thermal IR, with emphasis on the NIR, for eventual application to {direct-imaging} measurements made by WIRC+Pol (Tinyanont et al., 2018, in prep.) at Palomar Observatory.

In Section \ref{sec_model}, we briefly describe our 1D plane-parallel RTM, lay out our atmospheric setup, and provide the detailed formalism for simulating the disc-resolved Stokes parameters $I$, $Q$ and $U$ over an oblate (assumed ellipsoidal) surface, along with the integration procedure used to obtain the corresponding disc-integrated signal. We outline the main differences between our method and the current state-of-the-art.

A uniform cloud envelope, represented by a uniform vertical distribution along a given latitude, and constant optical thickness and grain size throughout the BD, would lead to a constant signal as the BD rotates about its axis. In contrast, {longitudinally asymmetric} fragmented clouds would feature patches that follow the BD's rotation around its axis, leading to a variable signal, allowing the identification of cloudy weather patterns.
Our RT framework can model uniform as well as patchy atmospheres. In this paper, we focus mainly on uniform atmospheres. Patchy clouds will be the focus of the next paper in this series. 

{As BDs age, they spin increasingly faster reaching exceptionally high rotational velocities due to a lack of efficient braking mechanisms \citep{mohanty2003rotation, bouvier2013observational}. 
 As a result, several BDs are expected to be highly oblate \citep{basri2000observations, sengupta2001probing}. {Since oblateness affects polarization, such BDs are of special interest to us.} Gravitational darkening (GD) \citep{von1924radiative} is an important feature of fast-rotating radiative stars, which has been shown by \cite{claret2000studies} to be applicable also in the low mass limit, {including BDs}. The original theoretical formalism of GD developed for radiative stars has been comprehensively revised by \cite{lara2011gravity} to be applicable to convective bodies like BDs.} In Sec.\,\ref{sec_gravdark}, we use this revised formalism to examine how GD causes the effective temperature to vary latitudinally with effective surface gravity, ${g}_\mathrm{eff}$, which increases as the radius of the body decreases from the equator towards the poles. We show for a uniform semi-infinite Rayleigh scattering atmosphere that the temperature inhomogeneity imposed by GD strongly amplifies the disc-integrated polarization of light emitted by the oblate body.
 
In Sec.\,\ref{sec_BD}, we simulate the disc-integrated photopolarimetric signal of the thermal emission of a BD with a uniformly cloudy atmosphere, and separately examine its dependence on oblateness, $\eta$, in Sec.\,\ref{sec_uni_oblate}, and the inclination of the view direction, $\theta_\mathrm{incl}$, with respect to the axis of rotation of the BD in Sec.\,\ref{sec_uni_incl}. We show in Sec.\,\ref{sec_orient} that for sufficiently oblate BDs, accurate measurements of both $Q$ and $U$ can, in principle, allow the determination of the absolute spatial orientation of the spin axis.

Sec.\,\ref{sec_cloud} deals specifically with clouds, making use of multiple individual wavelengths to analyze their scattering characteristics (Sec.\,\ref{sec_uni_spec_scat}), with a focus on the cloud optical thickness in Sec.\,\ref{sec_scat_tau} and grain size in Sec.\,\ref{sec_scat_r0}.

{ In order to make the insights obtained from our RT studies meaningful from an observational perspective, we have discussed our modeling results in the context of current observations in Sec.\,\ref{sec_obs}. We report the effect of the clouds in our simplified atmosphere on the color magnitude diagram of BDs under the aforementioned assumption of a grey atmosphere with freely varying atmospheric parameters in Sec.\,\ref{sec_LT}. Our modeling results are discussed in the context of current polarimetric observations in Sec.\,\ref{sec_pol}, which we compare with the upper limit of polarization for various uniform cloud scenes on BDs of varying oblateness, computed at the central wavelengths of commonly used filters, viz., R (658\,nm), I (806\,nm), Z (900\,nm), Y (1020\,nm), J (1220\,nm), H (1630\,nm) and K (2190\,nm). Since our assumptions have limited fidelity with the true atmospheres of BDs, these comparisons are not intended to be conclusive, but only reflect the current state of our model assumptions. They are expected to be progressively  updated as we incorporate new capabilites for a more accurate representation of BD atmospheres. We summarize our conclusions in Sec.\,\ref{sec_conc}.}

\section{Radiative transfer modeling}\label{sec_model}
{
Unpolarized light like the {photospheric} thermal emission of a BD is made up of mutually orthogonal electromagnetic fields of equal amplitude oscillating perpendicular to the direction of propagation. When light changes direction due to scattering (e.g., by gaseous atoms/molecules or cloud condensates), one component of the field, e.g., that oscillating perpendicular to the scattering plane in the case of small particles, is preferred relative to the other, leading to its polarization. The resulting difference in the intensities of the two components depends on the properties of the scattering particles, whose number density, geographical distribution, microphysical size and dielectric characteristics define the scattering field. It is thus conceivable that, in addition to photometry, polarimetric measurements using advanced instruments like SPHERE \citep{beuzit2008sphere}, GPI \citep{macintosh2008gemini}, WIRC+Pol (Tinyanont et al., 2018, in prep.), etc.\ would provide more information on the nature of clouds in the atmosphere of a BD.

{The polarimetric signature of BDs}, however, is affected not only by the properties of clouds, but also by the (flattened) shape of the BD itself. As a result of radiative cooling, BDs shrink with age \citep{burrows2011dependence}. To conserve angular momentum, the shrinkage is compensated for by rapid rotation rates, causing significant flattening about the rotation axis. It is known that scattering on oblate bodies leads to a residual polarization of its disc-integrated radiation \citep{chandrasekhar1946radiative, chandrasekhar1967ellipsoidal, harrington1968intrinsic}. This is also true for BDs and EGPs, as  demonstrated by the radiative modeling studies of \cite{sengupta2009multiple,sengupta2010observed,marley2011probing} and \cite{de2011characterizing}. Direct comparison with the results of these works is difficult because no explicit mention has been made in them of the optical thicknesses, size-distributions and vertical profiles of the scatterers involved. We have attempted here to tabulate all optical thicknesses and vertical distributions used. Our cloud grains have log-normal size-distributions with clearly-defined parameters mentioned in the text.

The approach of \cite{sengupta2009multiple,sengupta2010observed,marley2011probing} is based on a solution of the RT equation expressed in spherical coordinates in order to deal with oblateness. While this is likely the most elegant way of simulating oblate bodies with uniform atmospheres, a drawback of this approach is its inability to deal with arbitrary inhomogeneities. \cite{de2011characterizing} employ {a modification \citep{karalidi2012modeled} of the original method of \cite{stam2006integrating} involving angular integration over the atmospheric scattering matrix to deal with inhomogeneities in a spherical object}, but have not shown the use of the same method to deal with oblateness and inhomogeneity simultaneously. {As a result, their simulations of oblate bodies are restricted to uniform atmospheres.} The recently published work of \cite{stolker2017polarized} uses Monte Carlo simulations to model patchy clouds on oblate bodies in an edge-on geometry. 
In contrast to Monte Carlo simulations, the analytic nature of our modeling framework allows for high computational speed and efficiency across all modeled scenarios. 

Our conics-based approach allows us to carry out disc-integration of the 1D plane-parallel contributions of distinct points covering an object of any given oblateness. By combining the individual plane-parallel contributions of atmospheric inhomogeneities covering the observed part of spherical/oblate bodies for a given viewing geometry, we are able to represent patchy clouds of any given shape and distribution on BDs/EGPs. 

In this theoretical exploration, the first of a series that will {incrementally} examine the different contributions to observed polarization, we consider monochromatic photopolarimetric effects of BDs enveloped in a uniform sheath of pure scatterers (simulated using Mie theory) in the absence of gaseous absorbers. This establishes independence from the height of the cloud, and from spectroscopic effects like Doppler broadening, pressure broadening, temperature dependence, etc., of absorption lines, allowing us to focus entirely on the effects of key cloud parameters like optical thickness and grain size. To the same end, we assume no horizontal temperature gradients over the BD disc, apart from those due to gravitational darkening (GD). 
Cloud optical thickness and grain size have been treated as free parameters independent of the temperature of the BD, allowing their individual scattering effects to be isolated from those of gaseous/cloud absorption. The inclusion of absorbers would impart a temperature dependence to the resulting spectropolarimetric signal, as the relative abundance of different absorbing {species} in the atmosphere varies with the {effective} temperature of the BD. In future work, we will build upon the {generalized} studies presented here by adding {type-specific opacities} and related vertical features. 

In order to model the radiative properties of a BD, we construct a (LAT, LON) grid over its surface, and assume each gridpoint to bear a plane-parallel atmosphere, which can be simulated independently of its neighboring gridpoints. We use the RTM vSmartMOM \citep{sanghavi2014vsmartmom} at each of these gridpoints to compute thermal emission at the effective temperature of the BD. An effective temperature, $T_\mathrm{eff}=1500\,$K, has been considered for this part of our study, roughly representative of a BD in the L/T-transition range. GD leads to a temperature-dependent latitudinal variation of the BD emission, so that the disc-integrated polarization varies with effective temperature as will be shown for $T_\mathrm{eff} = 2000\,$K (roughly representative of an early L-dwarf) and $T_\mathrm{eff} = 500\,$K (roughly representative of a late T-dwarf).

The vertical structure and composition of the atmosphere at each gridpoint is described in Sec.\,\ref{sec_atm}, followed by a description of the local 1D plane-parallel RT characteristics in Sec\,\ref{sec_local} and our disc-integration approach in Sec.\,\ref{sec_da}. The reader who is more interested in the results of our RTM than the RT methodology may skip over Sections \ref{sec_local} and \ref{sec_da} without loss of continuity.}

\subsection{{The model atmosphere}}\label{sec_atm}
{
{ In previous works by \cite{sengupta2009multiple, sengupta2010observed, marley2011probing}, the atmospheric profiles are as described in \cite{marley2000role, ackerman2001precipitating}. We take a simplified approach that merely defines the pressure at the BOA for a given effective temperature, thereby allowing an estimation of the full column-integrated gaseous Rayleigh-scattering optical thickness of the BD atmosphere for a given effective temperature, $T_\mathrm{eff}$, and surface gravity, $g$. This is the pressure level below which the gas phase ceases to exist, and radiation at all wavelengths can be assumed to be unpolarized, with subsequent Rayleigh/cloud scattering imparting the transmitted radiation a polarization signature specific to the given viewing geometry and atmospheric composition.} The atmospheric column above the BOA is assumed to consist of gaseous matter in hydrostatic equilibrium. Assuming a BD 20 times the mass of Jupiter ($M_\mathrm{BD}=20M_\mathrm{Jup}$) and of the same size as Jupiter ($R_\mathrm{BD}=R_\mathrm{Jup}$), we use the relation
\begin{equation}\label{eq001}
H_\mathrm{scale}=\frac{RT_\mathrm{BOA}}{\overline{\mu}g}
\end{equation}
to obtain the scale height $H_\mathrm{scale}$, where $R$ is the ideal gas constant, $g$ is the surface gravity of the BD, and $\overline{\mu}\sim2.2\,$gram (assuming solar metallicity) is the mean molar mass of the BD atmosphere. We compute the surface gravity as $g=\frac{GM_\mathrm{BD}}{R^2_\mathrm{BD}}=595.8\,$m/s$^2$ ($\log{g}=4.775$) to obtain $H_\mathrm{scale}=11.43\,$km. 
Assuming the heat transport in the atmosphere to occur through adiabatic transport of air parcels from the BOA to the TOA, the pressure $p$ and temperature $T$ at any given atmospheric level are related as $p^{1-\gamma}T^\gamma=\mathrm{const.}$, where $\gamma$ is the adiabatic index. To determine the constant, we consider the pressure to be $p_0$ at a BOA temperature of $T_\mathrm{BOA}=T_\mathrm{eff}$. {(Note: This assumption allows us to have the same effective temperature of the photosphere, both when it is at $p_\mathrm{cutoff}\sim0$, and at $p_\mathrm{cutoff}=p_0$. Had atmospheric absorption been accounted for, the BOA temperature would need to be greater to yield the given $T_\mathrm{eff}$. For a grey, conservatively scattering atmosphere, the reduced $T_\mathrm{BOA}$ causes an underestimation of the full-column optical thickness of gaseous (Rayleigh) scattering, $\tau_\mathrm{Rayl}$, by a factor of $\sim2$. As we show further in this section, $\tau_\mathrm{Rayl}$ is small enough for such an underestimation to be of negligible impact for our qualitative study, especially in the presence of clouds.  Temperature-dependent absorption and cloud effects have been eliminated through our assumption of a grey atmosphere and free cloud parameters. GD is also not affected, since it is defined as a function of $T_\mathrm{eff}$, and not $T_\mathrm{BOA}$. Potential latitudinal variations in opacity due to GD will be considered in future work.)} {An adiabatic temperature profile is known from \cite{marley1996atmospheric} to closely resemble the thermal structure of BDs at deeper levels because of efficient convection, though they report that the lapse rate at atmospheric levels over $p\sim1\,$bar exceeds the adiabatic lapse rate, especially for low $g$ and $T_\mathrm{eff}<900\,$K.} We obtain the following vertical profile for atmospheric pressure (derived in Appendix \ref{app_adiabat}) 
\begin{equation}\label{eq001p1}
p(z) = p_0\left[1-\frac{\gamma-1}{\gamma}\frac{z}{H_\mathrm{scale}}\right]^\frac{\gamma}{\gamma-1}.
\end{equation}
The adiabatic profile has a clear upper boundary (which can be assumed to be the TOA for most BDs) at a maximum altitude of $z_\mathrm{max}=\frac{\gamma}{\gamma-1}H_\mathrm{scale}$. The temperature profile then corresponds to 
\begin{equation}\label{eq001p2}
T(z)=T_\mathrm{BOA}\left[1-\frac{z}{z_\mathrm{max}}\right].
\end{equation}

Given that the atmosphere is dominated by gaseous H$_2$, we consider $\gamma=1.4$. To estimate the pressure, $p_0$, at BOA, we use the density of liquid hydrogen $\rho_\mathrm{aq.H_2}=70.85\,$g/l as a reference. The van der Waals radius of an H$_2$ molecule  \citep[$r_\mathrm{W}\approx109\,$pm,][]{rowland1996intermolecular} is about 3 times greater than its covalent radius \citep[$r_\mathrm{c}=31\pm5$pm,][]{cordero2008covalent}, so that the H$_2$ atmosphere would cease to be an ideal gas at an estimated density of  $\rho_{\mathrm{gas. H}_2}=\rho_\mathrm{aq.H_2}/27=2.62\,$g/l. Applying the ideal gas equation, we obtain the pressure corresponding to this maximum atmospheric density as $p_0=\frac{\rho_{\mathrm{gas. H}_2}RT_\mathrm{BOA}}{\overline{\mu}}=145.8\,$bar when $T_\mathrm{eff}=1500\,$K. {The pressure at BOA, $p_0$, represents an upper limit of the depth of the atmospheric column at which scattering by gaseous species or cloud affect the BD radiation emerging at the TOA. In general, however, opacity due to gaseous absorption can cause the effective atmospheric column to be truncated at considerably lower pressures, beyond which the atmosphere can be assumed to have negligible albedo. This pressure level, already introduced as $p_\mathrm{cutoff}$, represents the photosphere at the given wavelength. While \cite{de2011characterizing} considered, {for illustrative purposes,} ad hoc $T-p$ profiles assuming hydrostatic equilibrium between the pressure ranges $10^{-6}\,$bar$\,<p<5\,$bar and $10^{-3}\,$bar$\,<p<1\,$bar, respectively, \cite{sengupta2009multiple} consider the contribution of molecular scattering only above $p_\mathrm{cutoff}\sim1-0.1\,$bar. In general, the depth of the atmosphere for RT modeling is assumed to lie between 0.05--10\,bar. {Alkali elements (mainly K$^+$) in the BD atmosphere have strong lines of absorption, whose pressure broadening would cause the single scattering albedo of gaseous scattering to be greatly diminished for pressures greater than $p_\mathrm{cutoff}$ in the wavelength range of 700-1000\,nm \citep{burrows2000near}, while CH$_4$ and collisionally induced absorption (CIA) due to H$_2$ dominate absorption at longer wavelengths. As a result, $p_\mathrm{cutoff}$ in actual BD atmospheres is variable, depending on the wavelength and the relative abundance of absorbing species, in addition to its thermodynamic dependence on $T_\mathrm{BOA}$ at low opacities.}


The lower limit of this photospheric depth, $p_\mathrm{cutoff}\rightarrow0$, corresponds to the special case for which the atmosphere is fully opaque below the TOA. As enumerated in Sec.\,\ref{sec_intro}, for generality, we consider only the extreme cases $p_\mathrm{cutoff}\sim0$ (high opacity due to gaseous absorption) and $p_\mathrm{cutoff}=p_0$ (zero-opacity due to gaseous absorption) in our analysis.} Clearly, in the case where $p_\mathrm{cutoff}\sim0$, there are nearly no gaseous scatterers available above the photosphere, so that the optical thickness of Rayleigh scattering is negligible. The only scattering species are clouds which condense above the photosphere {(similar to high cirrus clouds on Earth)}. 

The vertical profiles for pressure and temperature when $p_\mathrm{cutoff}=p_0$ have been illustrated in the upper panel of Fig.\ref{fig_atm_prof}. The column integrated gaseous number density of the atmosphere for any given $T-p$ profiles is $N_\mathrm{gas.column}=\frac{p_0N_\mathrm{Av}}{\overline{\mu}g}$, where $N_\mathrm{Av}$ is Avogadro's number. {It can be seen that $N_\mathrm{gas.column}$ depends on $T_\mathrm{eff}$ through $p_0$, and would hence be affected by changes in opacity and by GD (discussed in Sec.\,\ref{sec_gravdark}). {The GD effect of latitudinally increasing $T_\mathrm{eff}$ would, in principle, be counterbalanced by $g$ in the denominator of the expression for $N_\mathrm{gas.column}$, which also increases with latitude for oblate bodies.}  However, for our current study, we make the simplifying assumption that the Rayleigh optical thickness remains constant at all latitudes, as we focus primarily on clouds having optical thicknesses larger than that of the gaseous atmosphere}. 

The total Rayleigh scattering optical thickness of our gaseous distribution with $p_\mathrm{cutoff}=p_0$ is found to be $\tau_\mathrm{Rayl}\sim0.057$ (computed using Eq. 4 of \cite{sneep2005direct} with $\alpha_\mathrm{vol}$ for H$_2$ obtained from \cite{nir1973polarizability}) at 1000\,nm. The vertical layers considered in our model and their respective pressures, temperatures and Rayleigh scattering optical depths (at 1000\,nm) are listed in Table \ref{tab_vp}. The Rayleigh scattering optical thickness, $\tau_\mathrm{Rayl}$ has been plotted as a function of wavelength in the bottom panel of Fig.\,\ref{fig_atm_prof}, showing its diminishing role beyond optical wavelengths. All wavelengths used in the following sections, have been highlighted using blue dots.

\begin{figure}[t]
\vspace*{0mm}
\begin{center}
\includegraphics[width=\textwidth]{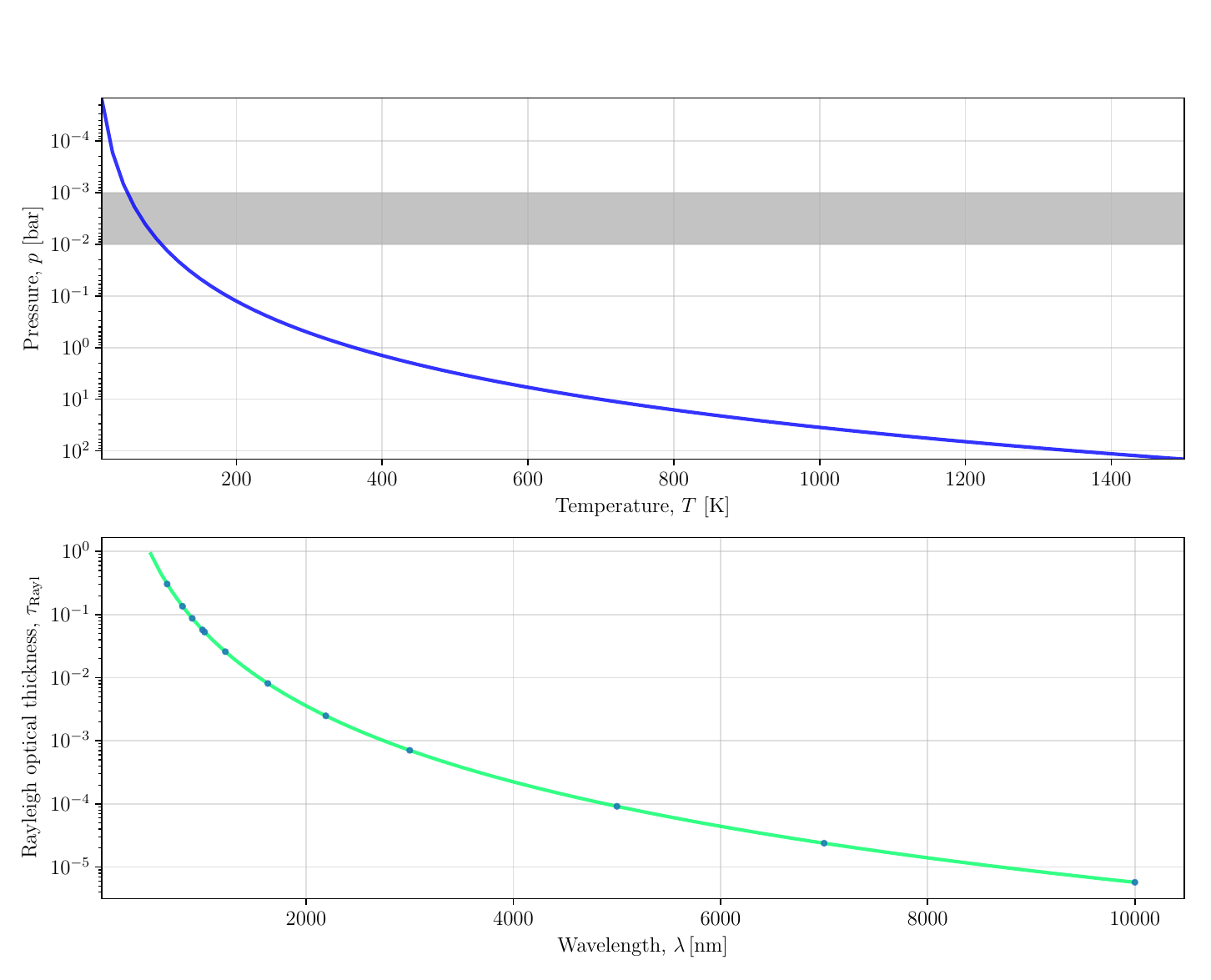}
\end{center}
\caption{\small{{\it Top:} Vertical profiles of $p$ and $T$ on a BD of surface pressure $p_0=145.8\,$bar, $T_\mathrm{eff}=1500\,$K, and a scale height of $H_\mathrm{scale}=11.43\,$km. A cloud distributed uniformly between the pressure levels $0.01$--$0.001\,$bar is illustrated by the grey-shaded area.\\
{\it Bottom:} Optical thickness due to full-column Rayleigh scattering in the BD atmosphere as a function of wavelength. The wavelengths 658, 806, 900, 1000, 1020, 1220, 1630, 2190, 3000, 5000, 7000 and 10000nm, used in the following sections, have been highlighted using blue dots.}}
\label{fig_atm_prof}
\end{figure}

In both limiting cases, $p_\mathrm{cutoff}\sim0$ and $p_\mathrm{cutoff}=p_0$, we consider a single layer of uniformly distributed haze/cloud extending between the pressure levels $p=10^{-2}-10^{-3}\,$bar  (illustrated as a grey-shaded area in Fig.\,\ref{fig_atm_prof}).  
{Haze/cloud grains have currently been assumed to be spherical so that Mie theory can be used for the computation of their scattering properties \citep{mie1908beitrage, hulst1957light, domke1975fourier, siewert1982phase, stratton2007electromagnetic, sanghavi2014revisiting}}. Throughout this work, cloud/haze grains are assumed to have a log-normal size-distribution with characteristic width $\sigma=1.13$, so that the probability $\rho(r)$ of finding a cloud grain in the interval $[r, r+\mathrm{d}r]$ is given by
\begin{equation}\label{eq6p1}
\rho(r)\mathrm{d}r = \frac{1}{\sqrt{2\pi}r{\sigma}}\exp{\left[-\frac{(\log{r}-\log{r_0})^2}{2{\sigma}^2}\right]}\mathrm{d}r,
\end{equation}
where $r_0$ is the characteristic radius of the grains.

Haze/cloud optical thicknesses, $\tau_\mathrm{cloud}=\{0.1,\,1.0,\,10.0,\,50.0\}$ and droplet sizes, $r_0=\{0.01,\,0.1,\,0.5,\,1.0,\,5.0,\,10.0,\,50.0\}\,\mu$m, have been chosen to cover plausible orders of magnitude in their respective parameter spaces. The haze and cloud are chosen to be purely scattering with a refractive index, $\mathbf{n}=1.66-0.0\imath$, at all wavelengths considered. The real part of the refractive index has been chosen to be typical of silicates \citep{kandler2007chemical}, but the imaginary part has been artificially set to zero in order to eliminate the effects of absorption for this general study. {The sphericity of these silicate grains is an artificial assumption currently made (as also by \cite{sengupta2009multiple, sengupta2010observed, marley2011probing, de2011characterizing, stolker2017polarized}) for the sake of simplicity even though the silicate cloud grains that mostly occur in solid state, especially at high altitudes, are more likely to be non-spherical. Polarimetric computations can be expected to be profoundly affected by particle non-sphericity, as has been demonstrated definitively by laboratory measurements \citep{perry1978experimental, munoz2011laboratory, munoz2012amsterdam} as well as by theory \citep{mishchenko1994light, mishchenko1999light, mishchenko2002scattering}. This issue can, in principle, be resolved using the single-scattering database created on the basis of numerically-exact T-matrix and improved geometrical-optics computations, as demonstrated by \cite{dubovik2006application}. However, RT computations of BD radiation are azimuthally symmetric, unlike those of \cite{dubovik2006application} which involve reflected sunlight. This alleviates some of the errors we incur through our assumption of sphericity, as long as the cloud grains can be assumed to be randomly oriented. It should be noted that in the case of aligned non-spherical grains \citep{wolf2002multiple, mishchenko1996t, matsumura1996extinction}, which may occur in the presence of strong winds or magnetic fields, this assumption would break down and cause error.}

\begin{table}[h!]
  \centering
  \caption{Simulated BD vertical profiles for the zero-opacity limit ($p_\mathrm{cutoff}=p_0$)}
  \label{tab_vp}
  \begin{tabular}{c|c|c|ccc}
    \hline
    Height $z$ [km] &{Pressure $p$ [Bar]}& {Temperature $T$ [K]} & $\tau_\mathrm{gas}$ at 1000\,nm\\
    \hline
40.01 &     0.00 &  0.0  & $0.0$\\
36.00 &    0.05 &  150.0  & $5.44\cdot10^{-5} $\\
32.00 &    0.52 & 300.0 &  $5.61\cdot10^{-4} $\\
28.00   &  2.16 & 450.0  & $1.93\cdot10^{-3} $\\
24.00   &  5.90 & 600.0  & $4.42\cdot10^{-3}$\\
20.00  &  12.89 & 750.0 &  $8.25\cdot10^{-3}$\\
16.00 &   24.39 & 900.0  & $1.36\cdot10^{-2}$\\
 12.00 &  41.50 & 1050.0  & $2.06\cdot10^{-2}$\\
 8.00 &  66.77 & 1200.0  & $2.94\cdot10^{-2}$\\
 4.00 &  100.83 & 1350.0   &$4.02\cdot10^{-2}$\\
 0.0  & 145.80 & 1500.00  & $5.31\cdot10^{-2}$\\
    \hline
  \end{tabular}
\end{table}
}
\subsection{Local 1D RT characteristics for a gridpoint}\label{sec_local}
A plane-parallel Stokes vector computation of the thermal emission for a given profile is azimuthally symmetric about the local surface normal, $\hat{\mathbf{n}}_{ij}$, at a gridpoint (LAT$_i$,LON$_j$), depending only on the angle, $\theta_{\mathrm{v},ij}$ made by the viewing direction $\hat{\mathbf{v}}$ with respect to $\hat{\mathbf{n}}_{ij}$ for a given atmospheric profile. While {photospheric} emission is isotropic and unpolarized, subsequent scattering by gaseous species and haze/cloud grains higher in the atmosphere lead to both anisotropy and polarization of light as a function of the local view angle $\theta_{\mathrm{v},ij}$. (Our assumption of perfect azimuthal symmetry about $\hat{\mathbf{n}}_{ij}$ is strictly valid only for a uniform atmosphere enveloping a perfect sphere, and is only an approximation for oblate BDs and/or patchy atmospheres. The plane-parallel assumption is an additional simplification.) 

\begin{figure}[t]
\vspace*{0mm}
\begin{center}
\includegraphics[width=\textwidth]{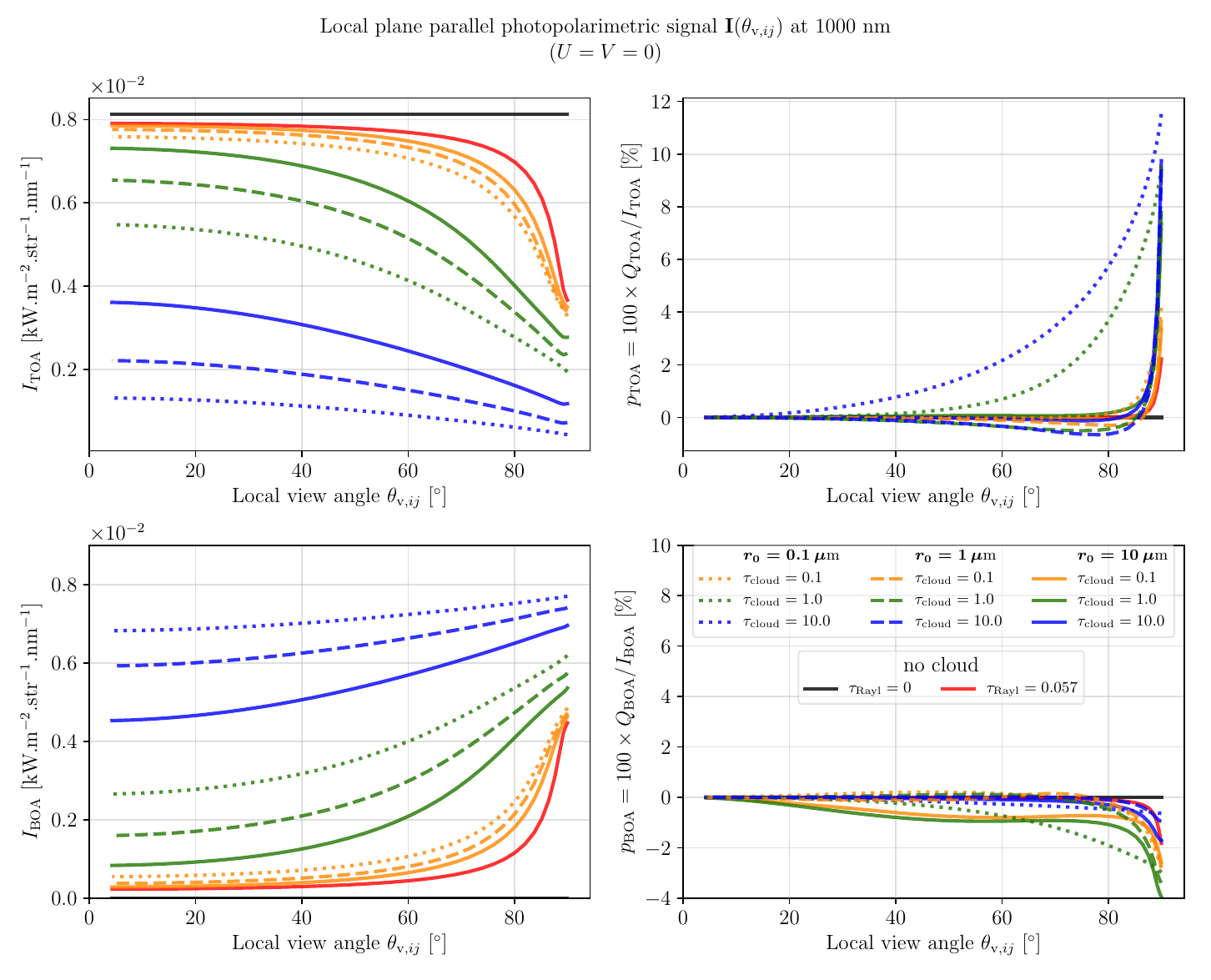}
\end{center}
\caption{\small{{\it Top left}:  Intensity $I$ of local thermal radiation emitted by a BD of effective temperature $T_\mathrm{eff}=1500\,$K at 1000\,nm. The observed signal at an angle $\theta_{\mathrm{v},ij}$ from the local normal depends on the transmission properties of the intermediate atmosphere. Full-column Rayleigh scattering (red) corresponding to $P_\mathrm{cutoff}=p_0$ causes an anisotropic redistribution of the isotropic radiation emerging from the BD {   photosphere} (black).  As scattering becomes more predominant in the atmosphere, the dependence on $\theta_{\mathrm{v},ij}$ becomes increasingly isotropic (orange, green, blue).}
\small{{\it Top right}:  The degree-of-polarization, $Q/I$, is greatest near the limb ($\theta_{\mathrm{v},ij}=90^\circ$), where $Q$ is large while $I$ is smallest. Completely unpolarized thermal emission at the {   photosphere} (black) becomes polarized in the presence of scatterers. Polarization due to a molecular optical thickness of $\tau_\mathrm{Rayl}=0.057$ is significant at the limb, so that the red curve is nearly coincident with the solid orange curve representing large cloud grains of low optical thickness. Polarization is most pronounced at $\theta_{\mathrm{v},ij}=90^\circ$ for scattering by small cloud grains (blue and green dotted lines). As grain size increases, increasing depolarization takes place due to changes in the phase matrix of the grain. Increasing optical thickness leads to more multiple scattering, which also has a depolarizing effect for all grain sizes.}
\small{{\it Bottom left}:  Intensity $I$ of thermal radiation reflected back at the bottom of its atmosphere at an angle $\theta_{\mathrm{v},ij}$ with respect to nadir (important for verification of flux conservation).}
\small{{\it Bottom right}:  The degree-of-polarization at BOA.}}
\label{fig_planeparallel}
 \end{figure}

We first consider the full-column Rayleigh scattering case, when $p_\mathrm{cutoff}=p_0$. The dependence on $\theta_{\mathrm{v}, ij}$ of the photopolarimetric thermal emission signal at the top-of-atmosphere (TOA) corresponding to the atmospheric profiles in Table \ref{tab_vp} is shown in the top panels of Fig.\,\ref{fig_planeparallel}, with intensity depicted on the left and the DoP on the right. The BD is assumed to emit radiation at its BOA as a perfect black body of effective temperature $T_\mathrm{eff}=1500\,$K. The signal due to a hypothetical BD completely devoid of an atmosphere (black line), only Rayleigh scattering (red line), a light haze ($\tau=0.1$, orange lines), a thick haze ($\tau=1$,  green lines) and a thick cloud ($\tau=10$, blue lines) have been depicted. Different grain sizes have been considered: small grains of $r_0=0.1\,\mu$m (dotted lines), medium-sized grains of $r_0=1\,\mu$m  (dashed lines) and coarse grains of $r_0=10\,\mu$m (solid lines). 

As depicted in the top left panel, the observed intensity $I$ at an angle $\theta_{\mathrm{v},ij}$ from the local normal depends on the transmission properties of the intermediate atmosphere. Molecular absorption is assumed negligible (single scattering albedo, $\varpi_\mathrm{gas}=1$). The red line shows that in the absence of a cloud, the BD is darkened by Rayleigh scattering, and appears locally darker as the local viewing angle becomes larger (limb darkening). 
As particulate scattering ($\varpi_\mathrm{cloud}=1$) becomes more predominant in the atmosphere, multiple scattering causes the dependence on $\theta_{\mathrm{v},ij}$ to become increasingly isotropic as seen by comparing the orange ($\tau_\mathrm{cloud}=0.1$), green ($\tau_\mathrm{cloud}=1.0$), and blue ($\tau_\mathrm{cloud}=10.0$) lines representing the cloudy case. For a given optical thickness, $\tau_\mathrm{cloud}$, larger grains can be seen to be more forward-scattering, as evident from the highest intensity $I_\mathrm{TOA}$ being associated with the solid lines and the lowest intensity corresponding to the dotted lines at all viewing angles.

The top right panel of Fig.\,\ref{fig_planeparallel} depicts the degree of DoP, $p=Q/I$ of the transmitted light ($U=V=0$ for thermal radiation scattered by spherical or randomly distributed non-spherical particles). The DoP, $p$, is greatest near the limb ($\theta_{\mathrm{v},ij}=90^\circ$), where generally $Q$ is large while $I$ is smallest. Completely unpolarized thermal emission (black line) at the BD photosphere becomes polarized as it undergoes transmission through a scattering atmosphere. Since polarization due to scattering is significant at large angles for gaseous scatterers of optical thickness $\tau_\mathrm{Rayl}=0.057$ compared to that due to large particles, the red curve is nearly coincident with the solid orange curve representing large cloud grains and optical thickness of $\tau_\mathrm{cloud}=0.1$. Polarization is most pronounced at $\theta_{\mathrm{v},ij}=90^\circ$ for scattering by small grains (dotted lines, note that the green and blue dotted lines are nearly coincident) of sufficient optical thickness. As grain size increases, the phase matrix of the grain becomes increasingly depolarizing (dashed and solid lines). Increasing optical thickness causes more emitted radiation to undergo scattering, while also reducing the intensity of transmitted light -- both effects that increase the DoP. On the other hand, more multiple scattering is depolarizing. As a result, the optical thickness has a nonlinear effect on the DoP, so that increasing $\tau_\mathrm{cloud}$ from 0.1 (orange) through 1 (green) to 10 (blue) successively amplifies $p_\mathrm{TOA}$, but an increase beyond $\tau_\mathrm{cloud}=1$ (green) yields a considerably smaller change in $p$.
Small grains are generally more polarizing than larger grains, as can be seen by comparing the dotted ($r_0=0.1\,\mu$m), dashed ($r_0=1\,\mu$m) and the solid ($r_0=10\,\mu$m) lines, with the grains being most polarizing in the Rayleigh scattering limit (when the size parameter, $\xi=\frac{2\pi r_0}{\lambda}$, of the grains is much smaller than unity, i.e., $\xi\ll1$). 

The bottom panels of Fig.\,\ref{fig_planeparallel} show the reflected radiation $I_\mathrm{BOA}$ (left panel) and the corresponding degree-of-polarization $p_\mathrm{BOA}$ (right panel) at the bottom-of-atmosphere (BOA) as a function of angle $\theta_{\mathrm{v},ij}$ measured from the nadir (same orientation as the zenith, but pointing down instead of up). {(Note that we do not consider any source of radiation other than the BD. The reflection implied here takes place internally within the BD atmosphere.)} Complementary to the transmitted radiation $I_\mathrm{TOA}$, $I_\mathrm{BOA}$ increases with increasing cloud optical thickness, which showcases the light-trapping nature of clouds. For a given optical thickness, smaller grains scatter back more light than the strongly forward-scattering large grains. Although BOA values are not used in subsequent simulations, the complementary nature of this reflected signal to the transmitted radiation at the TOA is evident, as for pure scattering ($\varpi=1$), conservation of flux yields the relation
\begin{equation}
\frac{1}{2}\int_{0}^{1}\mathrm{d}\mu_\mathrm{v}\mu_\mathrm{v}(I_\mathrm{BOA}(\lambda;\mu_\mathrm{v})+I_\mathrm{TOA}(\lambda;\mu_\mathrm{v})) = B(\lambda;T_\mathrm{eff}),
\end{equation}
 which we use to verify our RT computations. $\mu_\mathrm{v}$ denotes the cosine of the respective view angle $\theta_{\mathrm{v},ij}$ with respect to the local zenith/nadir.

A comparison of the overall response to clouds between the left and right panels of Fig.\,\ref{fig_planeparallel} reveals that $I_\mathrm{TOA}$ has a robust sensitivity to clouds at all viewing angles, but especially smaller viewing angles closer to the zenith. $p_\mathrm{TOA}$, on the other hand, has a more muted response at smaller viewing angles, and shows the greatest sensitivity near the limb ($\theta_{\mathrm{v},ij}\sim90^\circ$) with a marked preference for small grains. This property of $p_\mathrm{TOA}$ allows polarization measurements to provide valuable additional constraints orthogonal to those available from measurements of intensity alone.

 \begin{figure}[t]
\vspace*{0mm}
\begin{center}
\includegraphics[width=\textwidth]{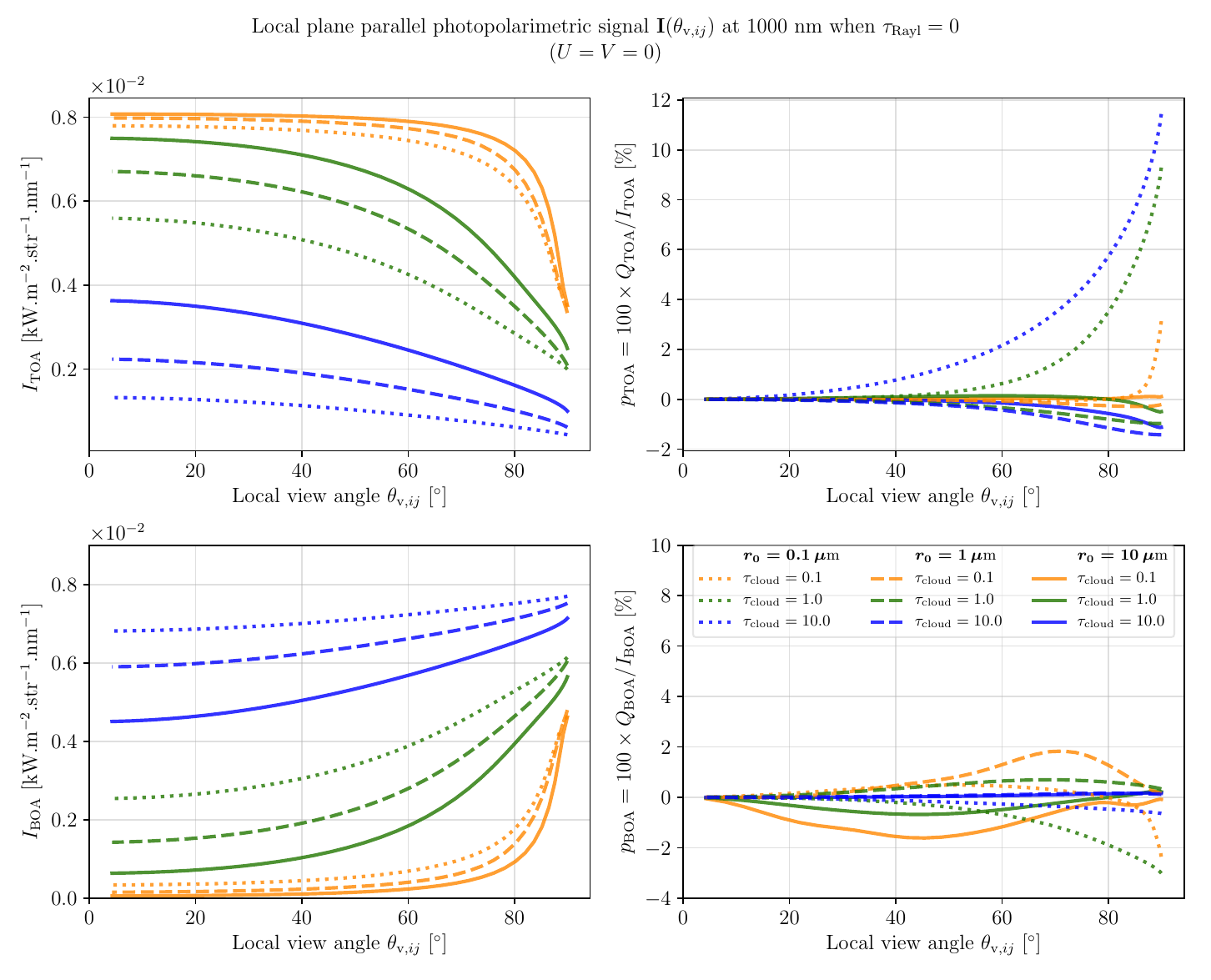}
\end{center}
\caption{\small{Same as Fig.\,\ref{fig_planeparallel} but with negligible contribution due to gaseous scattering ($p_\mathrm{cutoff}\sim0$).}}
\label{fig_planeparallel_c30}
 \end{figure}
 
 { Fig.\,\ref{fig_planeparallel_c30} represents the plane-parallel local scattering response of clouds due to an atmosphere devoid of gaseous scatterers, as expected when $p_\mathrm{cutoff}\sim0$. Comparison with corresponding cloud cases in Fig.\ref{fig_planeparallel} reveals that the emitted intensity is only marginally different. The DoP, $p_\mathrm{TOA}$, however, differs more significantly, especially at larger scattering angles: while the behaviour of clouds remains { practically unchanged} when the cloud grain size is small, the depolarizing effect of clouds with larger grains is no longer overshadowed by the polarization due to gaseous Rayleigh scatterers. This effect has implications for the interpretation of polarimetric measurements of BDs, as will be described in Sections\,\ref{sec_asym_p} and \ref{sec_opt}.}

 The above discussion was focused on the local photopolarimetric contribution of a BD. BD observations, however, are made using ground-based or space telescopes in orbit around the earth, and measure the photopolarimetric contribution of the entire face of the BD exposed to the telescope. This calls for integration over the local contributions of all points on the exposed BD disc, as examined in the next section.

\subsection{Disc-integration}\label{sec_da} 

\begin{figure}[t]
\vspace*{0mm}
\begin{center}
\includegraphics[width=\textwidth]{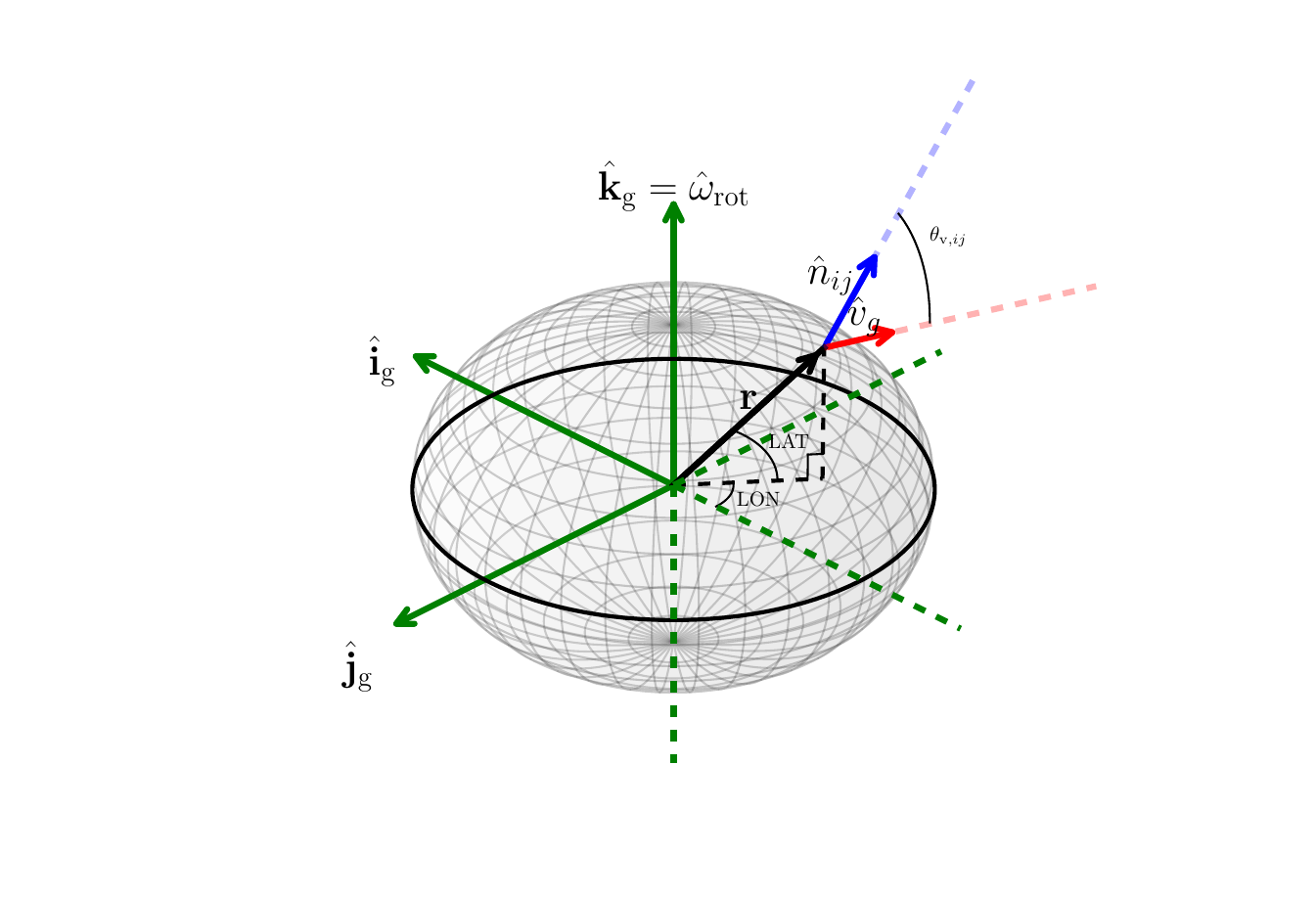}
\end{center}
\vspace*{-20mm}
\caption{\small{(LAT, LON) coordinates and local vectors defined in our setup.}}
\label{fig_local}
\end{figure}

\begin{figure}[t]
\begin{center}
\includegraphics[width=\textwidth]{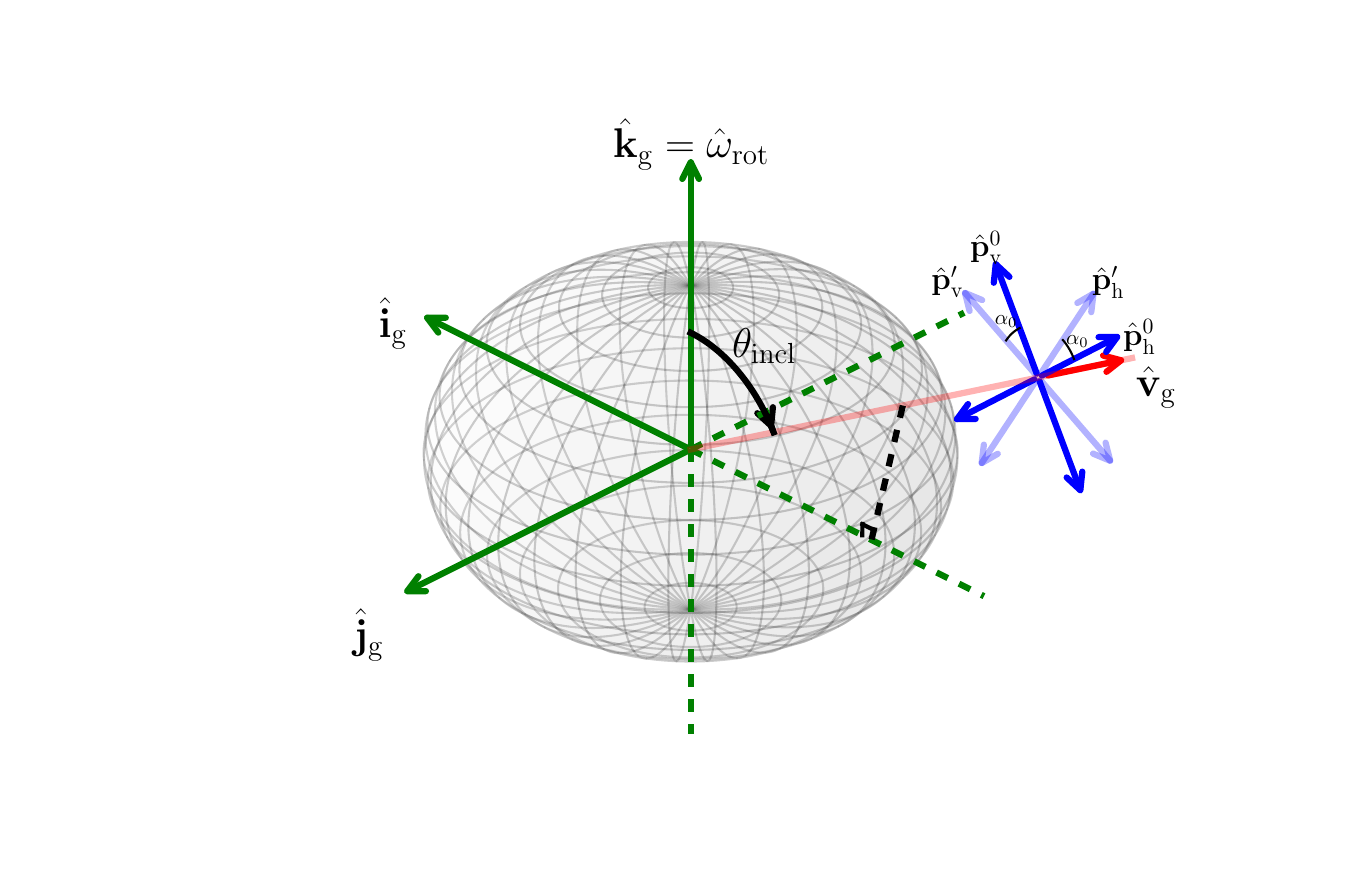}
\end{center}
\vspace*{-20mm}
\caption{\small{The special coordinate system $\{\hat{\mathbf{p}_h^0},\hat{\mathbf{p}_v^0}\}$ in the viewing plane, relative to an arbitrary instrumental reference system $\{\hat{\mathbf{p}_h^\prime},\hat{\mathbf{p}_v^\prime}\}$.}}
\label{fig_global}
\end{figure}
 
To facilitate the computation of the disc-integrated photopolarimetric radiation of the BD from locally computed $I_\mathrm{TOA},\,Q_\mathrm{TOA}$ and $U_\mathrm{TOA}$, we construct our (LAT, LON) grid using Gauss-Legendre quadrature. We use $N_\mathrm{lon}^\mathrm{quad}$ and $N_\mathrm{lat}^\mathrm{quad}$ quadrature points for longitudinal and latitudinal integration, respectively. Generally taking $N_\mathrm{lon}^\mathrm{quad}=2N_\mathrm{lat}^\mathrm{quad}$, we choose $-90^\circ<\mathrm{LAT}_i<90^\circ$, where $i=1,2,3\ldots N_\mathrm{lat}^\mathrm{quad}$, and $-180^\circ<\mathrm{LON}_j<180^\circ$, where $j=1,2,3\ldots N_\mathrm{lon}^\mathrm{quad}$. 

Assuming the rotation axis, $\hat{\bm{\omega}}_\mathrm{rot}$, to point along the positive z-axis, and the x-axis pointing from the observer towards the center of the BD when viewing along the equator, we define a ``global'' reference frame denoted henceforth by the subscript `g' (see Figures\,\ref{fig_local} and \ref{fig_global}), whose xz-plane contains the viewing direction  
\begin{equation}\label{eq0}
\hat{\mathbf{v}}_\mathrm{g}=-\sin{\theta_\mathrm{incl}}\hat{\mathbf{i}}_\mathrm{g}+\cos{\theta_\mathrm{incl}}\hat{\mathbf{k}}_\mathrm{g}
\end{equation}
pointing from the center of the BD towards the observer at an angle $\theta_\mathrm{incl}$ from the rotation axis. 
Choosing the null meridian ($\mathrm{LON}=0^\circ$) to lie in the global xz-plane directed towards the observer, the (LAT, LON) coordinates of $\hat{\mathbf{v}}$ become $(90^\circ-\theta_\mathrm{incl},0^\circ)$. The inclination $\theta_\mathrm{incl}$ is $0^\circ$ when viewing along the north pole of the BD, and $\theta_\mathrm{incl}$ is $90^\circ$ when viewing along the equator. Thus, the unit vectors of the global reference frame have the (LAT, LON) coordinates $\hat{\mathbf{i}}_\mathrm{g}\rightarrow(0^\circ,180^\circ)$, $\hat{\mathbf{j}}_\mathrm{g}\rightarrow(0^\circ,-90^\circ)$ and $\hat{\mathbf{k}}_\mathrm{g}\rightarrow(90^\circ,0^\circ)$, respectively.

Thus, other directions of interest take the following form 
\begin{eqnarray}\label{eq0p1}
\hat{\bm{\omega}}_\mathrm{rot}&=&\hat{\mathbf{k}}_\mathrm{g},\text{ and}\\\nonumber
\hat{\mathbf{n}}_{ij,\mathrm{g}} &=& \frac{-(1-\eta)\cos{\mathrm{LAT}_i}\left[\cos{\mathrm{LON}_j}\hat{\mathbf{i}}_\mathrm{g}+\sin{\mathrm{LON}_j}\hat{\mathbf{j}}_\mathrm{g}\right]+\sin{\mathrm{LAT}_i}\hat{\mathbf{k}}_\mathrm{g}}{\left[\sin^2{\mathrm{LAT}_i}+(1-\eta)^2\cos^2{\mathrm{LAT}_i}\right]^{1/2}},
\end{eqnarray}
where, the oblateness $\eta=1-R_\mathrm{p}/R_\mathrm{e}$, $R_\mathrm{p}$ and $R_\mathrm{e}$ being the polar and equatorial radii, respectively, of the BD, corresponding to the semi-minor and semi-major axes of an ellipsoid with its minor axis along the axis of rotation, $\hat{\bm{\omega}}_\mathrm{rot}$. The form for $\hat{\mathbf{n}}_{ij,\mathrm{g}}$ for an ellipsoid is derived in Appendix \ref{App1}. Substituting $\eta=0$ yields the familiar form for a spherical body. The modified form for the ellipse is caused by the fact that the radial direction for an ellipse is no longer coincident with the local surface normal at every point as in the case of a sphere.
  
The local view angle $\theta_{\mathrm{v},ij}$ at (LAT$_i$,LON$_j$) is now simply obtained using the scalar product of $\hat{\mathbf{v}}_\mathrm{g}$ and $\hat{\mathbf{n}}_{ij,\mathrm{g}}$ in Eqs.(\ref{eq0}) and (\ref{eq0p1}) as
\begin{equation}\label{eq1}
\theta_\mathrm{v, ij} = \cos^{-1}\left(\frac{(1-\eta)\cos{\mathrm{LAT}_i}\sin{\theta_\mathrm{incl}}\cos{\mathrm{LON}_j}+\sin{\mathrm{LAT}_i}\cos{\theta_\mathrm{incl}}}{\left[\sin^2{\mathrm{LAT}_i}+(1-\eta^2)\cos^2{\mathrm{LAT}_i}\right]^{1/2}}\right),
\end{equation} 

Thus, we can cover the entire surface of the BD, subsequently using the criterion $\hat{\mathbf{v}}_{g}\cdot\hat{\mathbf{n}}_{ij,\mathrm{g}}<0$ to mask out the side of the BD facing away from the observer, since a given location $(\mathrm{LAT}_i, \mathrm{LON}_j)$ on the surface contributes to the disc-integrated signal only when the angle between the local normal $\hat{\mathbf{n}}_{ij}$ and the viewing direction $\hat{\mathbf{v}}$ does not exceed a right angle. 

Now, we carry out disc-integration of the Stokes parameter, using the following equation
\begin{eqnarray}\label{eq2}
\mathbf{I}_\mathrm{da} = \frac{1}{2\pi}\sum_{i=1}^{N_\mathrm{LON}^\mathrm{quad}}w_{\mathrm{LON},i}\sum_{j=1}^{N_\mathrm{LAT}^\mathrm{quad}}w_{\mathrm{LAT},j}\mathrm{d}S_{ij}\mathbf{L}(\beta_{\mathrm{g},ij})\mathbf{I}_{ij}(\theta_{\mathrm{v}, ij}),
\end{eqnarray}
where the area element $\mathrm{d}S_{ij}$ for an ellipsoid of oblateness $\eta$ is simply given by
\begin{equation}\label{eq3}
\mathrm{d}S_{ij}=\cos{\mathrm{LAT}_i}\frac{\left[\sin^2{\mathrm{LAT}_i}+(1-\eta)^2\cos^2{\mathrm{LAT}_i}\right]^{1/2}}{1-\eta\cos^2{\mathrm{LAT}_i}}.
\end{equation}
Again, for $\eta=0$, we get the familiar form $\mathrm{d}S_{ij} = \cos{\mathrm{LAT}_i}$ for a perfect sphere.
The rotation matrix $\mathbf{L}(\beta_{\mathrm{g},ij})$ \citep{hovenier1969symmetry} given by
\begin{equation}\label{eq4}
  \mathbf{L}(\beta_{\mathrm{g},ij}) = 
 \begin{bmatrix}
       1 & 0 & 0 & 0           \\[0.3em]
       0 & \cos{2\beta_{\mathrm{g},ij}} & \sin{2\beta_{\mathrm{g},ij}} & 0\\[0.3em]
       0 & -\sin{2\beta_{\mathrm{g},ij}} & \cos{2\beta_{\mathrm{g},ij}} & 0\\[0.3em]
       0 & 0 & 0 & 1
     \end{bmatrix}
\end{equation}
is the rotation matrix that transforms the Stokes parameters, $\mathbf{I}_{ij}(\theta_{\mathrm{v}, ij})$,  determined in the local frame of reference (with respect to the plane containing $\hat{\mathbf{n}}_{ij}$ and $\hat{\mathbf{v}}$) to their corresponding values in the global frame of reference `g' (in which we define the parallel and perpendicular polarization directions with respect to the plane containing $\hat{\mathbf{k}}_\mathrm{g}$ and $\hat{\mathbf{v}}$). $\beta_{\mathrm{g},ij}$ is the angle through which $\hat{\mathbf{n}}_\mathrm{local}(\mathrm{LAT}_i, \mathrm{LON}_j)$ must be turned to coincide with the rotation axis $\hat{\mathbf{k}}$ while looking along the viewing direction $\hat{\mathbf{v}}$. In order to retain the exact signs of $\cos{2\beta_{\mathrm{g},ij}}$ and especially $\sin{2\beta_{\mathrm{g},ij}}$ as $\beta_{\mathrm{g},ij}$ varies in the interval $[0,2\pi]$, we use the trigonometric identities 
\begin{eqnarray}\label{eq5}
\cos{2\beta_{\mathrm{g},ij}}&=&2\cos^2{\beta_{\mathrm{g},ij}}-1\text{ and}\\\nonumber
\sin{2\beta_{\mathrm{g},ij}}&=&2\sin{\beta_{\mathrm{g},ij}}\cos{\beta_{\mathrm{g},ij}}
\end{eqnarray}
using 
\begin{eqnarray}\label{eq5a}
\cos{\beta_{\mathrm{g},ij}}&=&\frac{1}{D}\left(\sin{\theta_\mathrm{incl}}\sin{\mathrm{LAT}_i}-(1-\eta)\cos{\mathrm{LAT}_i}\cos{\mathrm{LON}_j}\cos{\theta_\mathrm{incl}}\right)\text{ and}\\\nonumber
\sin{\beta_{\mathrm{g},ij}}&=&\frac{1}{D}((1-\eta)\cos{\mathrm{LAT}_i}\sin{\mathrm{LON}_j})
\end{eqnarray}
where
\begin{eqnarray*}\label{eq6}
D = &\left[(1-\eta)^2\cos^2{\mathrm{LAT}_i}(1-\sin^2{\theta_\mathrm{incl}}\cos^2{\mathrm{LON}_j})+\sin^2{\mathrm{LAT}_i}\sin^2{\theta_\mathrm{incl}}\right.\\\nonumber
&\left.-\frac{1}{2}(1-\eta)\sin{2\mathrm{LAT}_i}\sin{2\theta_\mathrm{incl}}\cos{\mathrm{LON}_j}\right]^{1/2}.
\end{eqnarray*}
Our method of determining  $\cos{\beta_{\mathrm{g},ij}}$ and $\sin{\beta_{\mathrm{g},ij}}$ is summarized in Appendix \ref{App2}.

\begin{figure}
 \centering
\vspace*{0mm}
\begin{center} 
\begin{minipage}{\textwidth}
  \centering
  \includegraphics[width=\linewidth]{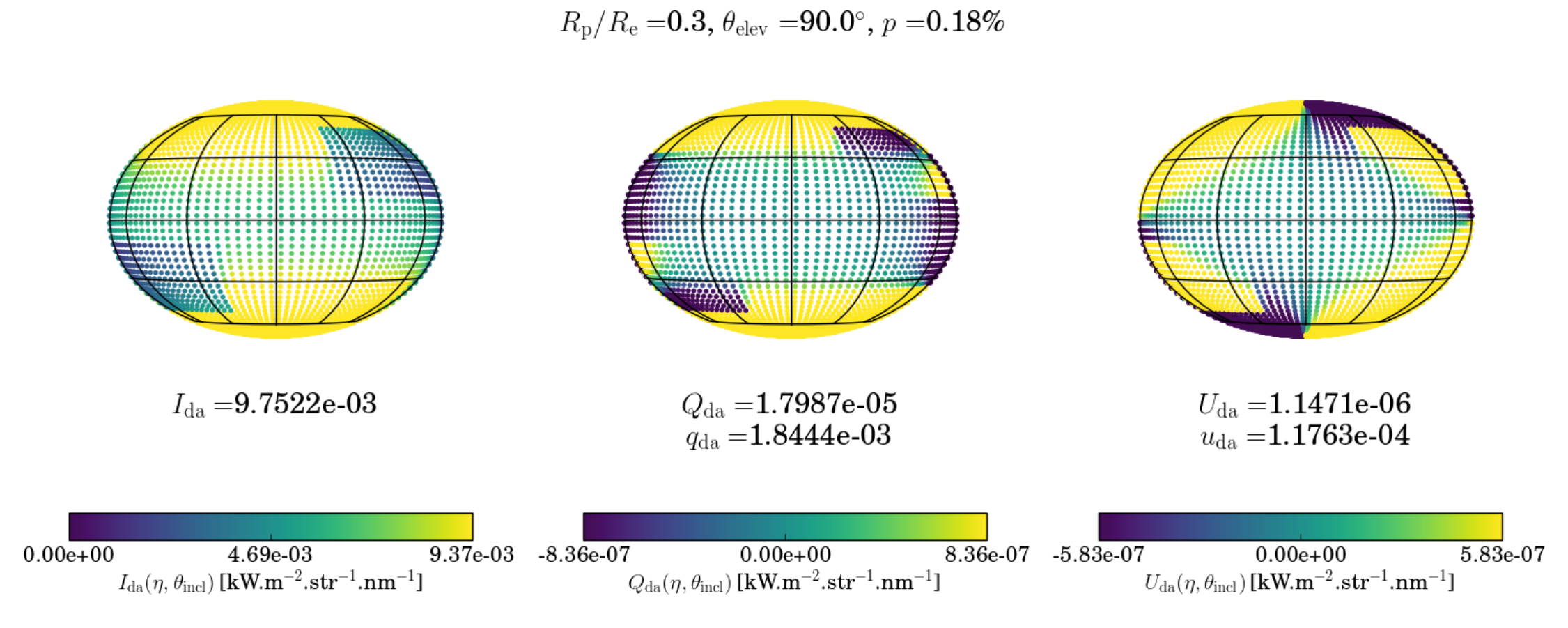}
\end{minipage}%
\vspace*{0mm}
\begin{minipage}{\textwidth}
  \centering
  \includegraphics[width=\linewidth]{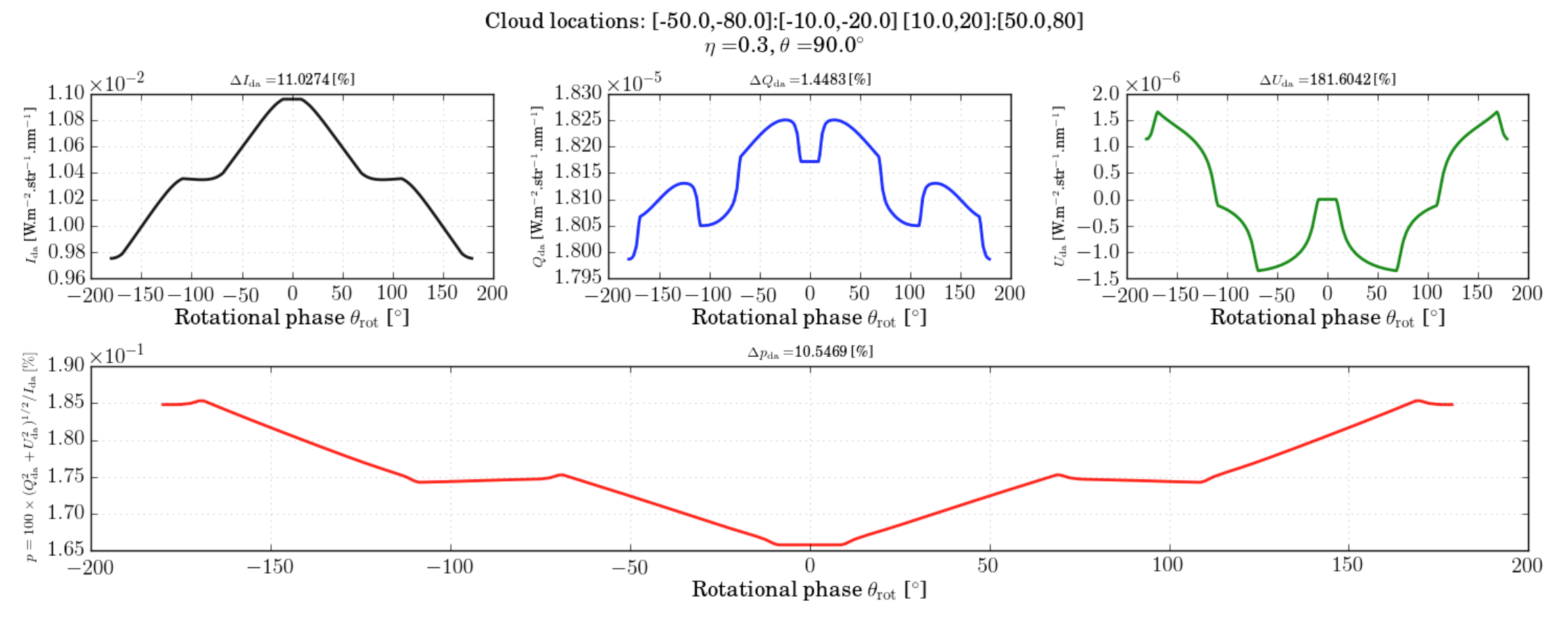}
\end{minipage}%
\end{center}
\caption{\small{{\it Top}: A snapshot of two clouds of longitudinal width $\Delta\mathrm{LON}=60^\circ$ each between the latitudes $10^\circ-50^\circ$ of the northern and southern hemispheres, respectively.\\{\it Bottom}: Light-curves illustrating the evolution of each Stokes vector component: $I_\mathrm{da}$ (top left), $Q_\mathrm{da}$ (top middle) and $U_\mathrm{da}$ (top right), and the DoP, $p_\mathrm{da}$ (bottom).}}
\label{fig_2patches}
 \end{figure}
 
From a measurement perspective, the global reference system outlined in this section is recreated by choosing the coordinate system in the viewing plane such that
the projection of the rotation axis $\hat{\mathbf{k}}$ on the viewing plane is the nominal direction $\hat{\mathbf{p}}^0_\mathrm{v}$ of vertical polarization, orthogonal to the coplanar horizontal direction $\hat{\mathbf{p}}^0_\mathrm{h}$ (see Fig.\,\ref{fig_global}). 

The preferred status of the reference frame given by $\hat{\mathbf{p}}^0_\mathrm{h}$ and $\hat{\mathbf{p}}^0_\mathrm{v}$ becomes evident in Sec.\,\ref{sec_BD} where we show how the disc-averaged Stokes parameter $U_\mathrm{da}$ can be instrumental in identifying $\hat{\mathbf{p}}^0_\mathrm{h}$ and $\hat{\mathbf{p}}^0_\mathrm{v}$ for a given measurement. 

{The above equations describle the methodology we use to simulate any given cloud scenario, whether uniform or patchy, in the atmosphere of an oblate BD. As an example, the evolution of Stokes vector components resulting from two patchy clouds over a period of rotation are illustrated in Fig.\,\ref{fig_2patches}.  We will, however, focus only on uniform cloud decks in the following.}

\section{Gravitational darkening}\label{sec_gravdark}
\begin{figure}[t]
\vspace*{0mm}
\begin{center}
\includegraphics[width=\textwidth]{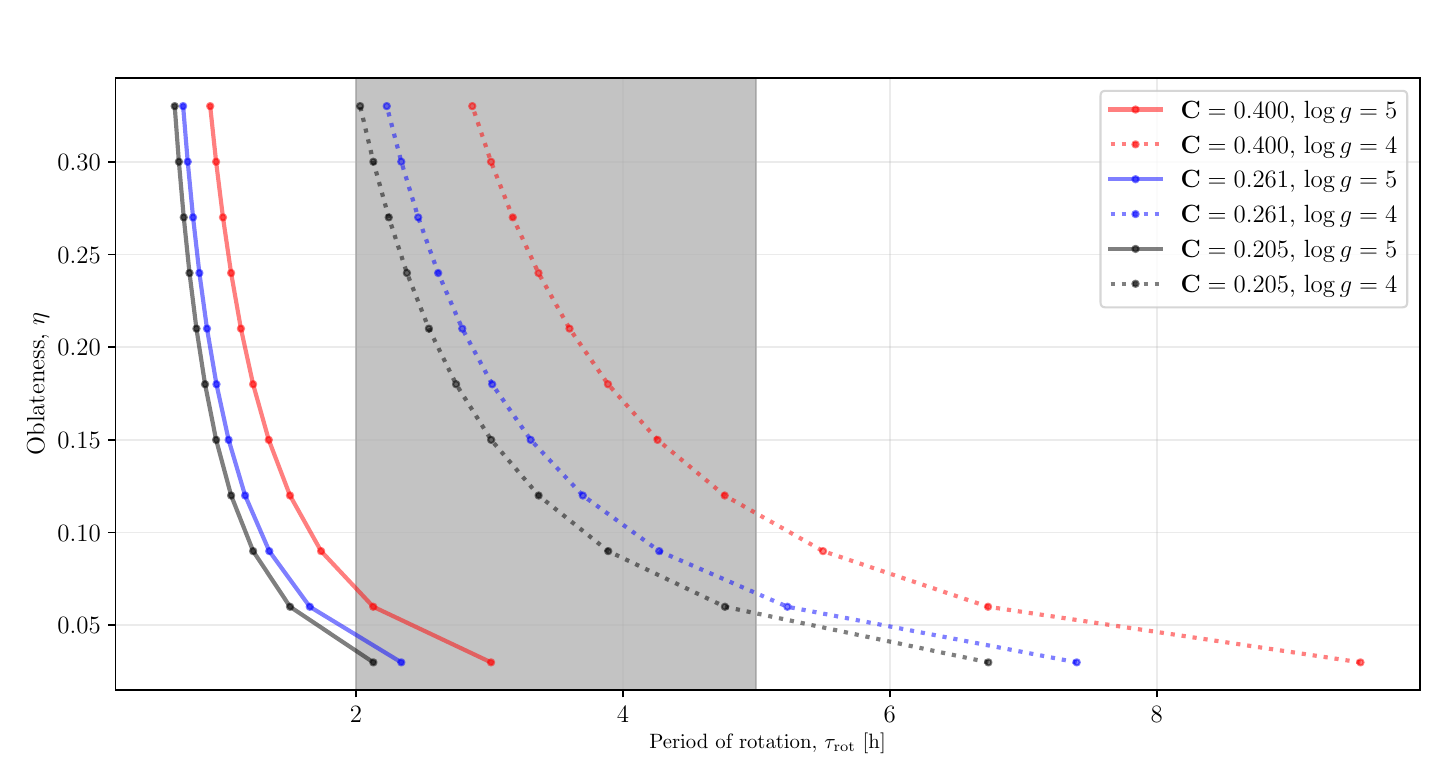}
\end{center}
\caption{\small{Oblateness as a function of rotational period of a brown dwarf. The black lines represent a brown dwarf of typical polytropic ratio $n=1.5$ ($\mathbb{C}=0.205$), while the blue line represents a gas giant planet of typical polytropic ratio $n=1$ ($\mathbb{C}=0.261$). The red lines correspond to a polytropic ratio of $n=0$, ($\mathbb{C}=0.4$) representing a uniform density body. The dotted and solid lines represent log$g$=4 and 5, respectively. The range \citep{osorio2003photometric, scholz2004rotation2, scholz2004rotation1, scholz2005rotation} of rotational periods, $\tau_\mathrm{rot}=2$--$5\,$h has been highlighted in grey.}}
\label{fig_rot_obl}
\end{figure}

\begin{figure}[t]
\vspace*{0mm}
\begin{center}
\includegraphics[width=\textwidth]{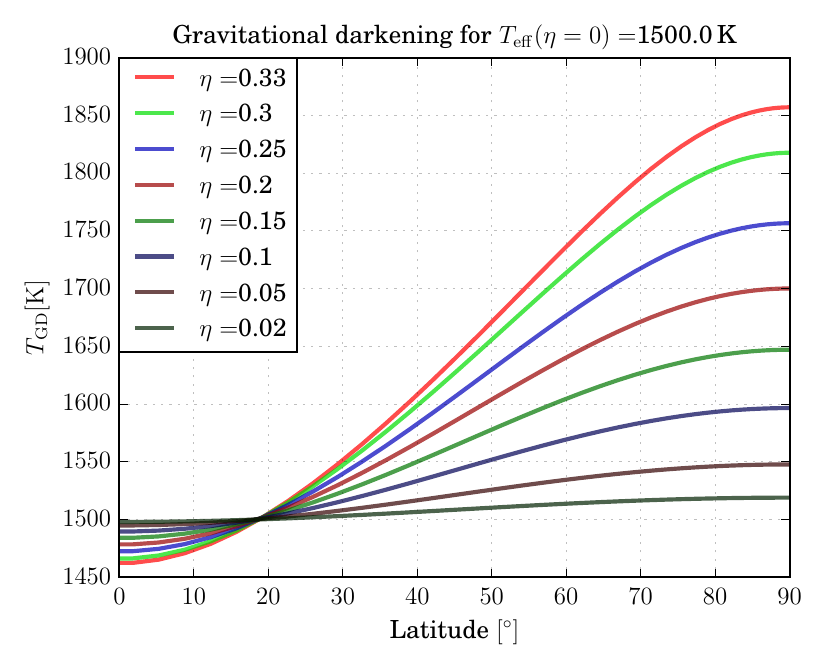}
\end{center}
\caption{\small{Effective temperature of a BD due to gravitational darkening (GD) as a function of latitude. The extent of GD depends on the rotation rate, and hence, on the oblateness, $\eta$, of the BD. The BD considered here has a spherical-equivalent effective temperature $T_\mathrm{eff}(\eta=0)=1500\,$K.}}
\label{fig_GD}
\end{figure}

\begin{figure}[t]
\vspace*{0mm}
\begin{center}
\includegraphics[width=\textwidth]{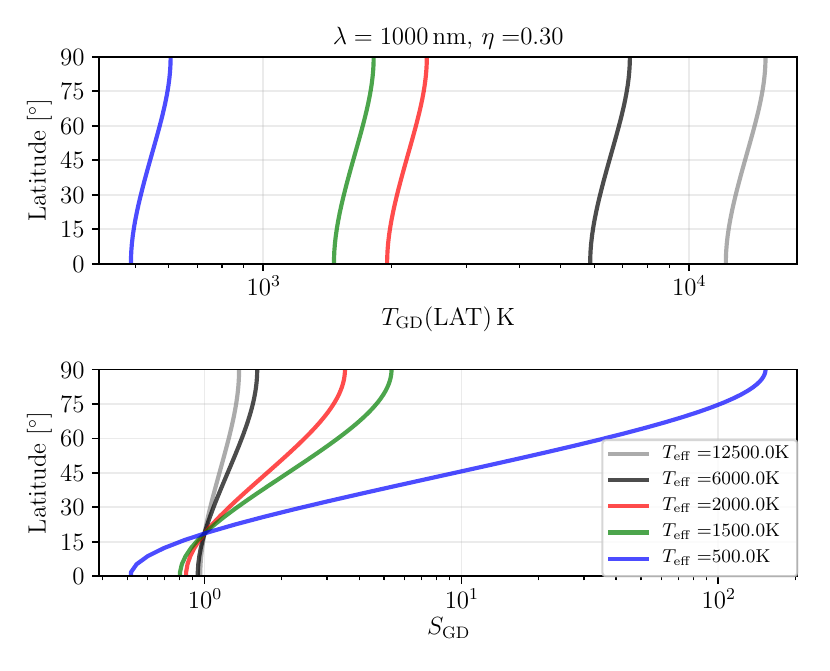}
\end{center}
\caption{\small{{\it Top}: Latitudinal variation of temperature due to GD for bodies of different effective temperatures. It is evident that the relative variation in latitudinal temperature, viz. $T_\mathrm{GD}(\mathrm{LAT})/T_\mathrm{eff}$, is independent of $T_\mathrm{eff}$.\\
{\it Bottom}: Latitudinal variation of the flux scaling factor due to GD for bodies of different effective temperatures. It can be seen that the latitudinal differences in emitted flux becomes increasingly pronounced as $T_\mathrm{eff}$ decreases.}}
\label{fig_SGD_vs_Teff}
\end{figure}

\begin{figure}[t]
\vspace*{0mm}
\begin{center}
\includegraphics[width=\textwidth]{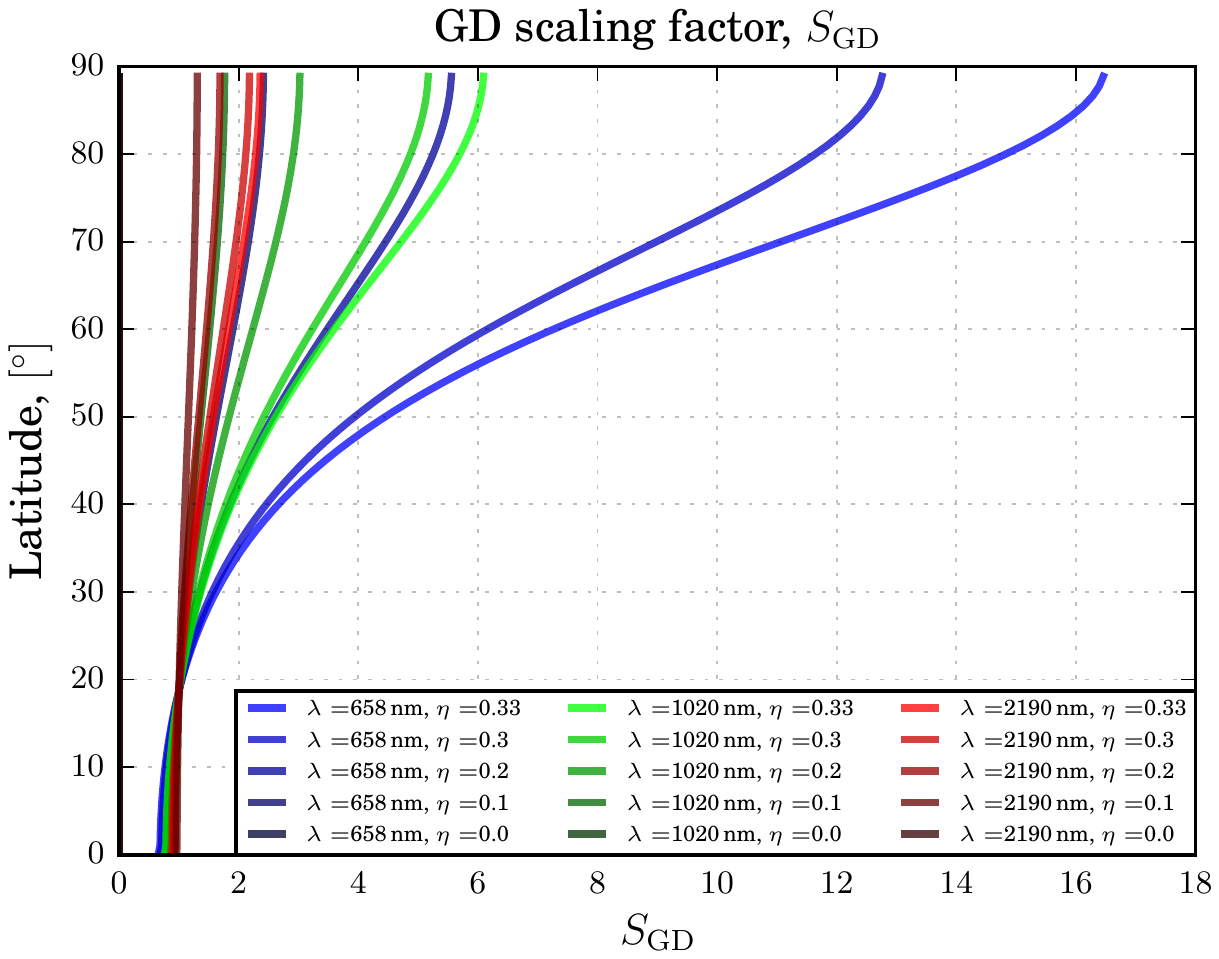}
\end{center}
\caption{\small{Scaling factor due to gravitational darkening (GD) for the latitude-dependent { photospheric flux} of a BD of spherical equivalent effective temperature {  $\mathbf{T_\mathrm{eff}=1500}\,$K}.
The wavelength-dependent $S_\mathrm{GD}(\mathrm{LAT})$ {  also varies with} the oblateness, $\eta$, and effective temperature, $T_\mathrm{eff}$, of the BD.}}
\label{fig_scale_GD}
\end{figure}

\begin{figure}[t]
\vspace*{0mm}
\begin{center}
\includegraphics[width=\textwidth]{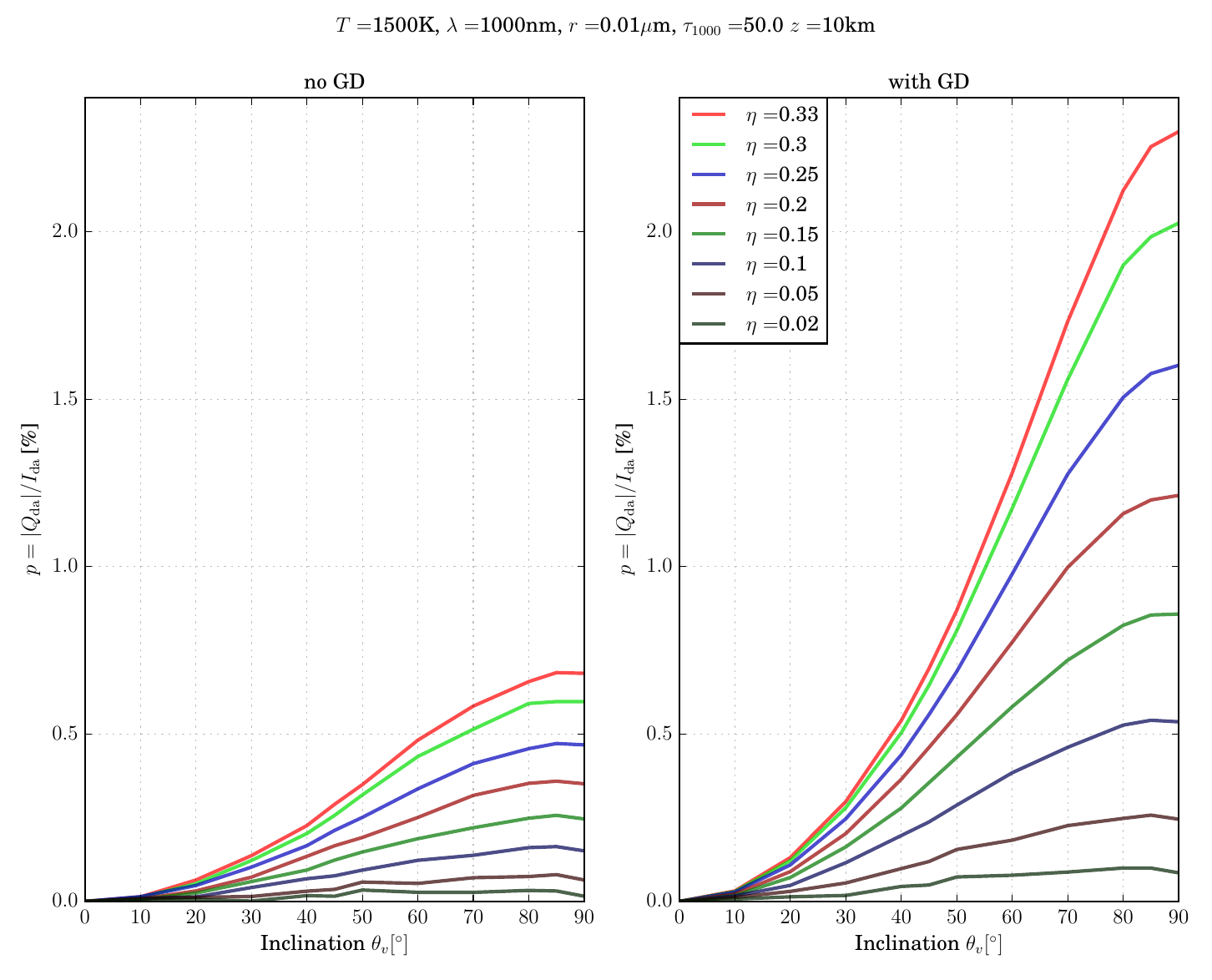}
\end{center}
\caption{\small{{\it Left}:  Degree-of-polarization as a function of inclination angle due to a BD of oblateness $\eta$ when GD is neglected. {\it Right}: Same as the left panel, except that GD is taken into account. The BD considered here has a spherical-equivalent effective temperature $T_\mathrm{eff}(\eta=0)=1500\,$K. { Around such temperatures, GD is seen to have a significantly amplifying effect on BDs of all oblatenesses excepting $\eta=0$.}} {Since we have considered a semi-infinite Rayleigh scattering atmosphere for these computations, the polarization shown here represents the asymptotic upper-limit for any given oblateness and inclination angle, so that any further increase in the number of scatterers does not change its polarization.}}
\label{fig_GDpol}
\end{figure}

Before proceeding to examine the disc-averaged photopolarimetric signal of a BD uniformly enveloped by scattering particles, 
we note that the high rotation rates that cause BDs to become oblate also lead to a latitudinal variation of effective temperature due to GD. {This effect was theoretically predicted by \cite{von1924radiative} for radiative stellar atmospheres, and extended to convective atmospheres of very low mass (VLM) stars and BDs by \cite{lucy1967gravity,claret2000studies,reiners2003effects,lara2011gravity}. {It can be argued that even though \cite{claret2000studies} acknowledged applicability to BDs, they only considered M-stars up to a minimum effective temperature of $T_\mathrm{eff}=2000\,$K, and \cite{lara2011gravity} expanded the theoretical framework of GD to convective VLM stars but not explicitly to BDs. However, BDs are very interesting candidates for exploring the effects of GD on account of their high surface gravities and high rotation rates. 
GD on stars has been observationally verified by \cite{de2005gravitational, peterson2006resolving, peterson2006vega, masuda2015spin}. Observational evidence for GD on BDs, however, does not currently exist.}

{ \cite{bouvier2014angular} show that VLM objects  at  young  ages  show  a  wide range of periods, similar to more massive stars, from a few 
hours up to at least 2 weeks. However, they observe a consistent trend of faster rotation toward lower masses. They cite work by \cite{osorio2003photometric, scholz2004rotation2, scholz2004rotation1, scholz2005rotation} based on high-precision photometric observations which report BDs rotating with time periods of 2--5 hours. This is very close to the breakup limit, where centrifugal  and gravitational forces just balance  each other. Though contested when compared to the results of \cite{cody2010precision} whose sample contains only one period shorter than 14 hours, their results appear to be vindicated by photometric/polarimetric/spectroscopic measurements of \cite{osorio2005optical, apai2013hst, crossfield2014global, radigan2014strong, Zap2011ApJ}, {which either directly record short rotation periods or are predicated on high oblateness (as in the case of highly polarized BD measurements)}.  

The oblateness has been chosen to vary between its physical limits: a minimum at $\eta=0$ for a perfect sphere, and its upper limit $\eta=0.33$ as given by the Roche model, beyond which a rotating body is expected to break apart as the centrifugal force at its equator exceeds the gravitational pull at its surface \citep{chandrasekhar1967ellipsoidal,seager2002constraining}. As mentioned in Sec.\,\ref{sec_model}, we assume a mass of M=20$M_{\mathrm{Jup}}$ and the equatorial radius of Jupiter, $R=R_{\mathrm{Jup}}$. We use the Darwin-Radau relationship given by \cite{barnes2003measuring} to generate a mapping between angular velocity, $\Omega$, and oblateness, $\eta$,
\begin{equation}
\Omega=\sqrt{\frac{\eta GM}{R^3}\left[\frac{5}{2}\left(1-\frac{3}{2}\mathbb{C}\right)^2+\frac{2}{5}\right]},
\end{equation}    
where $\mathbb{C}$ (shorthand for $C_\mathrm{BD}/M_\mathrm{BD}R_\mathrm{e}^2$, where $C_\mathrm{BD}$ is the moment of inertia) for BDs can be expected to lie between $\mathbb{c}\sim0.205$ \citep{stevenson1991search, sengupta2010observed} and $\mathbb{C}=0.261$ \citep{sengupta2010observed}. This allows us to plot the oblateness as a function of rotational period $\tau_\mathrm{rot}=2\pi/\Omega$ in Fig.\,\ref{fig_rot_obl}. The range $\tau_\mathrm{rot}=2$--$5\,$h is highlighted in grey, showing that the full range of oblateness $\eta=0$--$0.33$ is relevant to BDs.}

GD causes surface temperatures on an oblate body to increase towards the poles: due to their rapid rotation, the equator-pole difference in radius is large enough for the surface gravity at the poles to be significantly higher. This translates into higher polar temperatures and brightness.

Using the method developed by \cite{lara2011gravity}, Fig.\,\ref{fig_GD} shows our  computation of the latitudinal variation in the effective temperature of a BD whose spherical equivalent temperature is $T_\mathrm{eff}=1500\,$K. {As a result, the photospheric flux of the BD varies latitudinally. If $F(\mathrm{LAT})$ is the {photospheric flux} of the BD at latitude $\mathrm{LAT}$, and if we refer to the {photospheric flux} of the same BD in the absence of GD as $F_\mathrm{eff}$, then the scaling factor $S_\mathrm{GD}$ due to GD is a direct function of wavelength and temperature, and an indirect function of oblateness and latitude through $T_\mathrm{GD}$, expressed as
\begin{equation}\label{eq_SGD}
S_\mathrm{GD}(\mathrm{LAT}) = \frac{F(T_\mathrm{GD}(\mathrm{LAT}))}{F(T_\mathrm{eff})}=\frac{e^\frac{h\nu}{kT_\mathrm{eff}}-1}{e^\frac{h\nu}{kT_\mathrm{GD}(\mathrm{LAT})}-1},  
\end{equation}
where $h$ and $k$ are the Planck and Boltzmann constants, respectively, $\nu=\frac{c}{\lambda}$ is the frequency of emitted light,  $T_\mathrm{eff}$ is the (spherical equivalent) effective temperature, and $T_\mathrm{GD}(\mathrm{LAT})$ is the effective temperature at latitude $\mathrm{LAT}$ due to GD. 

{GD is found to cause the same relative change in latitudinal temperature, viz., $T_\mathrm{GD}(\mathrm{LAT})/T_\mathrm{eff}$ for a given oblateness, $\eta$, surface gravity, ${g}$, and mass distribution, $\mathbb{C}$, irrespective of the effective temperature $T_\mathrm{eff}$. However, the effect on the black-body flux emitted due to these latitudinal variations is nonlinear: as $T_\mathrm{eff}$ increases, the difference between the numerator and denominator in Eq.\,\ref{eq_SGD} decreases for a given $\eta$, so that $S_\mathrm{GD}$ approaches the lower limit of unity.
This is illustrated in Fig.\,\ref{fig_SGD_vs_Teff}, where the upper panel shows the latitudinal distribution of temperature due to GD over bodies of different effective temperatures ranging from 12500\,K (typical of bright stars like Regulus studied by \cite{cotton2017polarization}), 6000\,K (typical of our Sun), 2000\,K (L-dwarf), 1500\,K (L/T transition), to 500\,K (T-dwarf). It can be seen that the relative change in temperature with respect to latitude is independent of the effective temperature $T_\mathrm{eff}$. The resultant flux, however, varies as the flux scaling factor $S_\mathrm{GD}$ with latitude. It is clear from the bottom panel of Fig.\,\ref{fig_SGD_vs_Teff} that relative changes in $S_\mathrm{GD}$ become increasingly pronounced as $T_\mathrm{eff}$ decreases.}

A decrease in wavelength has the same effect as a decrease in $T_\mathrm{eff}$, as both affect the denominator in the term $\exp{\left(\frac{hc}{k\lambda T_\mathrm{GD}(\mathrm{LAT})}\right)}$ in Eq.\,\ref{eq_SGD}. This scaling factor has been illustrated in Fig.\,\ref{fig_scale_GD} for a BD at $T_\mathrm{eff}=1500\,$K for different levels of oblateness at wavelengths 658\,nm, 1020\,nm and 2190\,nm, {  showing that GD is more pronounced at shorter wavelengths}.} 

{Differences in flux between the poles and the equator add to the asymmetry causing polarization of the disc-integrated radiation in the presence of a scattering atmosphere. To examine this effect, we simulate }a semi-infinite, purely Rayleigh-scattering atmosphere by considering cloud grains of radius $0.01\,\mu$m and scattering optical thickness $\tau_{1000}=50$ at 1000\,nm, giving rise to different degrees of polarization for different levels of BD oblateness. Fig.\,\ref{fig_GDpol} shows the polarization expected in each case if GD is neglected (left panel), compared to when it is taken into account (right panel). {For any given oblateness and inclination angle, the polarization computed here represents the asymptotic upper-limit for those parameters, so that any further increase in the number of scatterers does not change its polarization.} The right hand panel of  Fig.\,\ref{fig_GDpol} is qualitatively comparable to Fig.\,4 of \cite{harrington1968intrinsic} who consider a semi-infinite Rayleigh-scattering atmosphere due to free electrons in a B-star. The temperature inhomogeneity due to GD causes a 3- to 4-fold increase in the DoP due to Rayleigh scattering at an inclination $\theta_\mathrm{v}=90^\circ$, and hence should not be neglected for rapidly rotating bodies like BDs. The degree-of-polarization increases with increasing oblateness as well as inclination angle, reaching a maximum near an inclination angle of $\theta_\mathrm{v}=90^\circ$. 
}
In the following, we correct for GD to study in detail the disc-resolved and disc-integrated dependence of the photopolarimetric BD signal on oblateness, inclination angle and cloud scattering properties. In Sections \ref{sec_BD} through \ref{sec_cloud}, we focus primarily on the case featuring full-column gaseous absorption, viz. $p_\mathrm{cutoff}=p_0$. 

\section{BD characteristics}\label{sec_BD}
\subsection{Dependence on oblateness}\label{sec_uni_oblate}
 
 \begin{figure}
 \centering
\vspace*{0mm}
\begin{center}
\begin{minipage}{\textwidth}
  \centering
    \includegraphics[width=\linewidth]{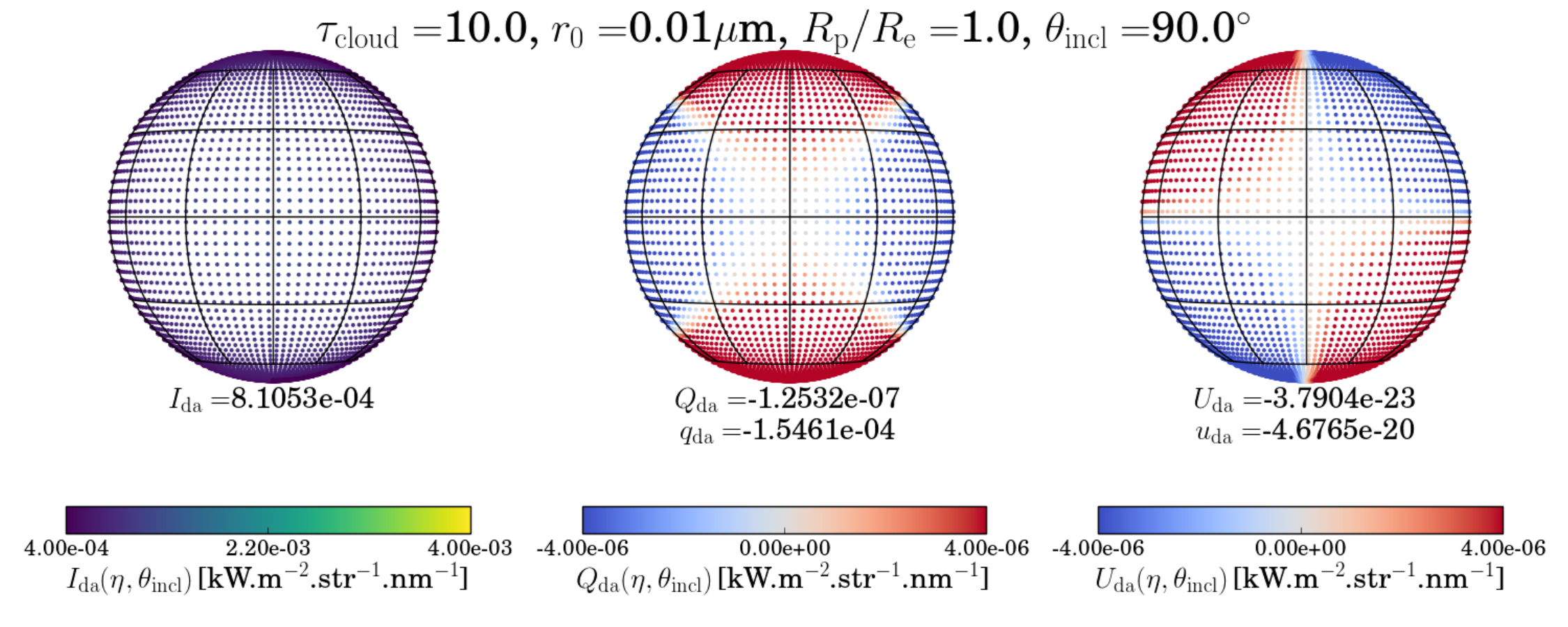}
\end{minipage}%
\vspace*{0mm}
\begin{minipage}{\textwidth}
  \centering
    \includegraphics[width=\linewidth]{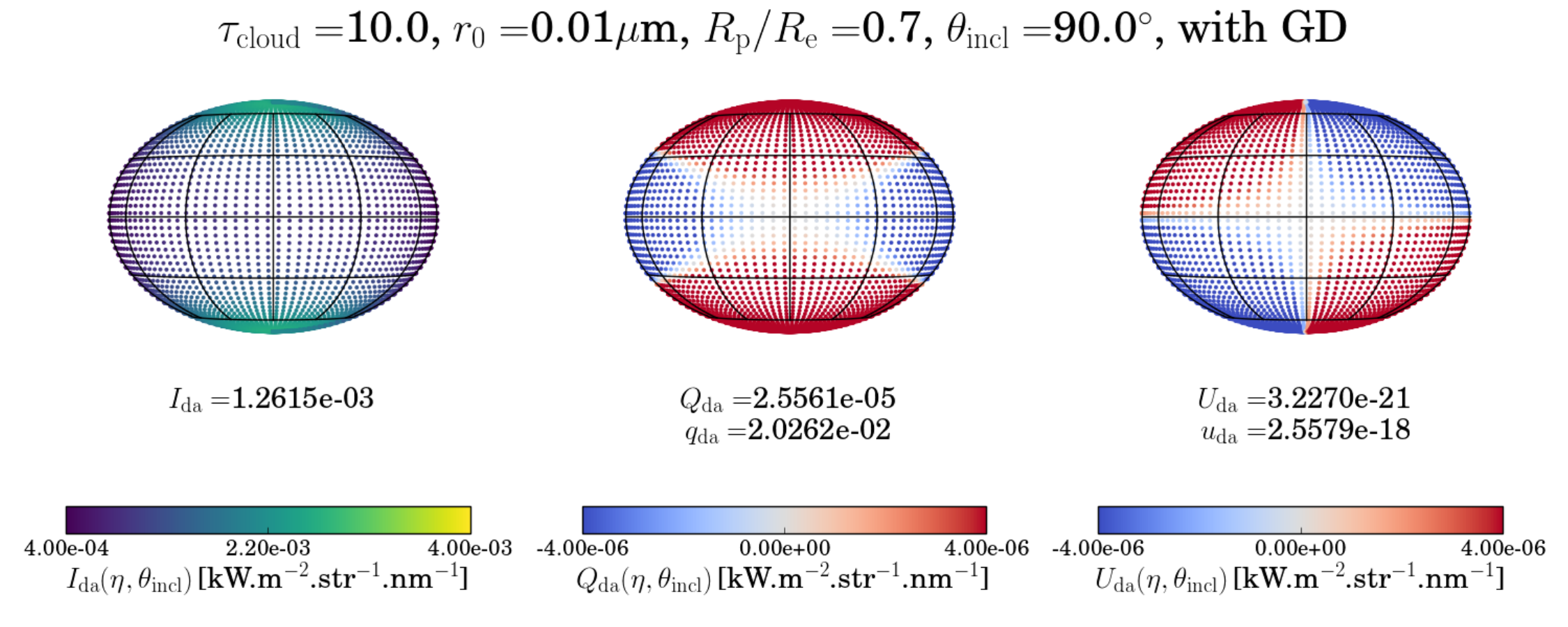}
\end{minipage}%
\end{center}
\caption{\small{{\it Top}: Disc-resolved Stokes elements $I$, $Q$ and $U$ of thermal radiation emitted by a spherical BD ($R_\mathrm{p}/R_\mathrm{e}=1$) of effective temperature $T_\mathrm{eff}=1500\,$K at 1000\,nm. The cloud envelope has an optical thickness of 10 and consists of small grains of radius $r_0=0.01\,\mu$m. \\
$I$ (left) shows limb darkening at the edge of the BD disc. The disc-integrated $I_\mathrm{da}$ is provided below the image of the disc.\\
The Stokes parameter $Q=\frac{1}{2}(I_\mathrm{v}-I_\mathrm{h})$ (middle) increases in the radial direction from $Q=0$ at the center towards a maximum near the limb ($\theta_\mathrm{v}=90^\circ$). $Q$ also varies as the cosine of twice the angular distance from the vertical axis $\hat{\mathbf{p}}^0_\mathrm{v}$. The disc-integrated $Q_\mathrm{da}$ and $q_\mathrm{da}=Q_\mathrm{da}/I_\mathrm{da}$ are provided below the image of the disc.\\
The Stokes parameter $U=\frac{1}{2}(I_\mathrm{135^\circ}-I_\mathrm{45^\circ})$ (right) has the same radial behavior as $Q$, but the reference direction for angular measurement is $\frac{1}{\sqrt{2}}(\hat{\mathbf{p}}^0_\mathrm{v}-\hat{\mathbf{p}}^0_\mathrm{h})$instead of $\hat{\mathbf{p}}^0_\mathrm{v}$.
The disc-integrated $U_\mathrm{da}$ and $u_\mathrm{da}=U_\mathrm{da}/I_\mathrm{da}$ are provided below the image of the disc.\\
(Note: The convergence to zero of the disc-integration is less exact for $Q$ than for $U$, because while the quadrature points are distributed symmetrically in the individual quadrants contributing to $U$, no such symmetry applies to $Q$.)\\
{\it Bottom}: Same as top panel, except for a BD of oblateness $\eta=0.3$ ($R_\mathrm{p}/R_\mathrm{e}=0.7$). Gravitational darkening has been taken into account.}}
\label{fig_uniform_dr_0p1}
 \end{figure}
 
\begin{figure}
 \centering
\vspace*{0mm}
\begin{center} 
\begin{minipage}{\textwidth}
  \centering
  \includegraphics[width=\linewidth]{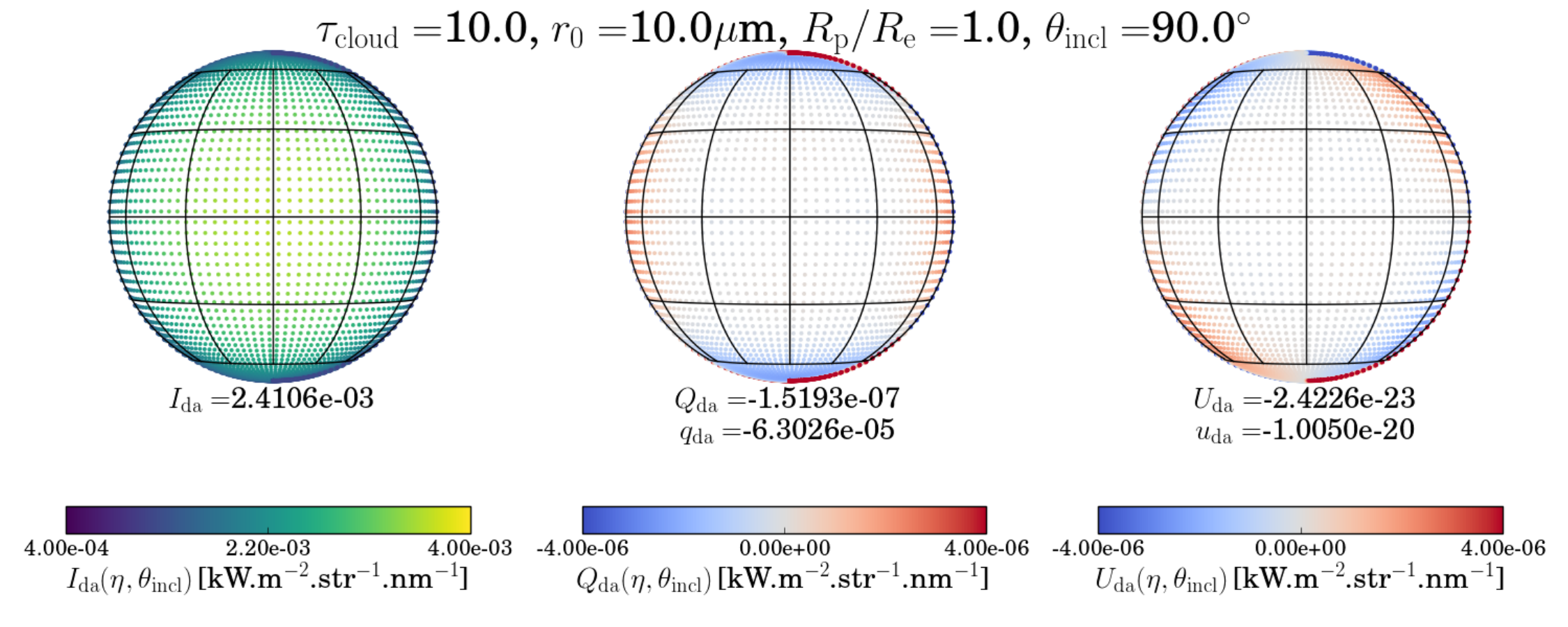}
\end{minipage}%
\vspace*{0mm}
\begin{minipage}{\textwidth}
  \centering
  \includegraphics[width=\linewidth]{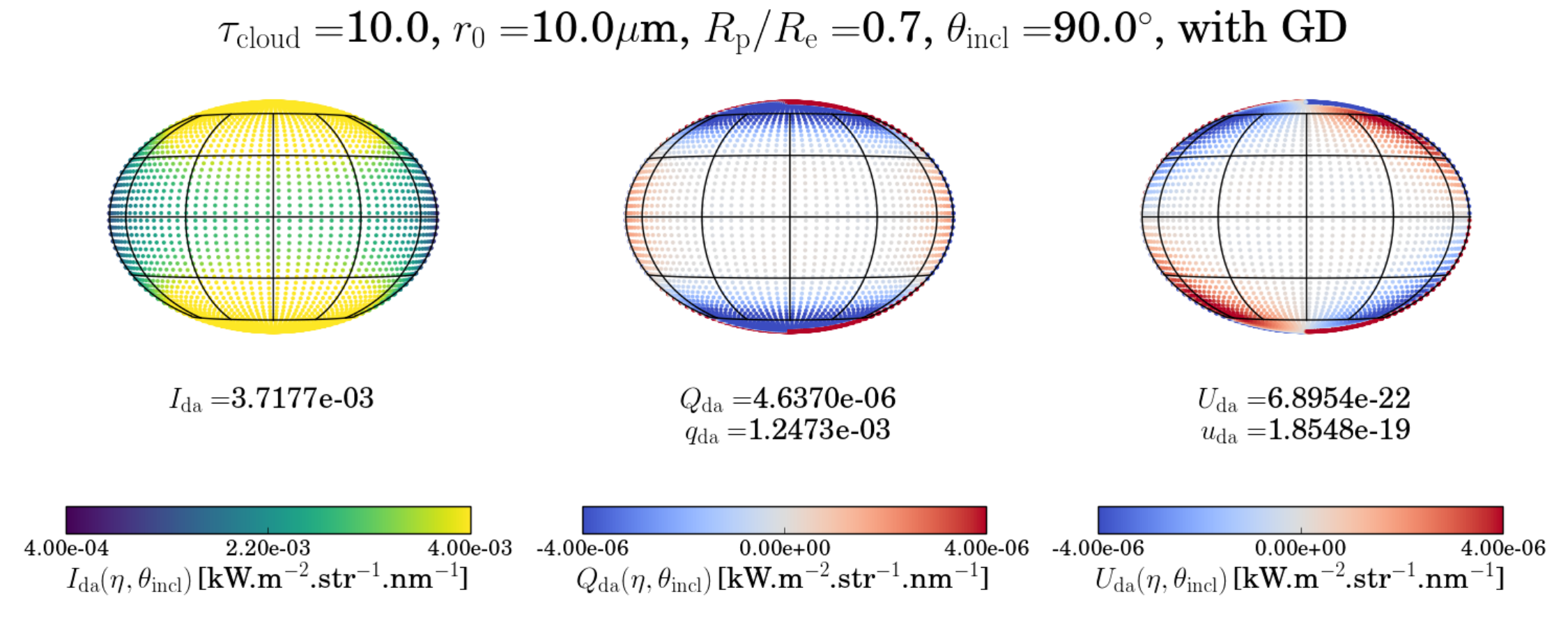}
\end{minipage}%
\end{center}
\caption{\small{Same as Fig.\,\ref{fig_uniform_dr_0p1}, except for a cloud having larger grains of radius $r_0=10\,\mu$m. Colorbars for $I$, $Q$ and $U$ have fixed ranges identical to those in Fig.\,\ref{fig_uniform_dr_0p1}.} Note that, for larger grains, the polarization is not only significantly weaker but has also changed sign.}
\label{fig_uniform_dr_10p0}
 \end{figure}

In this section, we examine the disc-resolved structure of polarized light emitted by a BD to understand the resulting disc-integrated signal. We consider two cloud cases, each of the same optical thickness, $\tau_\mathrm{cloud}=10$, but with small ($r_0=0.01\,\mu$m) and large ($r_0=10\,\mu$m) grains, as depicted in Figs.\,\ref{fig_uniform_dr_0p1} and \ref{fig_uniform_dr_10p0}, respectively. These sizes coincide with the lower (nucleation mode) and upper (condensation mode) limits of the mean particle size of the BD condensates modeled by \cite{helling2006dust}. The optical thickness has been chosen to be high enough to ensure sufficient atmospheric scattering.

The top panel represents a (hypothetical) spherical BD ($\eta=0$), viewed at an inclination angle $\theta_\mathrm{incl}=90^\circ$. It features the same temperature throughout its photosphere, by virtue of being unaffected by GD due to its spherical shape. The disc shows perfect radial symmetry in the intensity $I$, with limb darkening present in both cloud cases. However, the stronger forward-scattering characteristic of larger grains manifests itself in greater emission due to the cloud of grain size $r_0=10\,\mu$m depicted in Fig.\ref{fig_uniform_dr_10p0}, while the smaller particles of Fig.\,\ref{fig_uniform_dr_0p1} scatter light more isotropically. 

$Q$ and $U$ show variations in both sign and magnitude, with larger absolute values near the limb. Clearly, the smaller grains are more polarizing as is evident from the larger ranges covered by $Q$ and $U$ in the top panel of Fig.\,\ref{fig_uniform_dr_0p1} in contrast to those of Fig.\,\ref{fig_uniform_dr_10p0} (the same color-scale is used in both figures). 
In accordance with Eqs.\,(\ref{eq2}-\ref{eq4}), $Q$ (middle sub-panel) varies as the cosine of twice the angular distance $\beta_{\mathrm{g},ij}$ with respect to $\hat{\mathbf{p}}^0_\mathrm{v}$, while $U$ (right sub-panel) shows similar behavior with respect to the direction $\frac{1}{\sqrt{2}}(\hat{\mathbf{p}}^0_\mathrm{v}-\hat{\mathbf{p}}^0_\mathrm{h})$ ($45^\circ$ from $\hat{\mathbf{p}}^0_\mathrm{v}$). As the angle $\beta_{\mathrm{g},ij}$ varies from $0^\circ-360^\circ$ over the entire BD disc, quadrants of alternating signs are mapped out for $Q$ and $U$ each. For small grains as in Fig.\,\ref{fig_uniform_dr_0p1}, $Q$ maps northern and sourthern quadrants with positive values ($Q\ge0$) and western and eastern quadrants with negative values ($Q\le0$), while $U$ remains positive over the northwestern and southeastern quadrants ($U\ge0$) and negative over the northeastern and southwestern quadrants ($U\le0$).  The grains depicted in Fig.\,\ref{fig_uniform_dr_10p0}, however, are large enough to reverse the signs of $Q$ and $U$ relative to those in Fig.\,\ref{fig_uniform_dr_0p1}. 
 
When the BD has a spherical shape, as in the top panels of Figs.\,\ref{fig_uniform_dr_0p1} and \ref{fig_uniform_dr_10p0}, the projections of these angular distances $\beta_{\mathrm{g},ij}$ are perfectly symmetric about the center of the disc, and cancel out exactly to give vanishing disc-integrated values, $Q_\mathrm{da}=U_\mathrm{da}=0$. (It can be seen from the finite value of $Q_\mathrm{da}$ in the top panels of Figs.\,\ref{fig_uniform_dr_0p1} and \ref{fig_uniform_dr_10p0} that the convergence to zero of the disc-integration is less exact for $Q$ than for $U$. This is because while the quadrature points are distributed symmetrically in the individual quadrants contributing to $U$, no such symmetry applies to $Q$. We ameliorate this problem by using a large number of quadrature points, but it remains limited by computer memory storage constraints and the finite numeric precision of our computations).

The bottom panels of Figs.\,\ref{fig_uniform_dr_0p1} and \ref{fig_uniform_dr_10p0} represent the same cloud as in the top panel enveloping an oblate BD ($\eta=0.3$) viewed at an inclination angle $\theta_\mathrm{incl}=90^\circ$. 
Gravity darkening (GD), the absence of which would cause the oblate BD to have a constant effective temperature throughout its photosphere, leads to a symmetric increase in effective temperature away from the equator towards the poles. 
The modified effective temperature distribution due to GD, as shown in Fig.\,\ref{fig_GD}, is reflected in higher values of $I$, $Q$ and $U$ towards the poles and slightly lower values near the equator.

When the BD is oblate, the northern and southern quadrants identified in the disc-resolved mapping of $Q$ get ``stretched out'', while the western and eastern quadrants get ``squeezed in'', as can be seen in Figs.\,\ref{fig_uniform_dr_0p1} and \ref{fig_uniform_dr_10p0}, giving rise to finite residual values of $Q_\mathrm{da}$, with $|Q_\mathrm{da}|$ increasing with increasing oblateness. The distribution of $U$, however, still remains exactly anti-symmetrical about both the vertical axis, $\hat{\mathbf{p}}^0_\mathrm{v}$, and the horizontal axis, $\hat{\mathbf{p}}^0_\mathrm{h}$, so that $U_\mathrm{da}$ vanishes.

GD results in further accentuation of the differences in $Q$ between the northern and southern quadrants with respect to the western and eastern quadrants, leading to a larger disc-averaged value of $Q_\mathrm{da}$.  The vanishing $U_\mathrm{da}$, however, remains unchanged as exact anti-symmetry is maintained across the vertical axis  $\hat{\mathbf{p}}^0_\mathrm{v}$ (as well as the horizontal axis  $\hat{\mathbf{p}}^0_\mathrm{h}$ when viewing along the equator, i.e., $\theta_\mathrm{incl}=90^\circ$). 

Our RT modeling shows for both cloud types that GD leads to an increase not only in the absolute values of $I_\mathrm{da}$ and $Q_\mathrm{da}$, but also an increase in the disc-integrated degree-of-polarization given in this case by by $q_\mathrm{da}=Q_\mathrm{da}/I_\mathrm{da}$, since $U_\mathrm{da}=0$. However, $q_\mathrm{da}$ values in Fig.\,\ref{fig_uniform_dr_0p1} ($r_0=0.01\,\mu$m) are over 10 times larger compared to those in Fig.\,\ref{fig_uniform_dr_10p0} ($r_0=10\,\mu$m) for the oblate BD, underlining the relatively high polarizing efficiency of small scattering particles.

\subsection{Dependence on inclination angle}\label{sec_uni_incl}
 \begin{figure}[t]
\vspace*{0mm}
\begin{center}
\begin{minipage}{\textwidth}
  \centering
  \includegraphics[width=\linewidth]{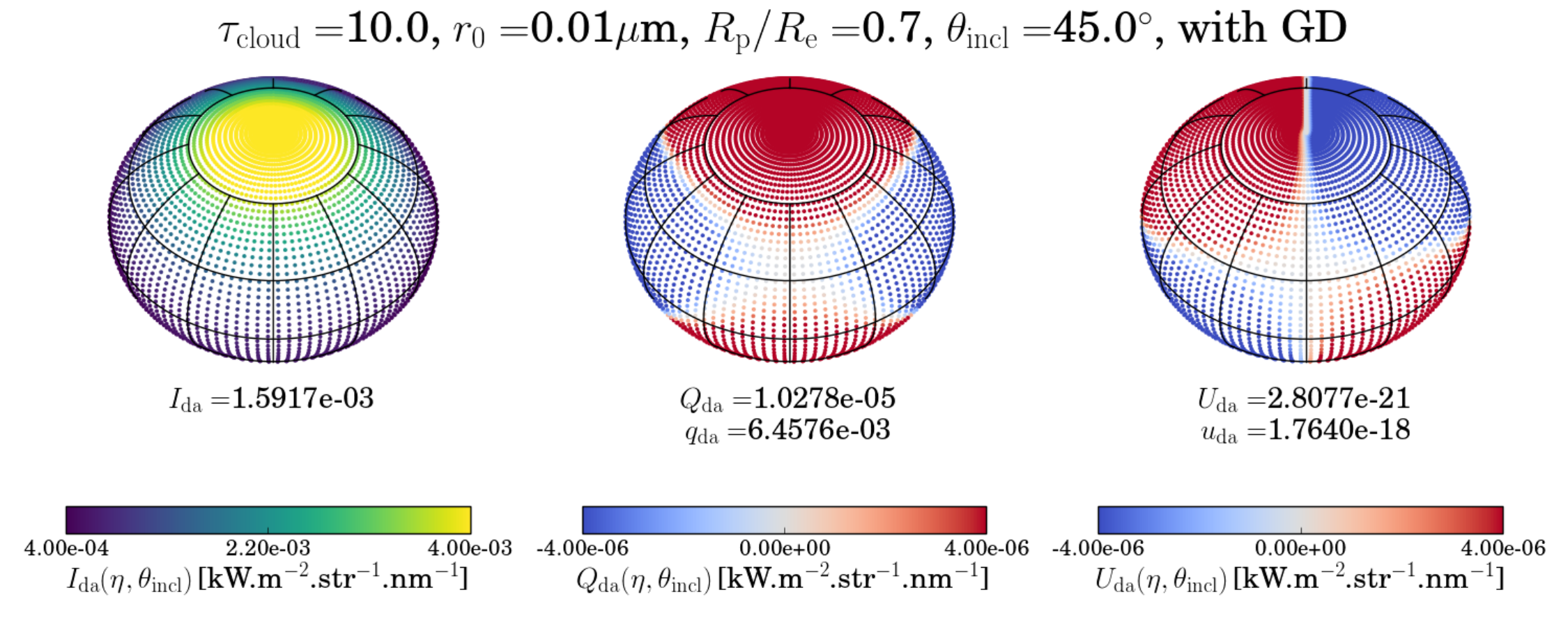}
\end{minipage}%
\begin{minipage}{\textwidth}
  \centering
  \includegraphics[width=\linewidth]{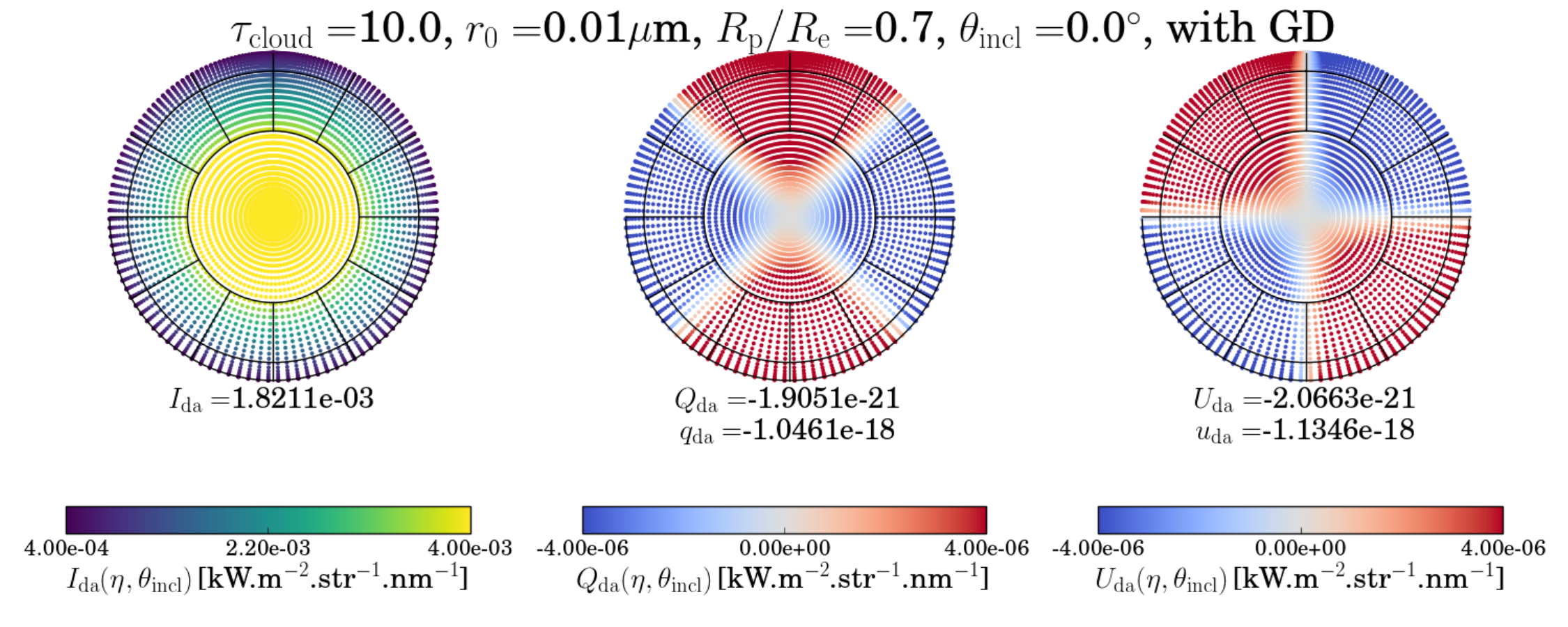}
\end{minipage}%
\end{center}
\caption{\small{{\it Top}: Same as the bottom panel of Fig.\,\ref{fig_uniform_dr_0p1}, except for an inclination of $45^\circ$ of $\hat{\mathbf{v}}$ with respect to the axis of rotation.}
\small{{\it Bottom}: Same as the bottom panel of Fig.\,\ref{fig_uniform_dr_0p1}, except for an inclination of $0^\circ$ of $\hat{\mathbf{v}}$ with respect to the axis of rotation.}}
\label{fig_uni_incl_dr_0p1}
 \end{figure}
 
  \begin{figure}[t]
\vspace*{0mm}
\begin{center}
\begin{minipage}{\textwidth}
  \centering
  \includegraphics[width=\linewidth]{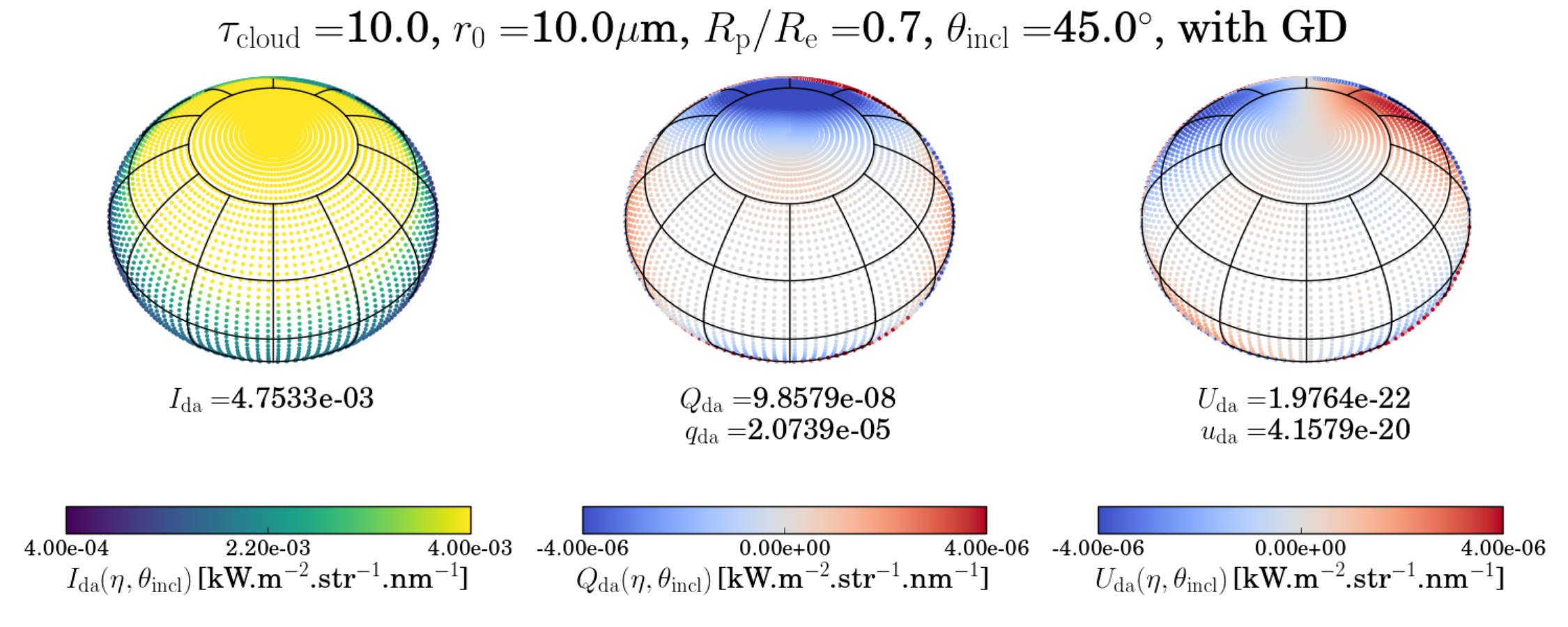}
\end{minipage}%
\begin{minipage}{\textwidth}
  \centering
  \includegraphics[width=\linewidth]{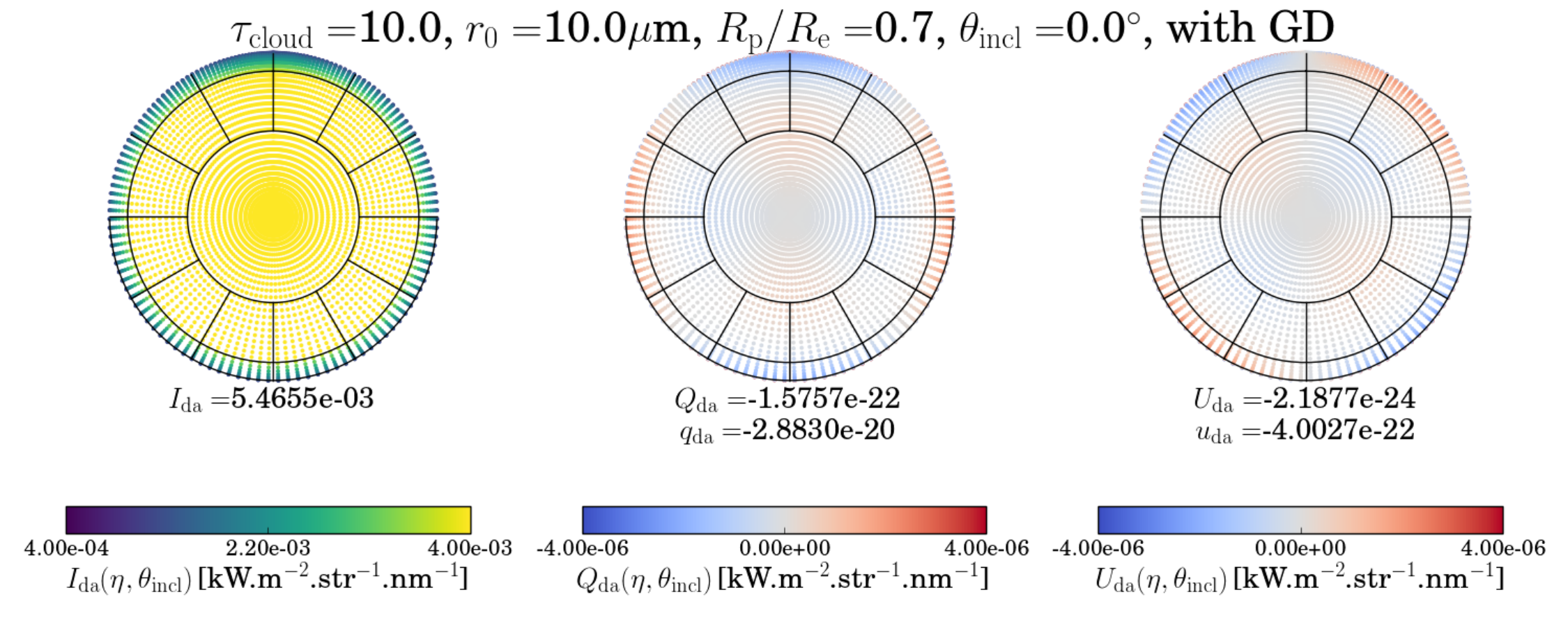}
\end{minipage}%
\end{center}
\caption{\small{{\it Top}: Same as the bottom panel of Fig.\,\ref{fig_uniform_dr_10p0}, except for an inclination of $45^\circ$ of $\hat{\mathbf{v}}$ with respect to the axis of rotation.}
\small{{\it Bottom}: Same as the bottom panel of Fig.\,\ref{fig_uniform_dr_10p0}, except for an inclination of $0^\circ$ of $\hat{\mathbf{v}}$ with respect to the axis of rotation.}}
\label{fig_uni_incl_dr_10p0}
 \end{figure}

When the viewing direction is not along the equator, i.e., $\theta_\mathrm{incl}\neq90^\circ$, there is no change in the polarization signal of { the uniformly cloudy spherical BD}, $Q_\mathrm{da}=U_\mathrm{da}=0$, owing to its perfect symmetry. However, the disc-integrated intensity $I_\mathrm{da}$ and $Q_\mathrm{da}$ vary for an oblate BD, as can be seen in the top and bottom panels of Figs.\,\ref{fig_uni_incl_dr_0p1} and \ref{fig_uni_incl_dr_10p0} representing the oblate BD of oblateness $\eta=0.3$ at inclinations $\theta_\mathrm{incl}=45^\circ$ and $0^\circ$, respectively. As in Figs.\,\ref{fig_uniform_dr_0p1} and \ref{fig_uniform_dr_10p0}, we consider clouds of the same optical thickness ($\tau_\mathrm{cloud}=10$) but small grains ($r_0=0.01\,\mu$m) in Fig.\,\ref{fig_uni_incl_dr_0p1} and large grains ($r_0=10\,\mu$m) in Fig.\,\ref{fig_uni_incl_dr_10p0}.

A change in inclination $\theta_\mathrm{incl}$ causes a redistribution of the angle $\theta_\mathrm{v,ij}$ (defined in Sec.\ref{sec_model}, Eq.\,\ref{eq1}) made by different facets of {the surface of an oblate BD} with respect to the viewing direction. As a result, the light propagating towards the observer gets redistributed, with more radiation arriving from parts where the surface normal makes a smaller angle with respect to the viewing direction (see Fig.\,\ref{fig_planeparallel}). For oblate bodies, the mean angle $\overline{\theta}_\mathrm{v,ij}$ made by the viewing direction with respect to the surface normal decreases with decreasing inclination, resulting in more forward-scattered light and hence increased disc-integrated brightness $I_\mathrm{da}$. Simultaneously, the projected oblateness of the ellipsoidal BD decreases with decreasing inclination $\theta_\mathrm{incl}$, approaching the limit of a circular disc at the poles ($\theta_\mathrm{incl}=0$). This reduced asymmetry diminishes the net polarization of the disc.  

Also, as the viewing direction moves poleward with decreasing $\theta_\mathrm{incl}$, the temperature hotspot due to GD at the nearer pole is displaced away from the limb towards the center of the BD, while that at the farther pole is removed from view to the opposite side. As a result, a smaller fraction of the observed light undergoes polarization at the limb, resulting in a smaller $q_\mathrm{da}$ with decreasing $\theta_\mathrm{incl}$, as is also evident from Fig.\,\ref{fig_GDpol}.

While there is an overall redistribution in $I$, $Q$ as well as $U$ over the BD disc, the disc-averaged quantity $U_\mathrm{da}$ remains zero as anti-symmetry in $U$ is still maintained across the western and eastern hemispheres under our choice of axes $\hat{\mathbf{p}}^0_\mathrm{v}$ and $\hat{\mathbf{p}}^0_\mathrm{h}$ in the viewing plane. 
Again, a comparison of $q_\mathrm{da}$ between Figs. \ref{fig_uni_incl_dr_0p1} and \ref{fig_uni_incl_dr_10p0} proves that for the cloud type considered, small grains are over 10 times more efficient polarizers than large grains ($r_0=10\,\mu$m), when viewed at $\theta_\mathrm{incl}=45^\circ$. Of course, at $\theta_\mathrm{incl}=0$, the projected disc becomes circular, so that its net polarization vanishes due to perfect radial anti-symmetry in both $Q$ and $U$.  

\subsection{Spatial orientation of the BD rotation axis}\label{sec_orient}
{ The reference frame $(\hat{\mathbf{p}}'_\mathrm{h},\,\hat{\mathbf{p}}'_\mathrm{v})$ in the viewing plane of the instrument observing a BD cannot, in general, be expected to be aligned with the special reference frame $(\hat{\mathbf{p}}^0_\mathrm{h},\,\hat{\mathbf{p}}^0_\mathrm{v})$  identified in Sec.\,\ref{sec_da}.}
Except in the case of $\theta_\mathrm{incl}=0^\circ$, for which every reference frame is equivalent, the Stokes vector elements corresponding to the reference frame defined by  $\hat{\mathbf{p}}^0_\mathrm{h}$ and $\hat{\mathbf{p}}^0_\mathrm{v}$ are easily obtained by using the matrix $\mathbf{L}(\alpha_0)$ (Eq.\,\ref{eq4}) to rotate the Stokes vector measured in an arbitrary instrument reference frame ($\hat{\mathbf{p}}'_\mathrm{h}$, $\hat{\mathbf{p}}'_\mathrm{v}$) by an angle $\alpha_0=-\frac{1}{2}\tan^{-1}(U'_\mathrm{da}/Q'_\mathrm{da})$, where all primed values correspond to measurements made in the instrument reference frame. Since the only requirement for $U_\mathrm{da}$ to disappear is symmetry across the eastern and western hemispheres, the above relation also holds true for every situation where the hemispheres to the left and right of the rotation axis are mirror images of each other, e.g., in the case of banded cloud structures, temperature bands, coaxial debris discs, etc.

\begin{figure}[t]
\vspace*{0mm}
\begin{center}
\includegraphics[width=13cm]{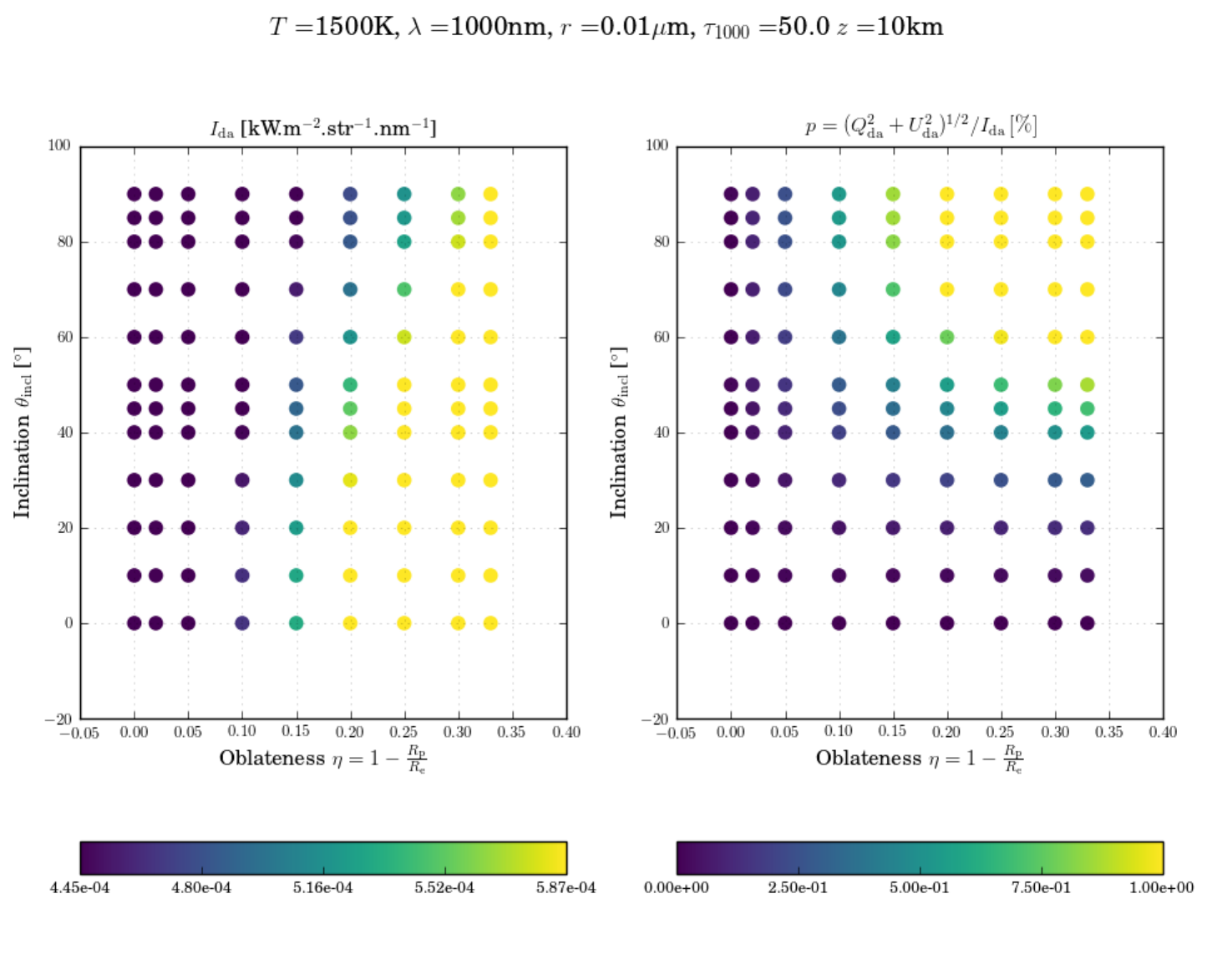}
\end{center}
\caption{\small{{\it Left:} Disc-averaged intensity $I_\mathrm{da}$ for a uniform cloud deck of optical thickness $\tau_\mathrm{cloud}=50$ at $\lambda=1000\,$nm and grain size $r=0.01\,\mu$m as a function of oblateness $\eta$ along the x-axis, and inclination $\theta_\mathrm{incl}$ of the viewing direction from the rotation axis along the y-axis.}
\small{{\it Right}: Same as the left panel, except for the disc-averaged degree-of-polarization $p_\mathrm{da}$.}}
\label{fig_map_rho_incl}
 \end{figure}

Since both the disc-averaged intensity, $I_\mathrm{da}$, and the degree-of-polarization, $p_\mathrm{da}=\sqrt{Q'^2_\mathrm{da}+U'^2_\mathrm{da}}/I'_\mathrm{da}=Q_\mathrm{da}/I_\mathrm{da}$, are mappable functions of the oblateness, $\eta$, as well as inclination, $\theta_\mathrm{incl}$, they can provide constraints on the exact orientation in space of the  axis of rotation $\hat{\bm{\omega}}_\mathrm{rot}$ ($\equiv\hat{\mathbf{k}}_\mathrm{g}$): the oblateness $\eta$ of the BD gives rise to finite values of $Q'_\mathrm{da}$ and $U'_\mathrm{da}$, which can be used to determine $\hat{\mathbf{p}}^0_\mathrm{h}$ and $\hat{\mathbf{p}}^0_\mathrm{v}$ in the viewing plane as explained above. {Since it is known that the projection of the rotation axis $\hat{\bm{\omega}}_\mathrm{rot}$ onto the viewing plane is coincident with $\hat{\mathbf{p}}^0_\mathrm{v}$, knowledge of the inclination $\theta_\mathrm{incl}$   (the polar angle) in addition to that of the angle of polarization $\alpha_0$ (the azimuthal angle) in the viewing plane completes our ability to determine the exact 3D spatial orientation of the rotation axis $\hat{\bm{\omega}}_\mathrm{rot}$.}

This bivariate dependence of the disc-integrated photopolarimetric signal is illustrated in Fig.\,\ref{fig_map_rho_incl}, for a uniform cloud deck of optical thickness $\tau_\mathrm{cloud}=50$ and a characteristic radius $r_0=0.01\,\mu$m. {This high optical thickness of a cloud consisting of very small grains has been chosen to simulate the limiting case of a semi-infinite, Rayleigh-scattering atmosphere.} $I_\mathrm{da}$ and $p_\mathrm{da}$ represented in the left and right panels, respectively, as functions of oblateness, $\eta$, along the x-axis and the inclination, $\theta_\mathrm{incl}$, along the y-axis. 

Due to its symmetry in all directions, the uniform spherical body ($\eta=0$) shows no dependence of $I_\mathrm{da}$ on the inclination, $\theta_\mathrm{incl}$, and is completely unpolarized, i.e., $p_\mathrm{da}=0$, for all $\theta_\mathrm{incl}$, as seen in the left and right panels of Fig.\,\ref{fig_map_rho_incl}, respectively. As oblateness increases along the x-axis from $\eta=0$ through $0.33$, the dependence of both $I_\mathrm{da}$ and $p_\mathrm{da}$ on $\theta_\mathrm{incl}$ becomes more pronounced. The brightness of the BD, $I_\mathrm{da}$, increases with increasing oblateness, $\eta$, at a given inclination due to the latitudinal temperature gradients caused by GD. For a fixed inclination angle, the degree-of-polarization, $p_\mathrm{da}$, also increases with increasing oblateness, as noted already in Sec.\,\ref{sec_uni_oblate}. This behavior of brightness and polarization is reflected in a monotonic increase in both $I_\mathrm{da}$ and $p_\mathrm{da}$  along the x-axis of Fig.\,\ref{fig_map_rho_incl}. The main difference is that the dependence of brightness on oblateness is strongest near  $\theta_\mathrm{incl}=0^\circ$ when the GD-induced polar hotspot is at the center of the disc, and weakest near $\theta_\mathrm{incl}=90^\circ$ when it occurs at the limb. The degree-of-polarization, $p_\mathrm{da}$, on the other hand, shows no dependence on oblateness at $\theta_\mathrm{incl}=0^\circ$ with a circular, radially symmetric projection of BD. The dependence of $p_\mathrm{da}$ on oblateness gradually grows with inclination angle, reaching a maximum when viewed along the equator, i.e., for $\theta_\mathrm{incl}=90^\circ$.
This is reflected in a monotonically decreasing brightness, $I_\mathrm{da}$, and a monotonically increasing degree-of-polarization, $p_\mathrm{da}$, along the y-axis as the inclination $\theta_\mathrm{incl}$ increases from $0^\circ$ along the poles to $90^\circ$ along the equator.

This complementary behavior of $I_\mathrm{da}$ and $p_\mathrm{da}$ with respect to the oblateness and inclination of a uniform BD could thus provide constraints {  (notwithstanding degeneracies due to variations in cloud properties as considered in Sec.\,\ref{sec_cloud})} to determine both the oblateness and the spatial orientation of the spin axis when measurements of the individual Stokes vector elements $I_\mathrm{da}$, $Q_\mathrm{da}$ and $U_\mathrm{da}$ are available. 


In the following section, we briefly examine the spectral dependence of the disc-averaged photopolarimetric signal on the optical thickness and grain size of atmospheric scatterers.

\section{Atmospheric scattering (cloud) characteristics}\label{sec_cloud}
\subsection{Wavelength dependence of scattering properties}\label{sec_uni_spec_scat}
\begin{figure}
\vspace*{0mm}
\begin{center}
\includegraphics[width=13cm]{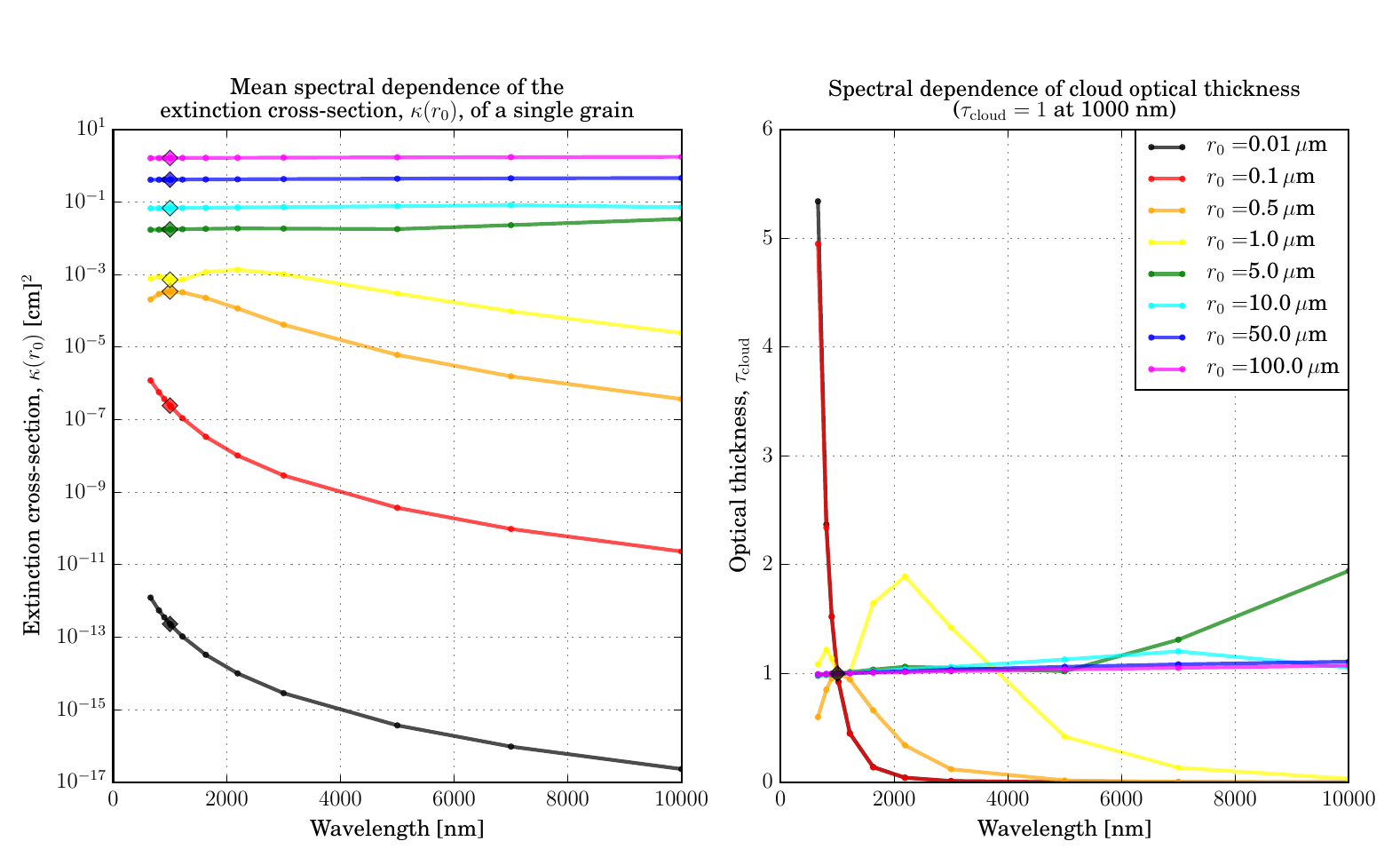}
\end{center}
\caption{\small{{\it Left:} Mean spectral dependence of the extinction cross-section $\kappa(r_0)$ of a single grain in a log-normal distribution of effective size $r_0$. {\it Right:} Spectral dependence of cloud optical thickness for different grain sizes. The optical thickness at 1000\,nm (diamond marker) is $\tau_\mathrm{cloud}=1$ for each size-distribution.}}
\label{fig_spec_AOD}
\end{figure}

\begin{figure}
\vspace*{-2mm}
\begin{center}
\includegraphics[width=0.7\textwidth]{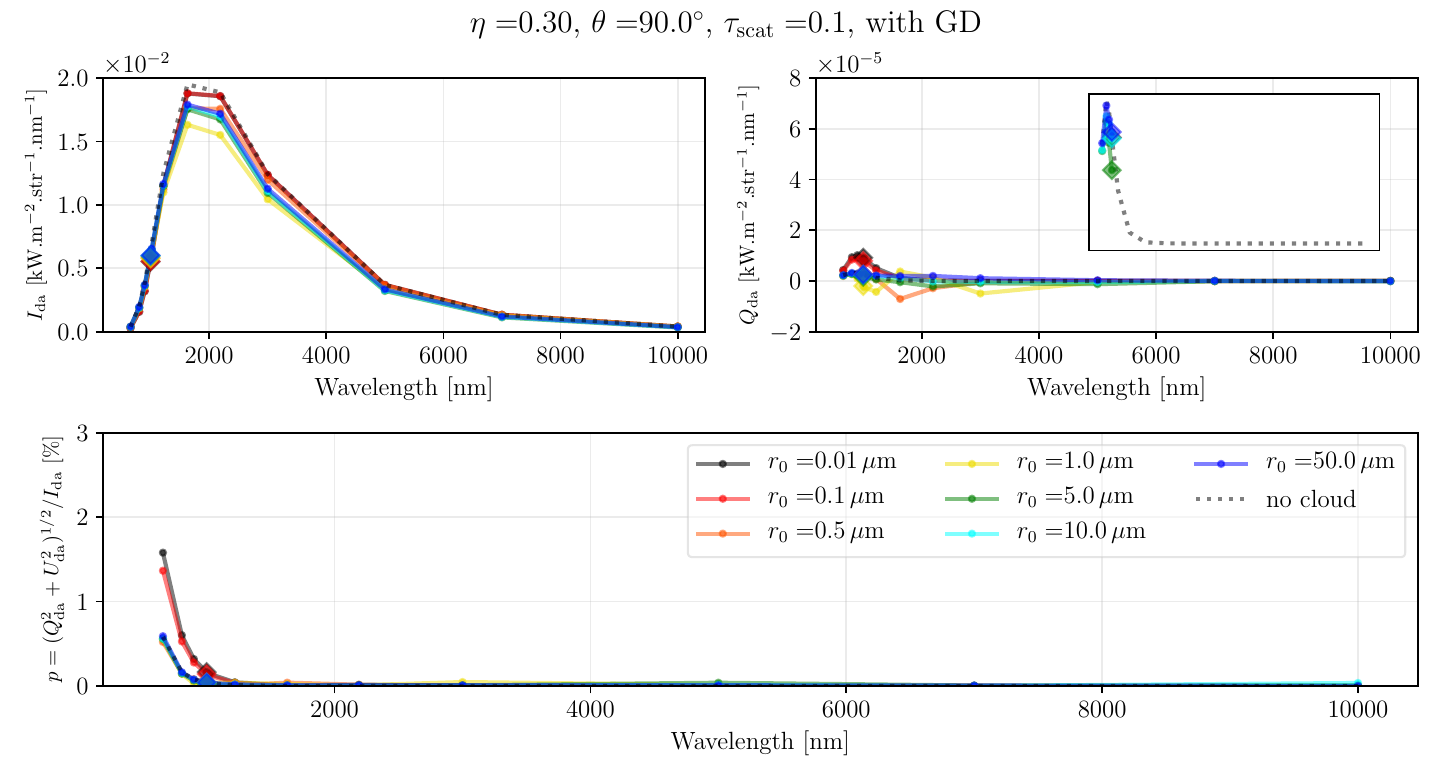}
\vspace*{-2mm}
\includegraphics[width=0.7\textwidth]{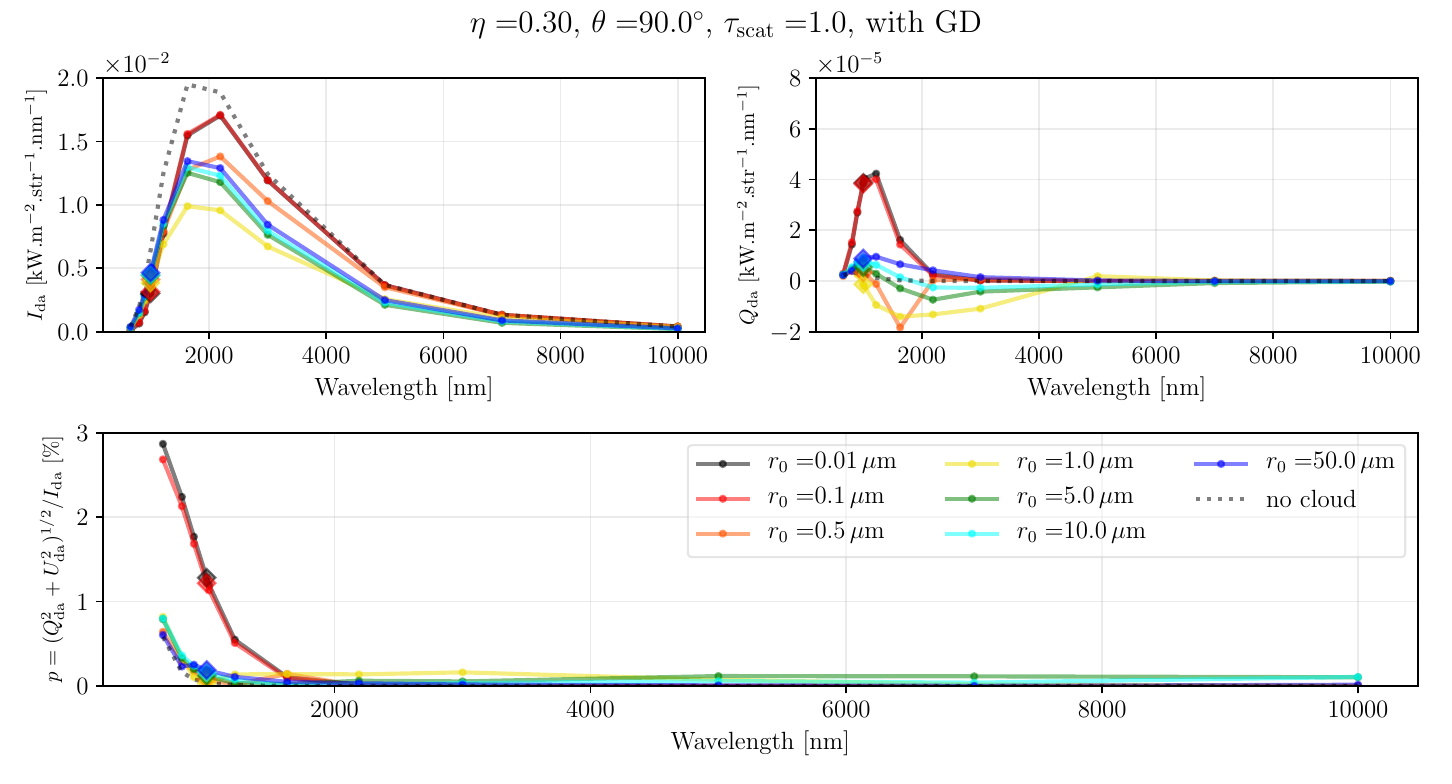}
\vspace*{-2mm}
\includegraphics[width=0.7\textwidth]{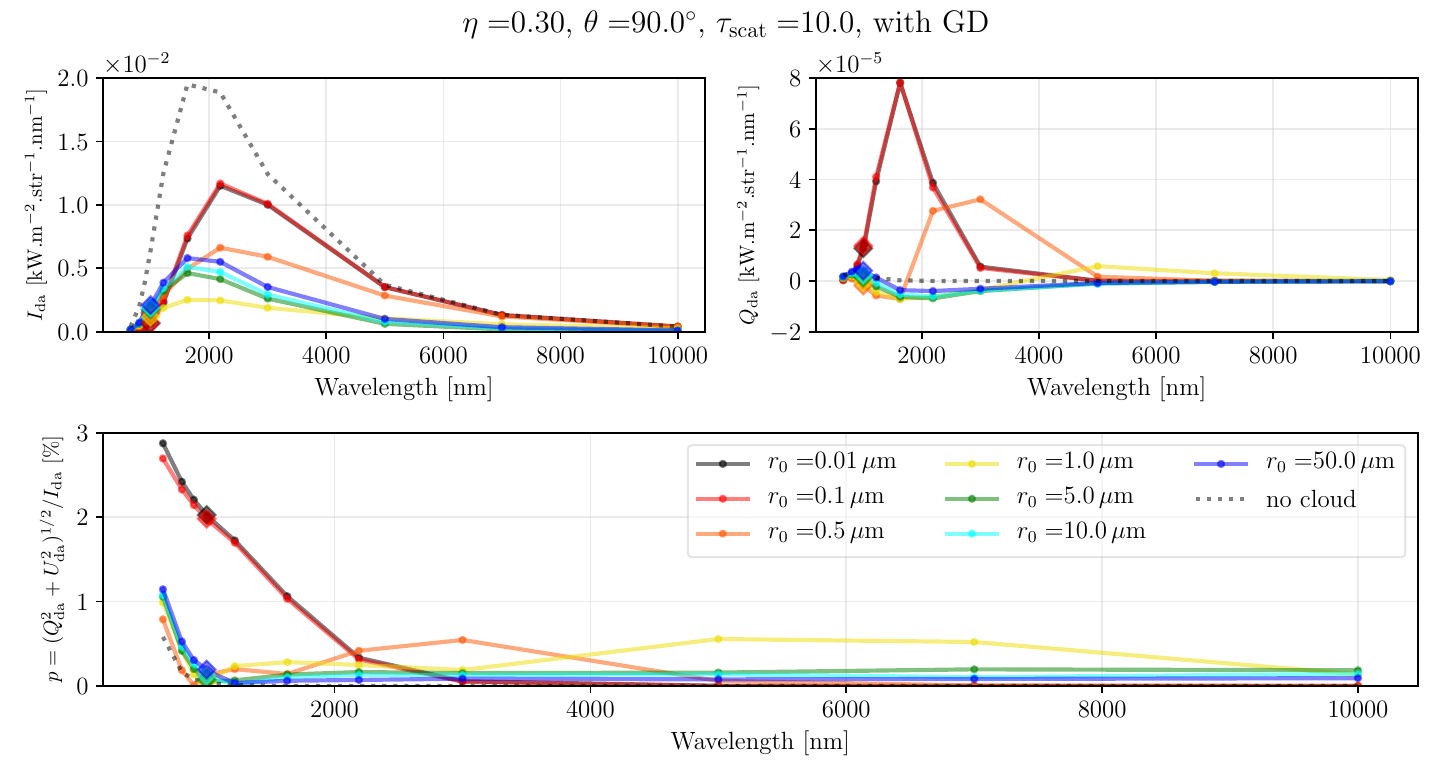}
\vspace*{-7mm}
\end{center}
\caption{\small{{\it Top}: Spectra of the disc-averaged $I_\mathrm{da}$, $Q_\mathrm{da}$ and $p_\mathrm{da}$ for a uniform cloud deck of optical thickness $\tau_\mathrm{cloud}=0.1$ at wavelength $1000\,$nm and varying grain sizes ranging between $r=0.01-300\,\mu$m. The oblateness is chosen to be $\eta=0.3$, and the inclination $\theta_\mathrm{incl}=90^\circ$. The inset shows $Q_\mathrm{da}$ for the large grains at wavelengths $\lambda\le1000\,$nm in relation to $Q_\mathrm{da}$ for purely gaseous Rayleigh scattering (for which the full spectral range up to $\lambda=10\,\mu$m is shown for reference).}
\small{{\it Middle}: Same as the top panel, except for an optical thickness $\tau_\mathrm{cloud}=1$ at $\lambda=1000\,$nm.}
\small{{\it Bottom}: Same as the top panel, except for an optical thickness $\tau_\mathrm{cloud}=10$ at $\lambda=1000\,$nm.} For clarity, $\lambda=1000\,$nm has been represented using a diamond-shaped marker.}
\label{fig_uni_spec_scat}
\end{figure}

\begin{figure}
\vspace*{0mm}
\begin{center}
\includegraphics[width=13cm]{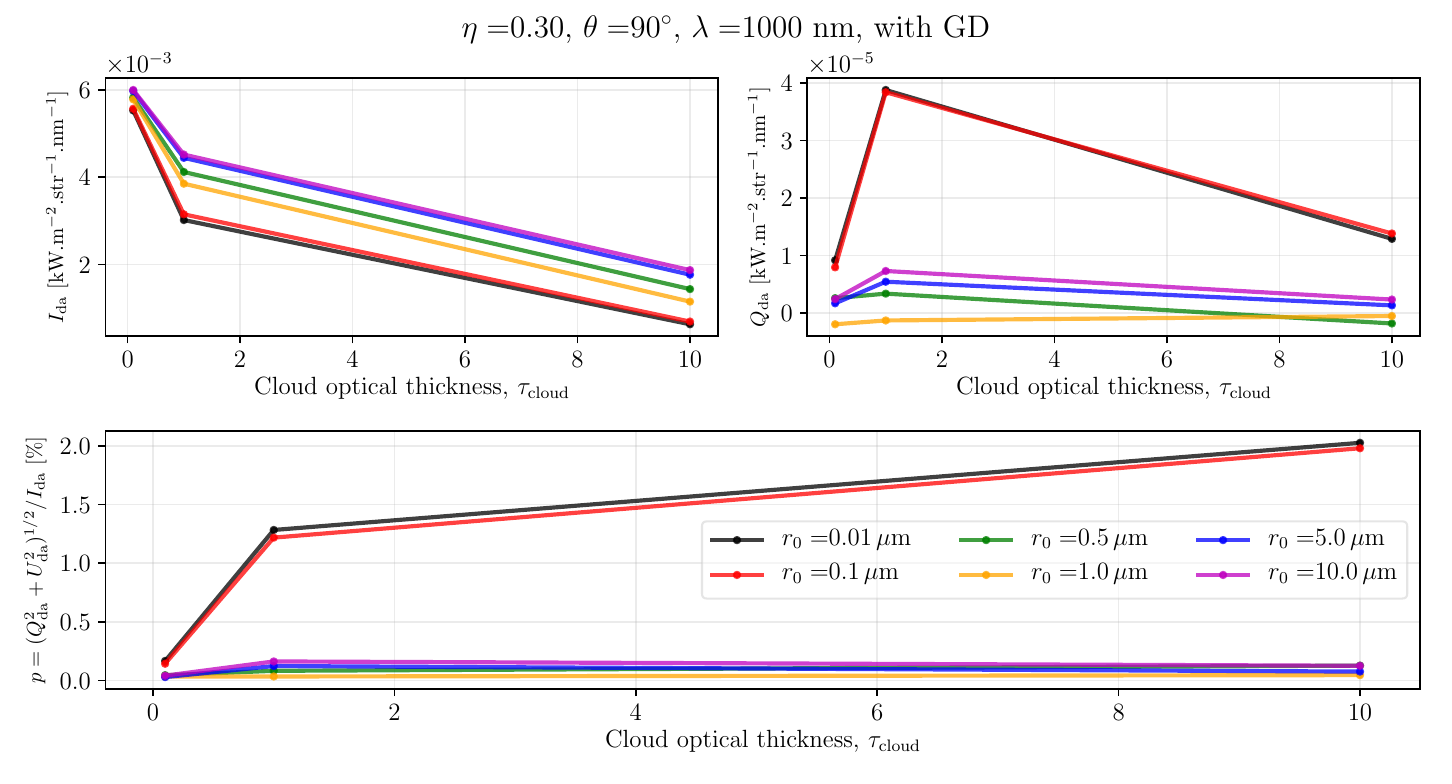}
\end{center}
\caption{\small{Dependence of the disc-integrated Stokes components $I_\mathrm{da}$ (top left panel), $Q_\mathrm{da}$ (top right panel), and the degree-of-polarization $p_\mathrm{da}$ (bottom panel) at the wavelength $\lambda=1000\,$nm on the cloud optical thickness, for different effective cloud grain sizes, $r_0$.}}
\label{fig_1000_AOD}
\end{figure}


To distinguish the polarimetric effects of atmospheric scattering properties from those of oblateness and inclination, it is useful to incorporate measurements at different wavelengths. For generality, we currently assume no spectral variation due to gaseous or particulate absorption. Since the atmosphere is purely scattering, it warrants an independence from cloud height, restricting our discussion mainly to cloud optical thickness and grain size.

We consider three cloud optical thicknesses, $\tau_\mathrm{cloud}=0.1,\,1$ and $10$ at a wavelength of 1000\,nm. In each case, we consider effective grain sizes ranging from $r_0=0.01\,\mu$m through $0.1,\,0.5,\,1,\,5,\,10$, and $50\,\mu$m for the lognormal distribution defined by Eq.\,\ref{eq6p1}. We compute the resulting photopolarimetric signal for a BD of effective temperature $T_\mathrm{eff}=1500\,$K at the central wavelengths of common filters used in the visible-NIR, viz., R(658\,nm), I (806\,nm), Z (900\,nm), Y (1020\,nm), J (1220\,nm), H (1630\,nm) and K (2190\,nm). Additionally, we have considered near- to mid-IR wavelengths of $1$, $3$, $5$, $7$ and $10\,\mu$m. 
The spectral dependence of cloud optical thickness reflects the spectral variation of the extinction cross-section, $\kappa$, for a given grain size-distribution. The mean extinction cross-section, $\kappa(r_0)$, of a single grain of effective size $r_0$ is shown in the left panel of Fig.\,\ref{fig_spec_AOD}. When $r_0\gg\lambda$, the extinction cross-section increases asymptotically with grain size according to the simple rule $\kappa\sim4\pi r^2_0$ (as seen for all large particles with $r_0\gtrsim10\,\mu$m). However, for $r_0\lesssim\lambda$, the spectral dependence is nonlinear, and can feature multiple maxima, as seen in the case of $r=1\,\mu$m at $\lambda\sim1000\,$nm and $\lambda\sim2000\,$nm. The extinction cross-section reaches an absolute maximum when the size parameter $\xi$ of the particulate distribution is close to unity, i.e., $\xi=\frac{2\pi r_0}{n_\mathrm{r}\lambda}\sim1$, as seen for $r_0=1\,\mu$m at $\lambda\sim2000\,$nm. It should be noted that the maximum at $\lambda\sim7000\,$nm for $r_0=10$ (cyan line) is not its absolute maximum, but a secondary one occurring at $\xi<1$. The spectral region where $\xi\sim1$ for small particles represented by $r_0=0.01\,\mu$m (black line) and $r_0=0.1\,\mu$m (red line) occurs at UV wavelengths. Consequently, we only see the monotonically decreasing spectral relationship of $\kappa$ at the wavelengths considered here.  

For a fixed optical thickness of $\tau_\mathrm{cloud}=1$ at $1000\,$nm, the right panel of Fig.\,\ref{fig_spec_AOD} shows the spectral dependence of cloud optical thickness for different grain sizes. It is evident that for the wavelengths considered here, the spectral dependence of the respective extinction cross-sections is reflected in spectrally flat optical thicknesses for large grains, non-monotonic behavior for grains of size $r_0=1$-$10\,\mu$m, and monotonically decreasing optical thicknesses for the smallest grains at $r_0=0.1\,\mu$m and $r_0=0.01\,\mu$m (as expected for Rayleigh scatters).

\subsection{Dependence on optical thickness}\label{sec_scat_tau}
The optical thickness of a given particulate distribution of characteristic radius, $r_0$, at a given wavelength, $\lambda$, can be expressed as
\begin{equation}\label{eqx1}
\tau(r_0,\lambda) = N_r\kappa(r_0,\lambda)\Delta z,
\end{equation}
where $N_r$ is the mean number density of the grains, $\kappa(r_0,\lambda)$ is their mean extinction cross-section, and $\Delta z$ is the vertical extent of the particulate distribution. For a given cloud species of effective grain size,  $r_0$, and vertical extent, $\Delta z$, the optical thickness at a given wavelength can only be varied by changing the number density of the cloud grains. Thus,
\begin{equation}\label{eqx1a}
\tau'(r_0,\lambda) = \frac{N'_r}{N_r}\tau(r_0,\lambda) 
\end{equation}
at all wavelengths $\lambda$ and grains sizes $r_0$.
Hence, to change the optical thickness from $\tau_\mathrm{cloud}=1$ at a given wavelength (as in Fig.\,\ref{fig_spec_AOD} at 1000\,nm) to $\tau_\mathrm{cloud}=10$ only involves increasing the number of particles by a factor of 10, all other parameters being unchanged. Obviously, reduction by an order of magnitude yields $\tau_\mathrm{cloud}=0.1$. 

Since $N_r$ is spectrally independent, the relative change in optical thickness due to a change in $N_r$ is spectrally uniform. The resulting photopolarimetric signal, however, has a nonlinear dependence on the optical thickness. This photopolarimetric response to an increasing number of particles is illustrated for different values of $r_0$ as we descend Fig.\,\ref{fig_uni_spec_scat} from an optical thickness $\tau_\mathrm{cloud}=0.1$ at 1000\,nm (indicated by a diamond-shaped marker) in the top panel through $\tau_\mathrm{cloud}=1$ in the middle to $\tau_\mathrm{cloud}=10$ in the bottom panel. We focus in this section on $\lambda=1000\,$nm (diamond-shaped marker) in order to compare the effect of increasing optical thickness for all grain sizes. {For more clarity, we show in Fig.\,\ref{fig_1000_AOD} the changes in $I_\mathrm{da}$, $Q_\mathrm{da}$, and $p_\mathrm{da}$ at 1000\,nm as the cloud optical thickness  increases from $\tau_\mathrm{cloud}=0.1$ through $\tau_\mathrm{cloud}=1$ to $\tau_\mathrm{cloud}=10$, for each of the different grain sizes considered in Fig.\,\ref{fig_uni_spec_scat}.}

Comparing the disc-averaged intensity, $I_\mathrm{da}$, down the three panels of Fig.\,\ref{fig_uni_spec_scat} shows a successive decrease with increasing optical thickness for all grain sizes, as decreasing fractions of the radiation emitted by the BD get transmitted by the increasing number of scatterers. Consequently, the BD which appears bright  at $\tau_\mathrm{cloud}=0.1$, becomes significantly fainter at $\tau_\mathrm{cloud}=10$. This is seen clearly in the top left panel of Fig.\,\ref{fig_1000_AOD}. 


In the absence of scattering, the radiation emitted by the BD is completely unpolarized. $Q_\mathrm{da}$ due to gaseous Rayleigh scattering alone (black dotted line) is large with a peak near $\lambda=1000\,$nm {(see inset in the top panel of Fig.\,\ref{fig_uni_spec_scat}). At shorter wavelengths than 1000\,nm,} the scant Rayleigh contribution accounts for practically the entire signal in $Q_\mathrm{da}$ when thin clouds (top panel) with grains larger than $r_0\sim5\,\mu$m are present.  (It is clear from our discussion in Sec.\,\ref{sec_BD} that $U_\mathrm{da}$ vanishes for all cases in the reference frame ($\hat{\mathbf{p}}^0_\mathrm{h}$, $\hat{\mathbf{p}}^0_\mathrm{v}$) chosen in the viewing plane, and is hence not shown.) $Q_\mathrm{da}$ is seen to assume both positive and negative values, showcasing the depolarizing character of larger cloud grains. 

 The absolute value of $Q_\mathrm{da}$ at 1000\,nm (see diamond marker) can be seen to be largest in the middle panel { of Fig.\,\ref{fig_uni_spec_scat} representing $\tau_\mathrm{cloud}=1$, as is also evident in the top right panel of Fig.\,\ref{fig_1000_AOD}.} The optical thickness $\tau_\mathrm{cloud}=1$ (at 1000\,nm) is optimal for high polarization: at low optical thicknesses, as illustrated by the top panel ($\tau_\mathrm{cloud}=0.1$ at 1000\,nm), there are not enough cloud grains to intercept and scatter the light emitted by the BD towards the observer. Since unscattered light constitutes the larger fraction of the measured signal, the polarized intensity $Q_\mathrm{da}$ is low for all $r_0$.
At a high optical thickness like $\tau_\mathrm{cloud}=10$, there are many scatterers along a given light path, so that multiple scattering contributes significantly to the measured signal. Multiple scattering has two effects: it causes much of the light to remain trapped in the atmosphere, reducing the total intensity of the emergent light, and hence also $Q_\mathrm{da}$. The other effect is to undo the polarization of light by the initial scattering events. This is especially true for small grains, which are more efficient scatterers, as a result of which they produce greatly diminished $Q_\mathrm{da}$ at $\lambda=1000\,$nm for $\tau_\mathrm{cloud}=10$.  When the optical thickness of scatterers is neither too high nor too low, as in the middle panel, single scattering is a more significant contributor to the measured signal, allowing for the largest values of $|Q_\mathrm{da}|$.

At $\lambda=1000\,$nm, the degree-of-polarization, $p_\mathrm{da}=|Q_\mathrm{da}|/I_\mathrm{da}$, shown in the bottom sub-panels of each panel { of Fig.\,\ref{fig_uni_spec_scat} and in the bottom panel of Fig.\,\ref{fig_1000_AOD}}, is smallest at low optical thickness ($\tau_\mathrm{cloud}=0.1$) due to small $|Q_\mathrm{da}|$ and maximum intensity $I_\mathrm{da}$. $|Q_\mathrm{da}|$ increases and the intensity $I_\mathrm{da}$ decreases as optical thickness increases to $\tau_\mathrm{cloud}=1$, so that $p_\mathrm{da}$ increases considerably (by an order of magnitude for the smallest grains). An increase in optical thickness to $\tau_\mathrm{cloud}=10$ at 1000\,nm in the bottom panel causes both $I_\mathrm{da}$ and $|Q_\mathrm{da}|$ to decrease considerably. However, the decrease with respect to the case where $\tau_\mathrm{cloud}=1$ is similar for both quantities, causing the degree-of-polarization $p_\mathrm{da}$ at all grain sizes to continue to increase, albeit more modestly than between the top two panels. {This is evident from the bottom panel of Fig.\,\ref{fig_1000_AOD}, where $p_\mathrm{da}$ can be seen to increase for small grains by an order of magnitude when the cloud optical thickness is increased from $\tau_\mathrm{cloud}=0.1$ to $\tau_\mathrm{cloud}=1$, but is less than doubled relative to the case of $\tau_\mathrm{cloud}=1$ when the optical thickness is increased to $\tau_\mathrm{cloud}=10$.}

This response of $p_\mathrm{da}$ shows that increasing fractions of radiation emerging from the BD atmosphere undergo scattering as the optical thickness of the cloud increases. Nevertheless, multiple scattering effects cause the DoP to approach a plateau at very high cloud optical thicknesses (for example, see Fig.\,\ref{fig_GDpol} representing $\tau_\mathrm{cloud}\sim50$ and small grains).

The spectral variation of the extinction coefficient is different for different grain sizes as shown in Fig.\,\ref{fig_spec_AOD}, leading to different spectral variations in their optical thickness. As the optical thickness at 1000\,nm increases, larger fractions of the wavelengths considered reach a sufficiently high optical thickness to yield significant $p_\mathrm{da}$. 
 



\subsection{Dependence on grain size}\label{sec_scat_r0}
In the previous section, we focused on the effect of increasing cloud optical thickness from the top to bottom panels in Fig.\,\ref{fig_uni_spec_scat}, equivalent to changing only the total number of cloud grains of a given size-distribution. In this section, we compare different grain sizes within each panel of Fig.\,\ref{fig_uni_spec_scat}, all of which have the same optical thickness at 1000\,nm.
The vertical extent, $\Delta z$, of the cloud remains unchanged, implying from Eq.\ref{eqx1} that the number density of grains of a given size varies inversely as their extinction cross section at 1000\,nm, so that 
\begin{equation}\label{eqx2}
N'_{r} = N_{r}\frac{\kappa(r_0,\lambda = 1000\, \mathrm{nm})}{\kappa(r'_0,\lambda=1000\, \mathrm{nm})}.
\end{equation}
At wavelengths other than 1000\,nm, the optical thicknesses diverge for different grain sizes as shown in the right panel of Fig.\,\ref{fig_spec_AOD}, as $\kappa(r_0,\lambda)$ (shown in the left panel) has a different spectral form for different grain sizes.

The sub-panels of Fig.\,\ref{fig_uni_spec_scat} representing brightness $I_\mathrm{da}$ show that clouds of characteristic radius $r_0=1\,\mu$m cause maximum darkening (yellow curve) of the BD. This is due to the fact that the extinction cross-section of the 1\,$\mu$m cloud is maximum at the BD emission peak near 2000\,nm. This maximum in extinction cross-section leads to more darkening (yellow curve) compared to all other grain sizes for all cloud optical thicknesses considered here ($\tau_\mathrm{cloud}=0.1$-$10$). For high $\tau_\mathrm{cloud}$ (especially when the cloud is not purely scattering), the darkening is sometimes sufficient to cause a dip in the spectral emission between 1000-2000\,nm, changing the overall shape of the BD spectrum. 

Extinction due to small grains is dominant at shorter wavelengths, owing to which the BD looks reddened in the NIR. Conversely, larger grains contribute more to the extinction of light at longer wavelengths, and allow more light to escape unimpeded at shorter wavelengths, making it appear bluer. This is most clearly observed in the middle and bottom panels of Fig.\,\ref{fig_uni_spec_scat} corresponding to $\tau_\mathrm{cloud}=1$ and $\tau_\mathrm{cloud}=10$ at 1000\,nm, respectively. Thus, the presence of cloud grains smaller than $r_0=1\,\mu$m causes an increasing shift of the transmitted light in the NIR towards longer wavelengths, while grains larger than $r_0=1\,\mu$m successively cause a color shift towards shorter wavelengths.  



{The top right panel of Fig.\,\ref{fig_1000_AOD} shows that $|Q_\mathrm{da}|$ is strongly determined by the optical thickness of the scattering grains.} {The size-dependent spectral variation of the optical thickness shown in the right panel of Fig.\,\ref{fig_spec_AOD} causes} $|Q_\mathrm{da}|$ for the smallest grains (black and red curves) to be dominant in the shortwave domain of the BD spectrum, as seen in all panels of Fig.\,\ref{fig_uni_spec_scat}. At longer wavelengths, their optical thickness and, thus, the ability to scatter radiation becomes negligible. The largest particles (blue curves) have a slowly increasing but essentially flat optical thickness across the spectral range considered. However, they are inefficient polarizers of scattered radiation, as a result of which $|Q_\mathrm{da}|$ mostly coincides with that due to gaseous Rayleigh scattering alone.  The orange and yellow curves representing $r_0=0.5\,\mu$m and $1\,\mu$m, respectively, have the strongest overall signal in $|Q_\mathrm{da}|$ for two reasons: they have relatively small size parameters between about $1000\,$nm-$4000\,$nm, but their optical thicknesses are still large compared to the smaller grains of size $r_0=0.01\,\mu$m or $0.1\,\mu$m. Smaller grains scatter light at wider angles than larger grains, while the latter produce more strongly forward-peaked scattering. Light scattered at wider angles, however, is more strongly polarized (especially in the case of the spherical particles considered here), causing smaller particles to be more polarizing at a given optical thickness, {as expected}. 
The peak $|Q_\mathrm{da}|$ for grains of size $r_0=0.01\,\mu$m (black curve), $r_0=0.1\,\mu$m (red curve), $r_0=0.5\,\mu$m (orange curve) and $r_0=1\,\mu$m (yellow curve) can be seen to occur at successively longer wavelengths with increasing optical thickness down the three panels of Fig.\,\ref{fig_uni_spec_scat}: as the optical thickness increases in the middle and bottom panels to $\tau_\mathrm{cloud}(\lambda=1000\,\text{nm})=1$ and 10, respectively, the optical thickness becomes large enough for a greater range of wavelengths to give rise to larger $|Q_\mathrm{da}|$. Multiple scattering, on the other hand, makes $|Q_\mathrm{da}|$ dimmer at wavelengths at which the grains have very high optical thickness ($\tau_\mathrm{cloud}\gtrsim10$). 

The bottom sub-panels depicting the degree-of-polarization $p_\mathrm{da}=|Q_\mathrm{da}|/I_\mathrm{da}$ in Fig.\,\ref{fig_uni_spec_scat} again make it clear that polarization is significant when the grain size is small compared to the wavelength ($\xi\ll1$). In the small-grain limit, $p_\mathrm{da}$ increases towards shorter wavelengths as the optical thickness of the scatterer increases (see Fig.\,\ref{fig_spec_AOD}). However, the spectral gradient of $p_\mathrm{da}$ is higher in the low optical thickness case of $\tau_\mathrm{cloud}=0.1$ (top panel) than at higher optical thicknesses corresponding to  $\tau_\mathrm{cloud}=1$ (middle panel) and $\tau_\mathrm{cloud}=10$ (bottom panel), where $p_\mathrm{da}$ in the small-particle limit can be seen to reach an asymptotic  plateau of $\sim2.03\%$ at 1000\,nm (comparable with $q_\mathrm{da}$ estimated in the bottom panel of Fig.\,\ref{fig_uniform_dr_0p1} representing a BD of oblateness $\eta=0.3$ with a cloud cover of $\tau_\mathrm{cloud}=10$ and $r_0=0.01\,$nm). That this value of $p_\mathrm{da}$ remains constant for all high optical thickness scattering cases at 1000\,nm in the small-grain limit can be verified by comparison with the green curve representing $\eta=0.3$ (at $\theta_\mathrm{incl}=90^\circ$) in the right panel of Fig.\,\ref{fig_GDpol}. 
This asymptotic limit for $p_\mathrm{da}$ depends both on wavelength, $\lambda$, and, because of GD, on the effective temperature, $T_\mathrm{eff}$, of the BD (explained in detail in Section\,\ref{sec_asym_p}).  

Rayleigh scattering at shorter (optical) wavelengths is important, constituting nearly the entire signal in $p_\mathrm{da}$ due to large particles at $\tau_\mathrm{cloud}=0.1$. However, the extinction due to these cloud grains at higher optical thickness, $\tau_\mathrm{cloud}$, (see bottom panel) reduces the transmitted intensity $I_\mathrm{da}$ of the BD with increasing optical thickness, causing $p_\mathrm{da}$ to be higher than the Rayleigh limit even at short wavelengths. 


{
\section{Preliminary comparisons with observations}\label{sec_obs}
{In the following we compare our model output with photometric and polarimetric observations in Sec.\,\ref{sec_LT} and Sec.\,\ref{sec_pol}, respectively. It is indeed our long-term goal to use develop a theoretical model using the ideas presented in the previous section to interpret actual observations. However, our current modelling results are by no means mature enough for a conclusive comparative analysis. The tentative nature of our comparisons are underlined by the fact that we have only considered the special case of a uniform cloud cover in a hypothetical atmosphere devoid of absorption. The comparisons motivate the need for further model development and will help track future model developments in terms of their ability to simulate spectropolarimetric measurements of BDs.}
\subsection{Photometric measurements: L/T transition}\label{sec_LT}
The L/T transition is defined as the region between late L-dwarfs and early T-dwarfs (roughly L8-T5) that is found to show a stark discontinuity in color-magnitude diagrams \citep{saumon2008evolution,burrows2006and}. While earlier L-dwarfs and later T-dwarfs feature J-H and J-K band reddening as they age and hence cool, BDs in the L/T transition domain demonstrate a relative brightening of the J-band.

{It is interesting to examine the relationship determined in Sec.\,\ref{sec_scat_r0} between the size of cloud grains and the spectral energy distribution (SED) of the absorber-less BD spectrum, because the blueward shift of the NIR colors in response to increasing grain size is relevant in the broader context of the L/T transition. What is the color-magnitude effect of changing merely the cloud grain size or cloud optical thickness of our hypothetical BD whose atmosphere is devoid of absorbing gaseous species? The answer to this hypothetical question helps isolate the photometric effect of cloud grains from other competing and compounding effects like changing effective temperature, surface gravity and metallicity, as shown by \cite{burrows2006and}. As we proceed through our series of papers, we will add more complexity at each stage. This will help us achieve an acceptable level of fidelity with the atmospheric state of BDs, while at the same time, allowing us to understand the individual effects of each successive phenomenon incorporated into our model. A deeper understanding of such individual effects would potentially help develop remote sensing strategies towards resolving degeneracies that restrict the interpretation of many measurement sets.} {To this end, we have refrained from assuming any prior correlations between our cloud parameters (optical thickness and grain size) and BD parameters (effective temperature, surface gravity, and metallicity), treating them all as independent free parameters.} 

By converting the intensity $I_\mathrm{da}$ as computed in the previous section into flux for a BD of Jupiter radius $R_\mathrm{Jup}$ at a distance of 10 pc, we can compute the absolute magnitudes of the simulated BDs to generate a synthetic color-magnitude diagram (CMD) shown in Fig.\,\ref{fig_CMD}. These  BDs have the full-column atmosphere defined in Section \ref{sec_atm}, and bear clouds of freely varying optical thickness ($\tau_\mathrm{cloud}=0-10$ at 1000\,nm) and grain size ($0.01\,\mu\mathrm{m}\le r_0\le50\,\mu\mathrm{m}$). BDs of effective temperature $T_\mathrm{eff}=2000\,$K, $1500\,$K, and $500\,$K have been considered to roughly represent an early L-dwarf, an L/T transition dwarf, and a late T-dwarf. Fig.\,\ref{fig_CMD} shows the absolute J-band magnitude of the BD plotted against its J-K color. We have included known BD magnitudes (grey circles) from Table 4 of \cite{burgasser2007binaries} for a general comparison.

\begin{figure}
\vspace*{0mm}
\begin{center} 
  \includegraphics[width=\linewidth]{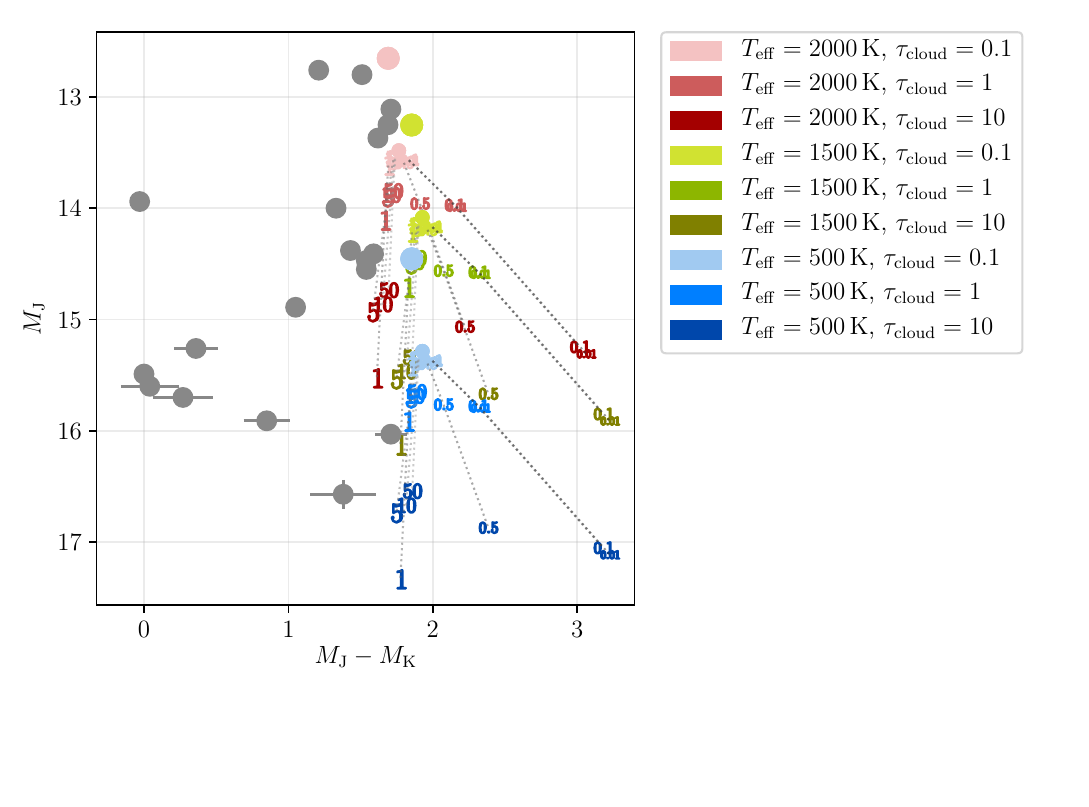}
\vspace*{-20mm}
\end{center}
\caption{\small{J-band magnitude plotted against its difference with respect to the K-band magnitude of a modeled BD devoid of atmospheric absorption. The large circles of each color denote BDs of the corresponding effective temperature without an atmosphere. The smaller circle denotes the corresponding cloudless atmosphere (full column gaseous scattering is included). Since each synthetic BD represents a grey atmosphere, these J-K colors are only caused by differences in cloud grain size. The color-coded numbers denote the cloud grain size (in microns) of the corresponding optical thickness. Dotted lines have been used to connect points representing the same grain sizes and effective temperatures but different cloud optical thickness. Observed BDs from \cite{burgasser2007binaries} are depicted as grey circles, as a reference for the actual BD measurements.}}
\label{fig_CMD}
 \end{figure}

{In our modeled BDs, our assumption of a BOA temperature of $T_\mathrm{eff}$ as defined in Sec.\,\ref{sec_model} abstracts all gaseous opacities of the BD to a reduced $T_\mathrm{BOA}$. Our assumption of a grey atmosphere further causes the gaseous opacities to lose their spectral form. The former assumption allows our modeled BDs to fall roughly in the expected brightness range represented by the y-axis, but the latter assumption causes them to lose their expected J-K colors.}
Under these assumptions, Fig.\,\ref{fig_CMD} shows that the presence of clouds in the modeled BD atmospheres causes not only a general dimming of the modeled BD in the J-band, but also a spread in the J-K colors at a given J-band brightness that is attributable to differences in cloud grain size. Comparison with the actual BD observations shows a general bias of the synthetic BDs towards larger J-K differences, clearly caused by our omission of atmospheric absorption. This bias is small for $M_\mathrm{J}\lesssim15$ typical of L-dwarfs, but for dimmer BDs near L/T transition, the difference between the synthetic and real BDs becomes remarkable. 


{It can be seen that the range $1.5\lesssim(M_\mathrm{J}-M_\mathrm{K})\lesssim2$ of Fig.\,\ref{fig_CMD} in which these L-dwarfs are found to lie is in {agreement} with the model results showcased in Fig.\,5 of \cite{burrows2006and} for L-dwarfs.  The {apparent} agreement in this range between our model results, those of \cite{burrows2006and} and the measurements \citep{burgasser2007binaries} in spite of different model assumptions {highlight one unifying factor}: that L-dwarfs are most likely to feature optically thin condensate clouds above the photosphere. This is evident in Fig.\,\ref{fig_CMD} from the paler shades (representing thin clouds of $\tau_\mathrm{cloud}=0.1$ and $\tau_\mathrm{cloud}=1$) of red  ($T_\mathrm{eff}=2000\,$K) and green ($T_\mathrm{eff}=1500\,$K) covering the range $M_\mathrm{J}\lesssim15$ of L-dwarfs. This is in  agreement with the observation of \cite{burrows2006and} {that ``although oxygen is abundant in early L-dwarfs, the elemental abundances of calcium and aluminum are not, as a result of which the first generation of clouds is rather thin, although the particle sizes in radiative zones can be small (0.5-5\,$\mu$m).  It is not until the appearance at slightly lower temperatures of condensates containing silicon, magnesium, and/or iron, with elemental abundances $\sim10$ times those of calcium and aluminum, that the areal mass density of refractory clouds can be respectable". The recent work of \cite{hiranaka2016exploring} also appears to be in agreement with such an interpretation.} Future addition of spectrally resolved gaseous opacities to our model will allow us to determine better constraints for fitting cloud parameters to simulate observations of L-dwarfs. 

L-dwarfs are largely unaffected by CH$_4$ absorption, and show only weak absorption due to H$_2$O in the NIR, while absorption in the NIR due to both H$_2$O and CH$_4$ becomes increasingly dominant towards the end of the L-dwarf sequence and down the T-dwarf sequence \citep{geballe2003transition, geballe2002toward}. 
As BDs become dimmer at the L/T transition ($15\lesssim M_\mathrm{J}\lesssim16$), a fragmentation of clouds near the photosphere, and greater variations in cloud optical thickness and grain size are expected, given that the atmospheres of these BDs experience intense cloud processes involving condensation and precipitation \citep{tsuji1996evolution, ackerman2001precipitating, allard2001limiting, burrows2006and, saumon2008evolution, cushing2008atmospheric}. 
This reveals deeper regions of the atmosphere, featuring highly pressure-broadened alkali absorption lines in the optical ($\lambda<1\,\mu$m), strong absorption due to CH$_4$ in the NIR, and collisionally induced absorption (CIA) by H$_2$ near the K-band. Given our assumption of a grey atmosphere, we cannot reproduce the relative J-band brightening characteristic of T-dwarfs. 

{However, the J-K color difference for both real and our synthetic BDs starts to increase below $M_\mathrm{J}\sim15$, albeit in opposite directions: our modeled J-K colors become redder than $(M_\mathrm{J}-M_\mathrm{K})\sim2$, while the observed colors lie blueward of it. Even so, the general trend with respect to particle size remains similar to that seen in Fig. 5 of \cite{burrows2006and}, with larger grains giving rise to bluer J-K colors.} It is thus likely that the cloud optical thickness is high and cloud grain size plays a substantial role in the observed variability of J-K colors in this domain. This has been theoretically shown by \cite{ackerman2001precipitating, burrows2006and, charnay2017self} and observationally by \cite{knapp2004near}. However, many other factors like surface gravity and atypical metallicities, also contribute to the spread in $M_\mathrm{J}-M_\mathrm{K}$. {Thus, although the cloud grain size is a parameter which can provide valuable information on the atmospheric state, {in the absence of sufficiently constraining measurements}, it can also be a source of degeneracy affecting the interpretation of cloudy scenes in the L/T transition.}

Disentangling the effects of these multiple causes may be made easier by the {synergistic} use of polarization for the identification of the grain size of atmospheric scattering species: large cloud grains would produce less polarization, while small grains would be highly polarizing.
Such considerations are examined in more detail in the next section.

The above discussion highlights the necessity of a spectrally resolved representation of absorption in BD atmospheres in conjunction with patchy clouds, which will be the focus of the next paper in this series.}

}

{
\subsection{Polarimetric measurements}\label{sec_pol}
\subsubsection{Effect of $T_\mathrm{eff}$ on the size dependence of $p_\mathrm{da}$}\label{sec_asym_p}
As noted in the previous sections, the degree-of-polarization $p_\mathrm{da}$ depends on several parameters: the oblateness and inclination of the BD, as well as the composition of its atmosphere, including clouds, which can vary in optical thickness and grain size. In addition to these, the effective temperature of an oblate BD also affects $p_\mathrm{da}$: 
Eq.\,\ref{eq_SGD} in Sec.\,\ref{sec_gravdark} revealed that the latitudinal temperature distribution on an oblate BD varies both with effective temperature (Fig.\,\ref{fig_SGD_vs_Teff}) and with wavelength (Fig.\,\ref{fig_scale_GD}).  

Considering a cloud optical thickness of $\tau_\mathrm{cloud}=10$ at 1000\,nm to represent clouds thick enough to yield the asymptotic upper limit of the disc-integrated polarization $p_\mathrm{da}$ in the NIR, and additionally assuming $\theta_\mathrm{incl}=90^\circ$ where $p_\mathrm{da}$ is maximum, we examine in this section the size-dependent spectral variation of $p_\mathrm{da}$ for BDs of different effective temperatures, viz., 2000\,K (typical of an early L-dwarf), 1500\,K (L/T transition dwarf), and 500\,K (representing a late T-dwarf). 

{Representing an early BD of $T_\mathrm{eff}=2000\,$K, the upper panel of Fig.\,\ref{fig_T2000} shows the disc-integrated polarization, $p_\mathrm{da}$, corresponding to different cloud grain sizes, viz., $r_0=0.01\,\mu$m (red),  $1\,\mu$m (green), and $10\,\mu$m (blue), and oblateness $\eta=0.33$ (solid lines), 0.3 (dashed lines), 0.2 (dash-dotted lines) and 0.1 (dotted lines).  Pure Rayleigh (cloudless) scattering computations have been represented in each case using black lines. At $p_\mathrm{cutoff}=p_0$, this represents an estimate of the full-column of gaseous scattering, thus forming an upper limit for gaseous scattering in the absence of clouds. The reference wavelength $\lambda=1000\,$nm has been depicted using diamond-shaped markers. The bottom panel shows the polarization that would be produced by the same clouds in the absence of gaseous Rayleigh scattering, representing the lower limit of the Rayleigh contribution at $p_\mathrm{cutoff}\sim0$. It can be seen that the latter case leaves the polarization signal {practically unchanged} for small-grained clouds shown in red, but {considerably modifies} the signal due to larger particles. Both the green and blue lines representing the polarization caused by clouds of respective grain sizes $r_0=1\,\mu$m and $r_0=10\,\mu$m are seen to become considerably subdued at short wavelengths relative to the full-column case. Also, the polarization signal remains relatively flat throughout the wavelength range, compared to the full-column case in the upper panel, where the polarizing effect of gaseous scattering dominates at short wavelengths. However, this gaseous contribution becomes weak enough at intermediate wavelengths ($\lambda\sim1000\,$nm for the green lines and $\lambda\sim1200\,$nm for the blue lines), to get cancelled out by the depolarizing effect of scattering by large cloud grains. Beyond this, the optical thickness of gaseous scattering ultimately becomes so negligible that no appreciable difference remains in polarization between the two cases represented by the upper and lower panels of the figure.}

Figs.\,\ref{fig_T1500} and \ref{fig_T500} similarly represent a BD in L/T transition and a late T-dwarf, respectively.  Comparing the three, it is evident that the coolest BD results in the strongest polarization for all cases considered. {We note that this result contrasts with the model results of \citep{sengupta2009multiple} who expect T-dwarf atmospheres to be largely devoid of condensates above the photosphere, and hence consider them unlikely to cause significant polarization.   \citep{sengupta2010observed} show a non-monotonic dependence with temperature for L-dwarfs. Their models consider cloud parameters in tight correlation with the effective temperature, while we treat them as free parameters. Also they account for opacities as a function of effective temperatures. Most significantly, however, \citep{sengupta2010observed} do not include GD in their computations.} {As discussed in Sec.\,\ref{sec_gravdark}, the effect of GD on the latitudinal flux distribution is nonlinear with respect to effective temperature. The flux scaling factor $S_\mathrm{GD}$ increases as $T_\mathrm{eff}$ decreases. Since BD polarization increases with increasing latitudinal differences in scattered flux, polarization increases with decreasing $T_\mathrm{eff}$ when an atmosphere is present. Wavelengths have a similar effect as temperature, so that the disc-integrated polarization of the BD is amplified more strongly at shorter wavelengths.}

This temperature-dependence of $p_\mathrm{da}$ in addition to its dependence on the grain size imparts it a characteristic spectral shape for a given BD. However, at longer wavelengths, $S_\mathrm{GD}$ continues to approach the lower limit of $S_\mathrm{GD}=1$, where the temperature-dependent divergence can be seen to subside considerably. This feature suggests that, especially at shorter optical wavelengths, a good estimate of the temperature of a BD could be afforded by polarization measurements made {in synergy with photo-/spectrometric measurements at different wavelengths}.


The following general outline can be inferred from the above discussion:
\begin{enumerate}
\item $p_\mathrm{da}$ always increases with increasing BD oblateness, $\eta$ (all other parameters remaining constant), 
\item At a given wavelength and cloud optical thickness, $p_\mathrm{da}$ becomes largest in the asymptotic limit of small particles defined by the size parameter $\xi=\frac{2\pi r_0}{n_\mathrm{r}\lambda}\rightarrow 0$,
\item For larger grains of size greater than $\sim1\,\mu$m, the relationship of $p_\mathrm{da}$ with respect to $r_0$ is nonlinear in the wavelength range considered,
\item When the atmosphere contains a mixture of small- and large-grained scattering species (as in the case of large-grained clouds with sufficient gaseous scatterers when $p_\mathrm{cutoff}=p_0$), the spectral gradient of $p_\mathrm{da}$ falls steeply at optical wavelengths, and is likely non-monotonic at longer wavelengths,
\item Due to GD, cooler BDs are more polarizing in general than hotter BDs {of a given oblateness under the same atmospheric conditions}.
\end{enumerate}
{In the following section, we consider how the above model results would affect our interpretation of actual polarization measurements of BDs. Note, again, that the following assessment is tenuous in view of our simplistic model assumptions. As we add further capabilities to our model, these interpretations will be progressively updated.}
\begin{figure}
 \centering
\vspace*{0mm}
\begin{center} 
\begin{minipage}{\textwidth}
  \centering
  \includegraphics[width=\linewidth]{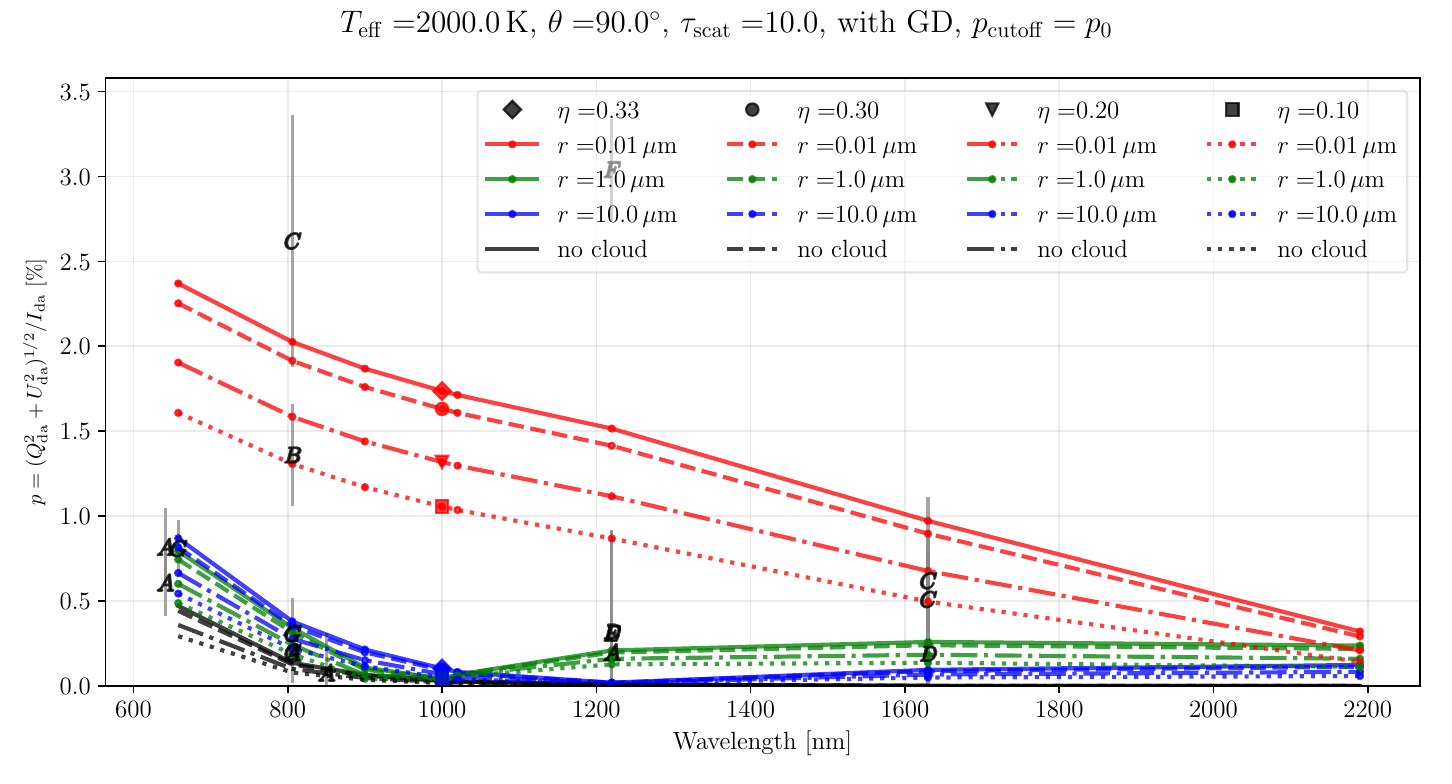}
  \end{minipage}%
\vspace*{0mm}
\begin{minipage}{\textwidth}
  \centering
  \includegraphics[width=\linewidth]{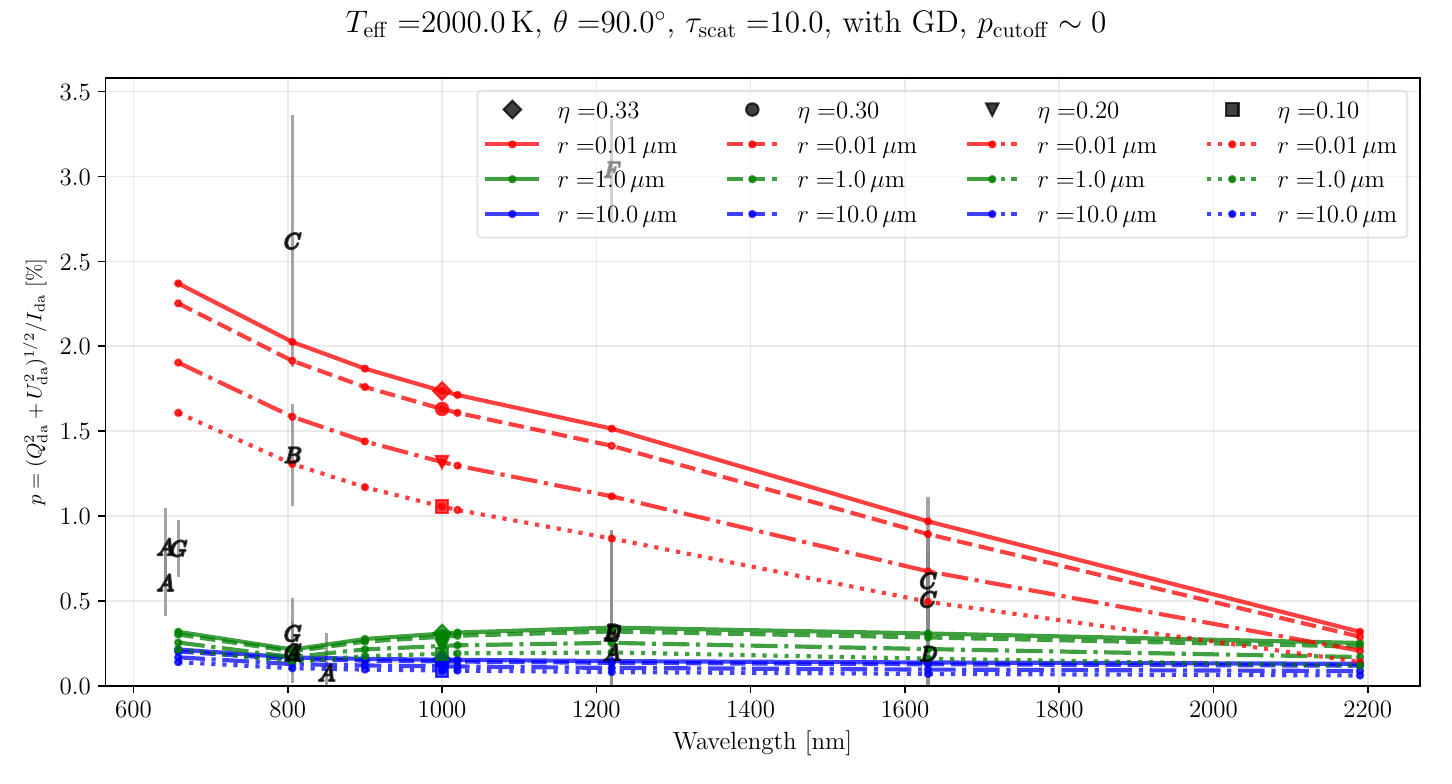}
  \end{minipage}%
\vspace*{0mm}
\end{center}
\caption{\small{Top: Spectral dependence of polarization due to a BD of $T_\mathrm{eff}=2000\,$K and $p_\mathrm{cutoff}=p_0$ observed in an edge-on geometry. The oblateness $\eta$ is varied from 0.1 (dotted lines) through 0.2 (dash-dotted lines) and 0.3 (dashed lines) to a break-up oblateness of $\eta=0.33$. The behavior of the disc-averaged DoP, $p_\mathrm{da}$, has been examined at the cloud high optical thickness of $\tau_\mathrm{cloud}=10$ at 1000\,nm with grain size varying from $r_0=0.01\,\mu$m (red lines) through 1\,$\mu$m (green lines) to 10\,$\mu$m (blue lines). Also shown is the cloudless, Rayleigh scattering atmosphere (black lines). The letters indicate the polarimetric measurements made by various observers as listed in Table \ref{tab_po}.\\
Bottom: Same as the top panel, but for an atmosphere devoid of gaseous scatters, i.e., $p_\mathrm{cutoff}\sim0$.}}
\label{fig_T2000}
 \end{figure}
\begin{figure}
 \centering
\vspace*{0mm}
\begin{center} 
\begin{minipage}{\textwidth}
  \centering
  \includegraphics[width=\linewidth]{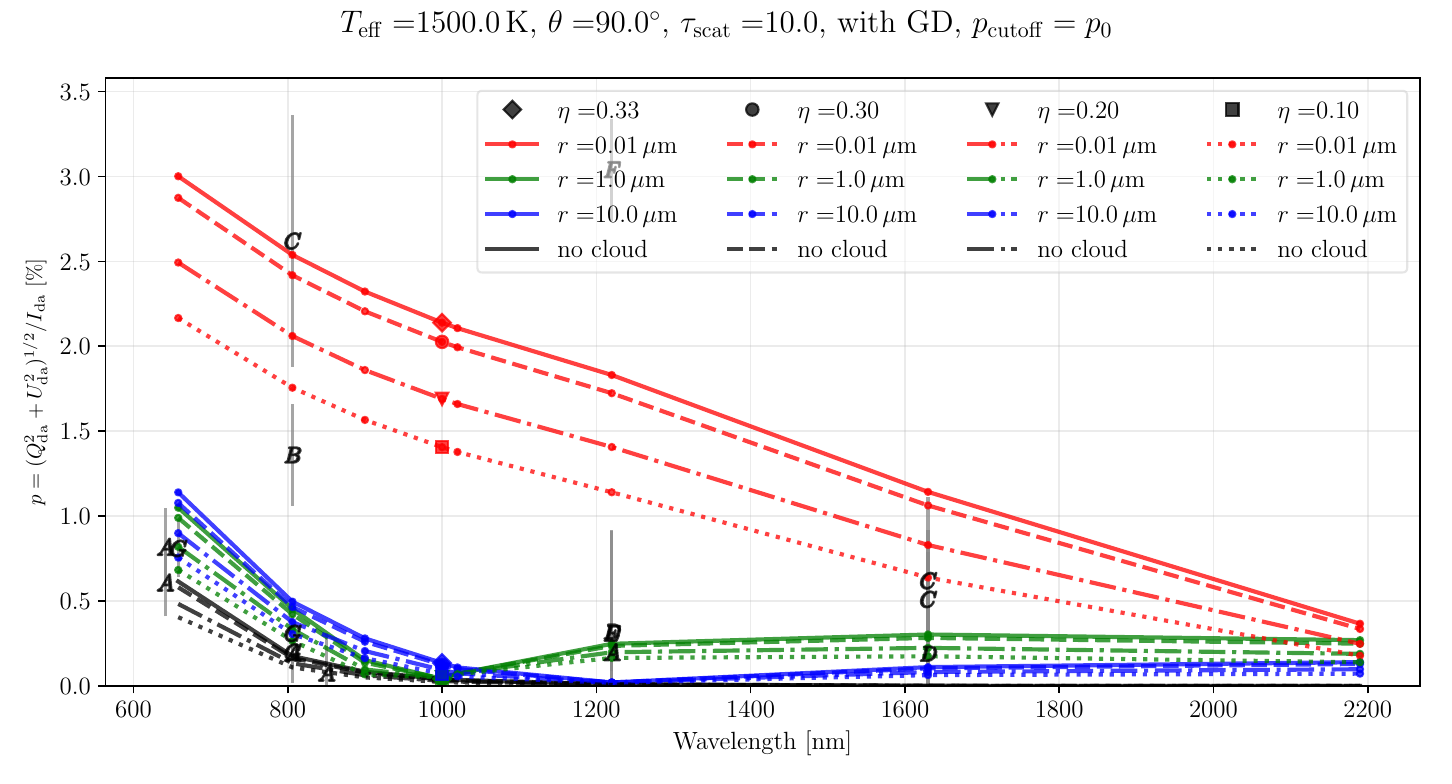}
  \end{minipage}%
\vspace*{0mm}
\begin{minipage}{\textwidth}
  \centering
  \includegraphics[width=\linewidth]{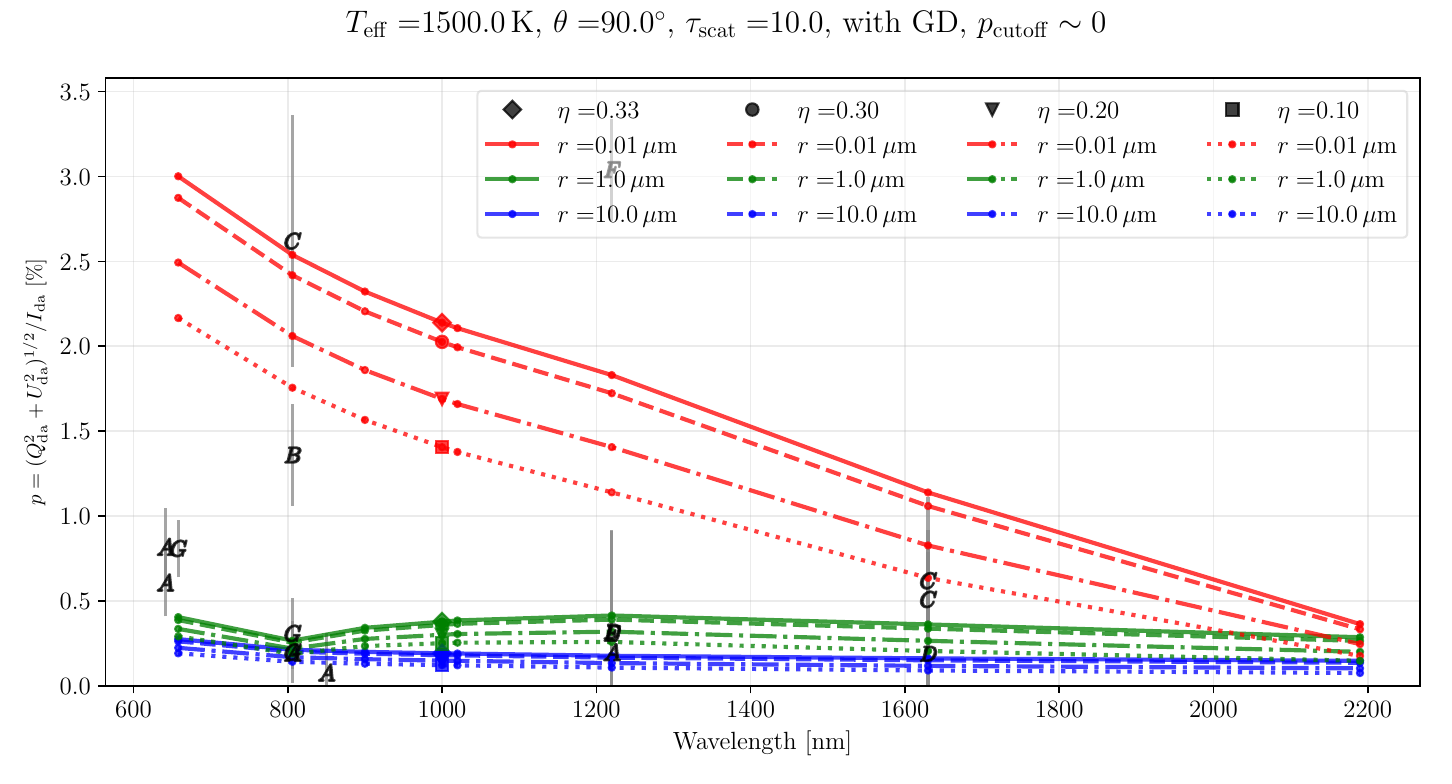}
  \end{minipage}%
\vspace*{0mm}
\end{center}
\caption{\small{Same as Fig,\,\ref{fig_T2000}, except for $T_\mathrm{eff}=1500\,$K}}
\label{fig_T1500}
 \end{figure}
 
\begin{figure}
 \centering
\vspace*{0mm}
\begin{center} 
\begin{minipage}{\textwidth}
  \centering
  \includegraphics[width=\linewidth]{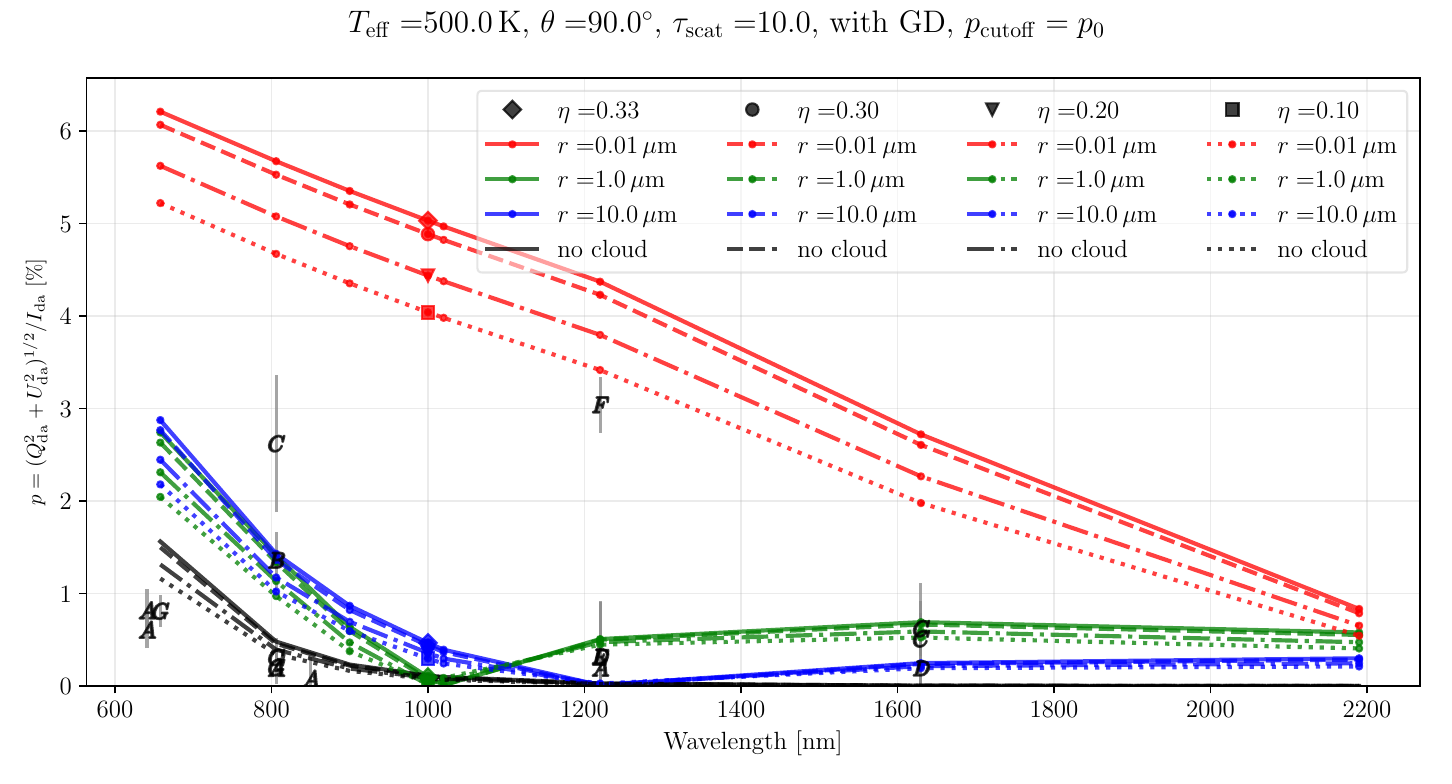}
  \end{minipage}%
\vspace*{0mm}
\begin{minipage}{\textwidth}
  \centering
  \includegraphics[width=\linewidth]{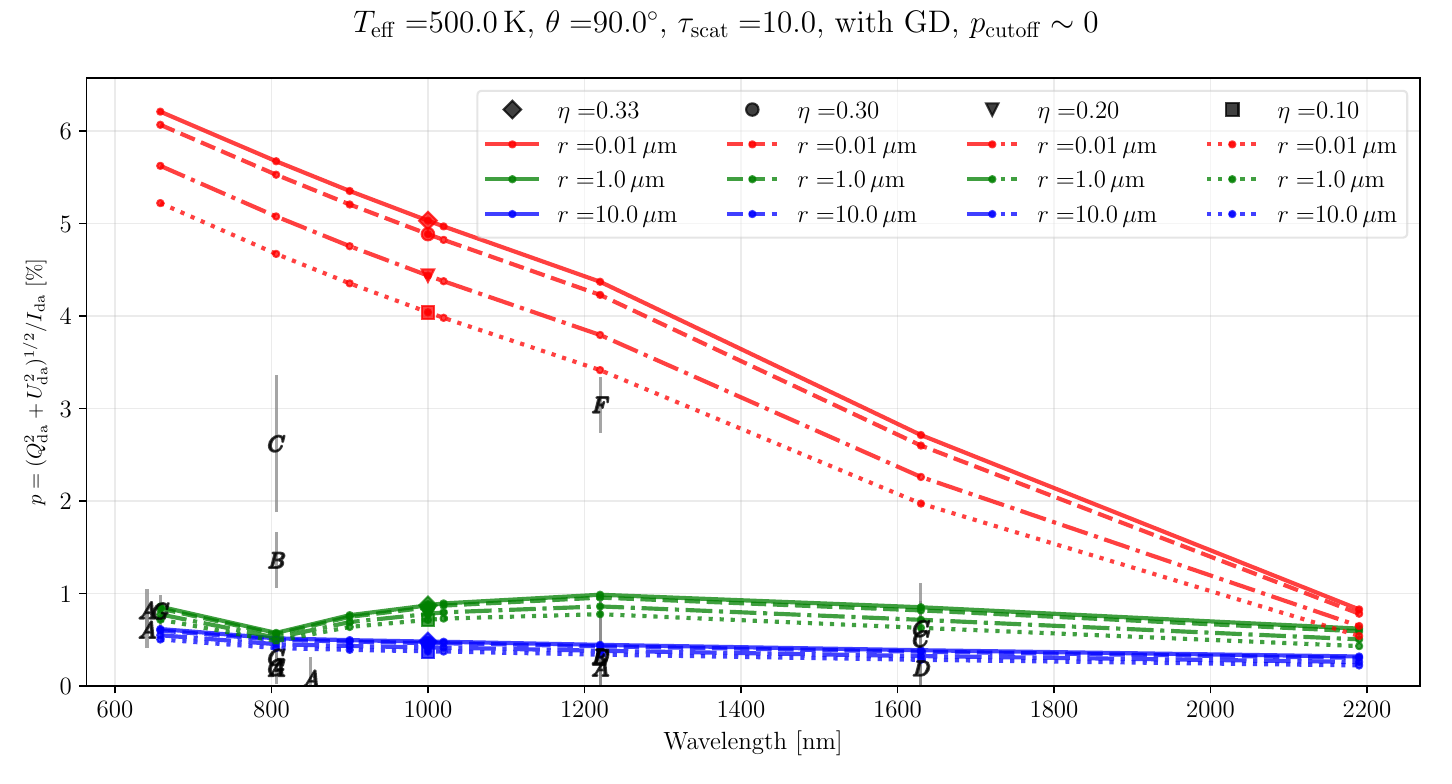}
  \end{minipage}%
\vspace*{0mm}
\end{center}
\caption{\small{Same as Fig,\,\ref{fig_T2000}, except for $T_\mathrm{eff}=500\,$K.}}
\label{fig_T500}
 \end{figure}
}
{
\subsubsection{Comparisons with observations}\label{sec_opt}
\begin{table}[h!]
  \centering
  \caption{Polarimetric observations}
  \label{tab_po}
  \begin{tabular}{c|c|c|c|c|c}
    \hline
    Label  & Name & SpT & Filter&$p_\mathrm{da}$\,[\%]&Reference\\
    \hline
A&2MASS J0036-1821 & L3.5& R  & $0.61\pm0.20$&\cite{osorio2005optical}\\
&&&R&$0.82\pm0.23$&\\
&&&I&$0.08\pm0.23$ (avgd.)&\\
&&&I&$0.197\pm0.028$&\cite{2017MNRAS.466.3184M}\\
&&&J&$0.2\pm0.11$&\\
B & 2MASS J1507-1627 & L5 & I & $1.36\pm0.30$&\cite{osorio2005optical}\\
C & 2MASS J2244+2043 & L6.5 & I & $2.62\pm0.74$ & \cite{osorio2005optical}\\
& & & H & $0.51\pm0.60$ &\cite{Zap2011ApJ}\\
& & & H & $0.62\pm0.30$ &\\
D & USco 128 & M7 & J & $0.32\pm0.60$ &\cite{Zap2011ApJ} \\
& & & H & $0.19\pm0.60$ &\\
E & USco 132 & M7 & J & $0.31\pm0.60$ &\cite{Zap2011ApJ}\\
F & 2MASS J0241-0326 & L0 & J & $3.04\pm0.30$ & \cite{Zap2011ApJ}\\
G & 2MASS J0422+1530 & M6$\gamma$ & I & $0.20\pm0.18$ &\cite{manjavacas2017testing}\\
& & & I & $0.31\pm0.21$ &\\
& & & R & $0.81\pm0.17$ &\\
    \hline
  \end{tabular}
\end{table}

{Figures\,\ref{fig_T2000}, \ref{fig_T1500}, and \ref{fig_T500} show that GD allows us to reproduce some of the extreme values of polarization observed on BDs, that could not be simulated by previous modellers. Using these figures as a reference, we discuss a sample of seven BDs for which polarization measurements are available. Given the limited selection of parameters (especially cloud optical thickness and inclination angles) represented in Figures\,\ref{fig_T2000}, \ref{fig_T1500} and \ref{fig_T500}, and in view of our simplifying model assumptions, the following  discussion is neither comprehensive not conclusive, and should be seen as exploratory speculation. As stated before, we will report enhancements of our model to achieve greater fidelity in the future, allowing us refine the following discussion to more conclusive interpretations.}

\begin{enumerate}
\item
\cite{menard2002optical} observed  seven L-dwarfs and one late M-dwarf, all having $p_\mathrm{da}<0.2\%$ (at 768\,nm using ESO/VLT+FORS1). If an assumption of uniform cloud cover may be made on the basis of low variability in these signals, then for the three cases involving $p_\mathrm{da}\sim0.2\%$, Fig.\,\ref{fig_T2000} suggests either a cloudless atmosphere for $p_\mathrm{cutoff}=p_0$ (upper panel), or clouds with large grain sizes when $p\sim 0$. However, it is likely that the inclination angle is less than $90^\circ$, which introduces significant degeneracies in the interpretation of these  measurements. {Though saddled with considerable uncertainty, their observation of increasing polarization with later spectral type appears to be in general agreement with the temperature dependence identified in Sec.\,\ref{sec_asym_p}.} 


\item
Whereas \cite{menard2002optical} observed less than 0.2\% polarization for the L-dwarf (L3.5) 2MASS J0036+1821 (A, see Table\,\ref{tab_po}) at 768\,nm, \cite{osorio2005optical} observed greater polarization ($0.61\pm0.20$\%, $0.82\pm0.23$\%) at 641\,nm, and negligible polarization at 850\,nm. {Subsequent measurements by \cite{miles2013linear,miles2016optical} reveal a linear polarization of $0.197\pm0.028$\% and $0.2\pm0.11$\% in the I and J-bands, respectively. The sharper spectral gradient between 641\,nm through 768\,nm and 850\,nm followed by finite $p_\mathrm{da}$ in the I- and J-bands suggests a mixture of large and small scattering species, similar to that seen in the upper panel of Fig.\,\ref{fig_T2000} for $r_0=1\,\mu$m. 

{The variations in measurements can be caused by the highly variable flux of 2M0036 as it rotates around its axis with a period of $3.08\pm0.05\,$h \citep{hallinan2008confirmation}. The high rotation rate favors oblateness, so that one of the primary conditions for linear polarization appears to be met (Our model studies of patchy clouds to be presented in the next part of this series indicate that oblateness has a much stronger polarizing effect than cloud patches/holes).}   

While the optical measurements cannot distinguish pure Rayleigh scattering and large-grain clouds, the measurements in the I and J-bands favor a cloudy atmosphere bearing large grains if we can assume edge-on viewing. For viewing geometries other than $\theta_\mathrm{incl}=90^\circ$ more degeneracies would need to be resolved in order to interpret the measurements. {Also, given that it is an early L-dwarf, the lower polarization may be caused by an actual cloud optical thickness that is lower than assumed in the figure ($\tau_\mathrm{cloud}=10$ at 1000\,nm).}}

\item \cite{osorio2005optical} detected {the polarization of} the L-dwarfs 2MASS J1507-1627 (L5, $p=1.36\%$ in the I-band) and 2MASS J2244+2043 (L6.5, $p=2.5\%\pm0.5\%$ in the I-band). {These two BDs have been labeled B and C (see Table\,\ref{tab_po}) for their representation in Figs. \ref{fig_T2000}, \ref{fig_T1500} and \ref{fig_T500}, respectively. 
From Fig.\,\ref{fig_T2000}, it is clear that such a high polarization across the I-band is probably produced by an optically thick, small-grained cloud for $T_\mathrm{eff}=2000\,$K. B (2M1507) can produce $p_\mathrm{da}=1.36\%$ for edge on viewing and an oblateness, $\eta$, between 0.1-0.2,  but also for higher oblateness if the inclination angle is smaller than $\theta_\mathrm{incl}=90^\circ$. However, since 2M1507 is a mid-late L-dwarf, Fig.\,\ref{fig_T1500} might describe it more accurately, suggesting either a small-grained cloud optically thinner than $\tau_\mathrm{cloud}=10$, or simply a non-edge-on viewing geometry if 2M1507 is sufficiently oblate: high rotation rates for 2M1507 have been confirmed by \cite{bailer2004spectroscopic} ($v\sin{i}\approx27\,$km/s, $T_\mathrm{rot}\approx3.5\,$h).  Larger grains or completely cloudless scenes on 2M1507 are unlikely for $T_\mathrm{eff}=2000-1500\,$K {(We note here that \cite{tata2009optical} report variability in the linear polarization of 2M1507 that is likely caused by cloud variability)}. 
Constraints provided by additional polarimetric measurements of 2M1507 in other bands could allow for more meaningful interpretation of its atmosphere. 

C (2M2244) can yield $p_\mathrm{da}=2.5\%$ in the I-band only if $T_\mathrm{eff}$ is lower than $\sim1500\,$K, under our model assumptions. Even so, this is only possible for edge-on viewing in the most oblate cases, and in the presence of sufficiently thick clouds bearing small grains. {This is in agreement with the effective temperature $T_\mathrm{eff}\sim1100\,$K ($1000\,$K-$1200\,$K) determined by \cite{stephens20090} (Table 4). {The edge-on viewing geometry is supported by the inclination angle of ${76^{+14}_{-20}}^{\circ}$ determined by \cite{vosmnras2018}.}

\cite{mclean2003nirspec} show that the low equivalent width of K I absorption on 2M2244 suggests low surface gravity, which would make it more susceptible to deformation, and hence extreme oblateness (see Fig.\,\ref{fig_rot_obl}). {This is confirmed by \cite{martin2017surface} who classify 2M2244 as a V-LG (low gravity) object.} \cite{mclean2003nirspec} show that 2M2244 exhibits normal spectral behaviour in the optical, but has an infrared spectrum in which the peak J-band flux is smaller than that in the H- or K-band. Our analysis in Sec.\,\ref{sec_scat_r0} suggests that this is likely caused by the presence above the photosphere of sufficiently thick clouds bearing grains in the size range of $r_0\sim0.01-0.5\,\mu$m, which, from the bottom panel of Fig.\,\ref{fig_uni_spec_scat} can be seen to be not only more reddening, but also more polarizing. In this respect, our conclusion contradicts that of \cite{osorio2005optical} who attribute the reddening and polarization to the presence of large-grained clouds below the photosphere.}


The measurements of \cite{osorio2005optical} for 2M2244 have been supplemented with measurements \citep{Zap2011ApJ} in the H-band. The lower polarization in the H-band for 2M2244 could suggest a slightly lower optical thickness than that assumed here. 

{Also, \cite{vosmnras2018} report a rotational period of $11\pm2\,$h that is considerably longer than the previous of $4.6\,$h by \cite{morales2006sensitive}. This adds a measure of uncertainly to our analysis: while the results of \cite{morales2006sensitive} would support an assumption of high oblateness, those of \cite{vosmnras2018} would either necessitate the V-LG categorization of \cite{martin2017surface} for sufficient oblateness or require the presence of atmospheric band structures, or a surrounding debris disk. However, we do not expect patchy clouds/holes to be responsible for the high degree-of-polarization, as will be considered in detail in the next part of this series.}}

\item {Measurements by \cite{Zap2011ApJ} of the late M-dwarf (M7) USco 128 (D, see Table\,\ref{tab_po}) in the I and J-bands seems to fit the signature of larger cloud grains at $T_\mathrm{eff}\sim2000-1500\,$K for edge-on viewing. For $\theta_\mathrm{incl}\ll90^\circ$, smaller grains could also be expected to fit this profile. However, given the early spectral type, lower polarization due to a smaller cloud optical thickness is more likely. 

{Also, USco 128 is not known to be a fast rotator \citep{Zap2011ApJ}, so that its polarization {($0.32\pm0.60$\% in the J-band and $0.19\pm0.60$\% in the H-band)} could be merely a result of a surrounding accretion disk \citep{jayawardhana2002probing, kraus2005multiplicity}.}}

{USco 132 (E, see Table\,\ref{tab_po}) observed by \cite{Zap2011ApJ} also yields a very similar polarization measurement to D in the J-band {($0.31\pm0.60$\%)}, {suggesting similar conditions as USco 128 \citep{kraus2005multiplicity}}.}

\item {The polarization of the early L-dwarf 2MASS J0241-0326 (F, see Table\,\ref{tab_po}) observed by \cite{Zap2011ApJ} in the J-band {($3.04\pm0.30$\%)} exceeds maximum values of linear polarization expected for $T_\mathrm{eff}$ between 2000\,K and 1500\,K under our model conditions, supporting the hypothesis of a surrounding dust envelope or disk \citep{Zap2011ApJ}.}

\item {Observations by \cite{manjavacas2017testing} of the M-dwarf (M6$\gamma$) 2MASS J04221530 (G, see Table\,\ref{tab_po}) in the I-band {($0.2\pm0.18$\% and $0.31\pm0.21$\%)} and R-band {($0.81\pm0.17$\%)}, coincide roughly with lines corresponding to clouds bearing large grains in Fig.\,\ref{fig_T2000}. This suggests large-grained clouds in an edge-on geometry for the assumed cloud optical thickness of $\tau_\mathrm{cloud}=10$. {However, this is unlikely for an M-dwarf, for which a plausible scenario fitting the measurements could be a cloud of low optical thickness, small grains and close to edge-on viewing.}}    
\end{enumerate}

{ Although we have plotted measurements along with our model simulations in Figs.\,\ref{fig_T2000}, \ref{fig_T1500}, and \ref{fig_T500}, measurements that are not extreme cases are {considerably more} difficult to interpret unambiguously from one plot alone. This is because multiple parameters influence the polarimetric measurement, presenting degeneracies. 

A robust interpretation of polarimetric measurements would benefit from concurrent {photometry and spectroscopy} in several different spectral bands. Given the multivariate dependence of the expected signal, a rigorous retrieval framework like MCMC or Optimal Estimation \citep{rodgers2000inverse} would be useful to quantify our estimates and their uncertainties, notwithstanding the need for a more detailed spectral representation of BD opacities in our forward model.}

}
\section{Conclusions}\label{sec_conc}
We have presented a conics-based RT scheme for computing of the disc-resolved and disc-integrated polarized infrared emission of an oblate BD or EGP. This provides an analytic basis for carrying out fast simulations of inhomogeneities like patchy clouds or temperature patterns on oblate bodies. In this part of our work, we have used this capability to model oblate BDs enveloped in a uniform, non-absorbing clouds at different orientations with respect to the observer, under the simplifying assumption of a grey, conservatively scattering atmosphere.

The dependence of the photopolarimetric signal on parameters like cloud optical thickness and grain size in addition to the oblateness of the BD and inclination of its axis has been examined. Corrections for horizontal temperature gradients due to gravitational darkening on oblate BDs are found to amplify its polarization signal significantly, {allowing us to simulate extreme levels of observed polarization that were not previously approachable, e.g., that of 2MASS J2244 in the I-band}. While both brightness and polarization increase monotonically with oblateness, an increase in inclination from the poles towards the equator causes the brightness of an oblate BD to fall while the degree-of-polarization increases continuously. We show that, when both oblateness and inclination angle can be determined, the measurement of both disc-integrated Stokes parameters $Q_\mathrm{da}$ and $U_\mathrm{da}$ can, in principle, allow an exact determination of the orientation in space of the rotation axis of the BD.

The polarization efficiency of small (Rayleigh-scattering) grains can be an order of magnitude larger than for grains of size comparable to or larger than the wavelength of observed radiation, showing that the presence of clouds is a necessary but not sufficient condition for a strong polarization signal. 

BDs have been shown to become dimmer as they get enveloped by scattering particles of increasing optical thickness. The color (which we have examined in isolation from the effects of molecular absorption) of BD emission is found to get bluer/redder as the size of the scattering particles increases/decreases, showing general agreement of our hypothetical study with most previous studies. However, opacities need to be included for conclusive comparisons with actual BDs, and will be presented in the next part of this series. 

{However, because of the extreme nature of the measurements of 2MASS J2244, and the large amount of information already available on its temperature, inclination angle, rotational period, and surface gravity, the two points noted above strongly suggest that the excess reddening of 2MASS J2244 in the J-band and its high polarization could be a consequence of thick clouds bearing grains in the size range of $r_0\sim0.01-0.5\mu$m. This contradicts Zapatero Osorio et al. (2005) who attribute both reddening and polarization to the presence of large-grained clouds below the photosphere.}

The spectral dependence of polarization on scattering properties is distinct from that of intensity-only measurements, so that photopolarimetric observations of BDs at different wavelengths can be expected to contain additional information to characterize clouds on BDs in terms of their optical thickness and grain size. Cloud grain size, while carrying valuable clues to the atmospheric state, can also be a source of degeneracy affecting the interpretation of cloudy scenes, e.g., in the L/T transition. The use of polarimetry in addition to photometric measurements would help resolve such cases: large cloud grains would produce less polarization, while small grains would be highly polarizing. {The effect of non-spherical cloud grains, which has been overlooked in our current work, will be addressed in the future, especially when cloud-grain alignment is expected}.

Our simulations show that the magnitude and spectral shape of disc-integrated polarization has a temperature dependence due to gravitational darkening, causing cooler dwarfs to be more polarizing, if oblateness and other atmospheric parameters are kept constant. 

{In follow-ups to this work, we will extend our analysis to include {spectrally resolved opacities, inhomogeneous clouds, and the latitudinal variation of the photospheric depth with temperature variations due to GD}. This will  allow us to update the discussion of Sec.\,\ref{sec_obs} with greater fidelity to the true atmospheric state of BDs. Simulations of patchiness will allow a closer examination of the temporal variability of BDs.}

\appendix
{
\section{Adiabatic atmosphere}\label{app_adiabat}
The pressure at altitude $z$ of an atmospheric column is given by
\begin{equation}\label{eqA0_1}
p(z)=\int_z^\infty \overline{\rho(z)}g\mathrm{d}z,
\end{equation}
where $\overline{\rho(z)}$ is the mean density of the atmosphere at level $z$ and $g$ is the acceleration due to gravity (assumed constant throughout the vertical extent of the atmosphere). If the molecular composition of the atmosphere is nearly constant, we can write
\begin{equation}\label{eqA0_2}
\overline{\rho(z)} = \overline{\mu}\frac{n}{V} = \overline{\mu}\frac{p(z)}{RT(z)},
\end{equation}
where $\overline{\mu}$ is the mean molar mass of the atmosphere, $\frac{n}{V}$ and $T(z)$ are, respectively, the molar density and temperature at $z$, and $R$ is the ideal gas constant. 
Substituting Eq.\,\ref{eqA0_2} in Eq.\,\ref{eqA0_1}, and taking the derivative of both sides with respect to $z$ yields
\begin{equation}\label{eqA0_3}
-\frac{\mathrm{d}p(z)}{\mathrm{d}z}=\frac{\overline{\mu}g}{R} \frac{p(z)}{T(z)}.
\end{equation}
But for an adiabatic atmosphere, 
\begin{equation}\label{eqA0_4}
T(z)=T_0\left(\frac{p(z)}{p_0}\right)^{\frac{\gamma-1}{\gamma}}.
\end{equation}
Substituting Eq.\,\ref{eqA0_4} in Eq.\,\ref{eqA0_3},  rearranging the terms and integrating yields
\begin{equation}\label{eqA0_5}
-\int_{p_0}^{p(z)}\frac{\mathrm{d}p(z)}{[p(z)]^{1/\gamma}} = \frac{\overline{\mu}g}{RT_0} p_0^{\frac{\gamma-1}{\gamma}} \int_0^z\mathrm{d}z,
\end{equation}
so that
\begin{equation}\label{eqA0_6}
p(z) = p_0\left[1-\frac{\gamma-1}{\gamma}\frac{\overline{\mu}gz}{RT_0}\right]^\frac{\gamma}{\gamma-1}.
\end{equation}
}
\section{Local surface normal of an ellipse}\label{App1}
The local normal vector $\hat{\mathbf{n}}$ on the surface of an ellipsoid generally differs from the radial vector $\hat{\mathbf{r}}=\{\cos{\mathrm{LAT}}\cos{\mathrm{LON}},\,\cos{\mathrm{LAT}}\sin{\mathrm{LON}},\,\sin{\mathrm{LAT}}\}$. Expressing the ellipsoid representing the oblate BD as
\begin{equation}\label{eqA1_1}
\frac{x^2+y^2}{a^2}+\frac{z^2}{b^2}=1,
\end{equation}
where $a=R_\mathrm{e}$ and $b=R_\mathrm{p}$, it can be seen that the surface point at $[\mathrm{LAT},\,\mathrm{LON}]$ is the intersection point of the ellipses
\begin{eqnarray}\label{eqA1_2}
\frac{x^2}{a^2}+\frac{z^2}{b^2}&=1-\cos^2{\mathrm{LAT}}\sin^2{\mathrm{LON}},\,\text{and}\\\nonumber
\frac{y^2}{a^2}+\frac{z^2}{b^2}&=1-\cos^2{\mathrm{LAT}}\cos^2{\mathrm{LON}},
\end{eqnarray}
in the $y$-$z$ and $z$-$x$ planes, respectively.
The tangents to these ellipses at $[\mathrm{LAT},\,\mathrm{LON}]$ are along
\begin{eqnarray}\label{eqA1_3}
{\mathbf{t}}_{yz}&=a^2b\sin{\mathrm{LAT}}\cdot\hat{\mathbf{j}}-ab^2\cos{\mathrm{LAT}}\sin{\mathrm{LON}}\cdot\hat{\mathbf{k}},\,\text{and}\\\nonumber
{\mathbf{t}}_{zx}&=a^2b\sin{\mathrm{LAT}}\cdot\hat{\mathbf{i}}-ab^2\cos{\mathrm{LAT}}\cos{\mathrm{LON}}\cdot\hat{\mathbf{k}}.
\end{eqnarray}
The normal vector is thus given simply by the cross product of $\mathbf{t}_{yz}$ and $\mathbf{t}_{zx}$, so that
\begin{eqnarray}\label{eqA1_4}
\hat{\mathbf{n}} &= \frac{\mathbf{t}_{yz}\times\mathbf{t}_{zx}}{|\mathbf{t}_{yz}\times\mathbf{t}_{zx}|}\\\nonumber
&=\frac{b\cos{\mathrm{LAT}}\cos{\mathrm{LON}}\cdot\hat{\mathbf{i}}
+b\cos{\mathrm{LAT}}\sin{\mathrm{LON}}\cdot\hat{\mathbf{j}}
+a\sin{\mathrm{LAT}}\cdot\hat{\mathbf{k}}}{\left[b^2+(a^2-b^2)\sin^2{\mathrm{LAT}}\right]^{1/2}}\\\nonumber
&=\frac{(1-\eta)\cos{\mathrm{LAT}}\cos{\mathrm{LON}}\cdot\hat{\mathbf{i}}
+(1-\eta)\cos{\mathrm{LAT}}\sin{\mathrm{LON}}\cdot\hat{\mathbf{j}}
+\sin{\mathrm{LAT}}\cdot\hat{\mathbf{k}}}{\left[\sin^2{\mathrm{LAT}}+(1-\eta)^2\cos^2{\mathrm{LAT}}\right]^{1/2}}.
\end{eqnarray}

According to the convention defined in Sec.\,\ref{sec_model} for the global reference frame, the longitude along the positive x-axis is $180^\circ$, whereas it has been taken to be $0^\circ$ in the discussion above. Adjusting for this convention yields the form used in Eq.\,\ref{eq0p1}.

\section{Angle of rotation between the local and global frames of reference}\label{App2}
Defining a plane by a unit vector perpendicular to it, the local reference plane containing the local surface normal, $\hat{\mathbf{n}}$, at $[\mathrm{LAT},\,\mathrm{LON}]$ and the viewing direction $\hat{\mathbf{v}}$ is given by
\begin{eqnarray}\label{eqA2_1}
\hat{\mathbf{L}} = \frac{\hat{\mathbf{n}}\times\hat{\mathbf{v}}}{|\hat{\mathbf{n}}\times\hat{\mathbf{v}}|}
\end{eqnarray}
Similarly, we define the global reference plane containing the axis of rotation $\hat{\mathbf{k}}$ and the viewing direction $\hat{\mathbf{v}}$ as
\begin{eqnarray}\label{eqA2_2}
\hat{\mathbf{G}} = \frac{\hat{\mathbf{k}}\times\hat{\mathbf{v}}}{|\hat{\mathbf{k}}\times\hat{\mathbf{v}}|}
\end{eqnarray}
Since both $\hat{\mathbf{L}}$ and $\hat{\mathbf{G}}$ are perpendicular to $\hat{\mathbf{v}}$, their vector product has the form 
\begin{eqnarray}\label{eqA2_3}
\frac{\hat{\mathbf{L}}\times\hat{\mathbf{G}}}{|\hat{\mathbf{L}}\times\hat{\mathbf{G}}|} = \sin{\beta}\hat{\mathbf{v}},
\end{eqnarray}
where $\beta$ is the angle of rotation between the two planes introduced in Section\,\ref{sec_model}.
The scalar product yields $\cos{\beta}$ as
\begin{eqnarray}\label{eqA2_4}
\frac{\hat{\mathbf{L}}\cdot\hat{\mathbf{G}}}{|\hat{\mathbf{L}}\cdot\hat{\mathbf{G}}|} = \cos{\beta}.
\end{eqnarray} 
Substituting the expressions for $\hat{\mathbf{v}}$, $\hat{\mathbf{n}}$ and $\hat{\mathbf{k}}$ from Sec.\,\ref{sec_model} in the above equations yields $\cos{\beta}$ and $\sin{\beta}$ as given by Eqns.\,\ref{eq5a}.



\section*{Acknowledgements}
Many thanks are due to three reviewers whose comments greatly improved the quality and expanded the scope of our work. We thank Prof.\,Heather Knutson for valuable discussions. S.\,S. thanks Prof.\,Yuk L. Yung and Prof.\,John Grotzinger of Caltech for their kind support. This work was partly funded by the ESI Project 01STCR, Task R.18.183.022 at JPL.
This research was carried out at the Jet Propulsion Laboratory, California Institute of Technology, under contract with NASA. Copyright 2018. All rights reserved. 

\bibliographystyle{elsarticle-harv}
\bibliography{bd_pol}
%
%



\end{document}